\documentclass[12pt]{article}
\usepackage{a4wide,epsfig,amsmath,amssymb,cite,scalefnt,graphicx}
\numberwithin{equation}{section}
\allowdisplaybreaks 
\usepackage{array}
\usepackage{epstopdf}
\usepackage{tikz}
\usetikzlibrary{intersections}
\usetikzlibrary{calc}
\usetikzlibrary{shapes}
\newcommand{\vertex}{\node[fill,circle,inner sep=0pt,minimum size=0pt]}
\newcommand{\coord}[1]{({sin(#1)},{cos(#1)})}

\def\be{\begin{equation}}
\def\ee{\end{equation}}
\def\nn{\nonumber\\}

\def\msbar{{\overline{\rm MS}}}

\def\hX{\hat X}

\def\frakk[#1#2{{{#1}\over{#2}}}
\def\go{\rightarrow}

\def\Xtil{\tilde X}

\def\dtill{\tilde d}

\def\Ncal{{\cal N}}
\def\Gcal{{\cal G}}

\def\chat{\hat c}

\def\pa{\partial}

\def\Scal{{\cal S}}
\input amssym.def
\input amssym
\def\pa{\partial}
\def\be{\begin{equation}}
\def\ee{\end{equation}}
\def\nn{\nonumber\\}

\def \D{{\cal D}}

\def \G{{\cal G}}

\def \S{{\cal S}}

\def \x{{\rm x}}
\def \y{{\rm y}}

\def\hhbet{{\hat \beta}}

\def\btil{\tilde \beta}

\def\tdelta{\tilde\delta}
\def\tepsilon{\tilde\epsilon}
\def\dhat{\hat d}
\def\os{\otimes_s}
\def\cirk{\,{\raise1pt \hbox{${\scriptscriptstyle \circ}$}}\,}

\def \olr{{\raise6.5pt\hbox{$\leftrightarrow  \! \! \! \! \!$}}}

\newcommand{\vbeta}{\node[fill=black,diamond,inner sep=3.5pt,minimum size=0pt]}
\newcommand{\vcross}{\node[draw=black,fill,line width=0.25mm,cross out,inner sep=5pt,minimum size=0pt]}

\font\ninerm=cmr9 \font\ninesy=cmsy9
\font\eightrm=cmr8 \font\sixrm=cmr6
\font\eighti=cmmi8 \font\sixi=cmmi6
\font\eightsy=cmsy8 \font\sixsy=cmsy6
\font\eightbf=cmbx8 \font\sixbf=cmbx6
\font\eightit=cmti8
\def\eightpoint{\def\rm{\fam0\eightrm}
  \textfont0=\eightrm \scriptfont0=\sixrm \scriptscriptfont0=\fiverm
  \textfont1=\eighti  \scriptfont1=\sixi  \scriptscriptfont1=\fivei
  \textfont2=\eightsy \scriptfont2=\sixsy \scriptscriptfont2=\fivesy
  \textfont3=\tenex   \scriptfont3=\tenex \scriptscriptfont3=\tenex
  \textfont\itfam=\eightit  \def\it{\fam\itfam\eightit}%
  \textfont\bffam=\eightbf  \scriptfont\bffam=\sixbf
  \scriptscriptfont\bffam=\fivebf  \def\bf{\fam\bffam\eightbf}%
  \normalbaselineskip=9pt
  \setbox\strutbox=\hbox{\vrule height7pt depth2pt width0pt}%
  \let\big=\eightbig  \normalbaselines\rm}
\catcode`@=11 %
\def\eightbig#1{{\hbox{$\textfont0=\ninerm\textfont2=\ninesy
  \left#1\vbox to6.5pt{}\right.\n@@space$}}}
\def\vfootnote#1{\insert\footins\bgroup\eightpoint
  \interlinepenalty=\interfootnotelinepenalty
  \splittopskip=\ht\strutbox %
  \splitmaxdepth=\dp\strutbox %
  \leftskip=0pt \rightskip=0pt \spaceskip=0pt \xspaceskip=0pt
  \textindent{#1}\footstrut\futurelet\next\fo@t}
\catcode`@=12 %
\def\today{\number\day\ \ifcase\month\or January\or February\or March\or
April\or May\or June\or July\or
August\or September\or October\or November\or December\fi, \number\year}

\input epsf

\begin{document}

\begin{titlepage}
\begin{flushright}
LTH1168\\
$\text{CP}^3$-Origins-2018-24 DNRF90\\
\end{flushright}
\date{}
\vspace*{3mm}

\begin{center}
{\Huge Scheme invariants in $\phi^4$ theory in four dimensions}\\[12mm]
{\bf  I.~Jack${}^a$ and 
C.~Poole${}^{a,b}$}\\

\vspace{10mm}

${}^a$Dept. of Mathematical Sciences,
University of Liverpool, Liverpool L69 3BX, UK\\
\vspace{5mm}
${}^b$CP${}^3$ Origins, University of Southern Denmark, Campusvej 55, 5320 Odense M, Denmark
\vspace{10mm}

{\it E-mail:} {\tt dij@liverpool.ac.uk}, {\tt c.poole@cp3.sdu.dk}

\end{center}

\vspace{3mm}

\begin{abstract}
We provide an analysis of the structure of renormalisation scheme invariants for the case of $\phi^4$ theory, relevant in four dimensions. We give a complete discussion of the invariants up to four loops and include some partial results at five loops, showing that there are considerably more invariants than one might naively have expected. We also show that  one-vertex reducible contributions may consistently be omitted in a well-defined class of schemes
which of course includes $\msbar$. 
\end{abstract}

\vfill

\end{titlepage}

\section{Introduction}
Beyond leading order it is well-known that the values of $\beta$-function coefficients are scheme-dependent, i.e. depend on the renormalisation scheme. On the other hand one would expect that statements with physical meaning should be expressible in a scheme-independent way. A notable recent example is the issue of the existence of an $a$-function; i.e. a function which generates the $\beta$-functions through a gradient-flow equation. For this to be feasible, the $\beta$-function coefficients must satisfy a set of consistency conditions, which must clearly be scheme-invariant; as has been verified for various field theories in three\cite{JJP, asusy, jp1}, four\cite{jp2} and six\cite{gap} dimensions. The number of scheme-independent combinations at each loop order would naively be expected to be given by the difference of the number of $\beta$-function coefficients and the number of independent variations of coefficients; however the number of independent invariants actually found is considerably larger. This may be understood in a pragmatic way in terms of the structure of the expressions for the scheme changes of the coefficients; however a possibly deeper insight is afforded by Hopf algebra considerations. A general discussion of scheme dependence with a particular focus on one-particle reducible (1PR) structures was recently given in Ref.~\cite{jo1}, and here the study of scheme-invariant combinations was initiated with reference to the $\Ncal=1$ scalar-fermion theory. The present paper is to be seen as a companion to a forthcoming article\cite{jo} where the ideas of scheme invariance and the relation to Hopf algebra will be explored in general and also exemplified for the case of $\phi^3$ theory in six dimensions; our purpose here is to extend the discussion to 
$\phi^4$ theory in four dimensions.  We shall summarise results of Ref.~\cite{jo} where necessary to render the present discussions self-contained. An additional complication in $\phi^4$ theory is due to the existence of one-vertex reducible (1VR) graphs. These are one-particle irreducible (1PI) graphs which may be separated into two distinct portions by severing a vertex. They have no simple poles when using minimal subtraction and dimensional regularisation, and hence a vanishing $\beta$-function coefficient in this scheme. It would be convenient to be able to omit these coefficients from our considerations. Indeed we shall show that although we may if desired include such coefficients, we may also consistently confine our attention to a well-defined subset of schemes in which these coefficients are absent.

The structure of the paper is as follows: in Section 2 we introduce the $\phi^4$ theory and give the results at one, two and three loops. Section 3 contains our main results, namely the full set of four-loop scheme invariants and a partial five-loop calculation. In Section 4 we show that one may straightforwardly restrict attention to a set of renormalisation schemes in which 1VR contributions are absent. In Section 5 we set our results for scheme invariants within the Hopf algebra framework. We summarise our results and give pointers to future work in the Conclusion. Some general theory which is developed in detail in Ref.~\cite{jo} and which underpins our work is summarised in Appendix A. Appendix B lists some Hopf algebraic cocommutative coproducts which arise in Section 5 but were too complex for inclusion in the main text. Finally, in Appendix C 
we show how to express scheme changes in terms of differential operators acting on the $\beta$-function coefficients.

\section{One, two and three loop calculations}
In this section we establish our notation and obtain the invariants up to three loop order (the first non-trivial case for $\phi^4$ theory). We consider the action 
\be
S=\int d^dx\left(\tfrac12\pa_{\mu}\phi^i\pa^{\mu}\phi^i - \tfrac12m^2\phi^i\phi^i -\tfrac{1}{4!}g_{ijkl}\phi^i\phi^j\phi^k\phi^l\right).
\ee
for the case $d=4$ which corresponds to a renormalisable theory.
The anomalous dimension $\gamma_{ij}$ may be expressed as a series of two-point 1PI diagrams with 4-point vertices connected by internal lines representing the contractions of couplings. Up to three loops we have
\be 
         2\gamma= d_2\tikz[baseline=(vert_cent.base)]{
  \node (vert_cent) {\hspace{-13pt}$\phantom{-}$};
 \draw (-0.4,0)--(0.1,0);
\draw (-0.4,0)--(1.3,0);
    \draw (0.7,0) ++(0:0.6cm and 0.6cm) arc (0:180:0.6cm and 0.6cm) node(n1) {}
             (0.7,0) ++(180:0.6cm and 0.6cm) arc (180:360:0.6cm and 0.6cm) node(n2){};
             \draw (1.3,0) -- (1.8,0); 
        }
+d_3\tikz[scale=0.6,baseline=(vert_cent.base),scale=1]{
	\node (vert_cent) {\hspace{-13pt}$\phantom{-}$};
	\draw (-1.5,0)--(-1,0) (1,0)--(1.5,0);
	\draw (0,0) circle[radius=1cm];
	\vertex at \coord{0} (A) {};
	\draw [bend right=60] (-1,0) to (A) [bend right=60] (A) to  (1,0);
}+\ldots,
\label{twogam}
\ee
where here and elsewhere we suppress indices as far as possible. We consistently neglect contributions from ``snail'' diagrams in which a bubble is attached to a propagator. Such contributions do not arise in minimal subtraction and will not be generated by redefinitions if the redefinitions themselves do not include such diagrams.
The $\beta$-function $\beta_{ijkl}$ may then be decomposed into 1PI pieces together with one-particle reducible  pieces determined by the anomalous dimension, in the form:
\be
 \beta=\btil+\Scal_4 \tikz[baseline=(vert_cent.base)]{
  \node (vert_cent) {\hspace{-13pt}$\phantom{-}$};
 \draw (-1.4,0.3)--(-1,0);
\draw (-1.4,-0.3)--(-1,0);
\draw (-1.4,0)--(-0.6,0);
 \draw (-0.3,0) circle [radius=0.3cm];
\draw (0,0)--(0.3,0);
\draw (-0.3,0) node {$\gamma$}
        }
\ee
with $\btil$ denoting the 1PI contributions and $\S_4$  the sum over the four terms where $\gamma$ is attached to each external line. Up to three loops the contributions to $\btil$ are given by
\begin{align}
          \btil^{(1)}=&c_1\Scal_3\tikz[baseline=(vert_cent.base)]{
  \node (vert_cent) {\hspace{-13pt}$\phantom{-}$};
 \draw (-0.4,0.3)--(0.1,0);
\draw (-0.4,-0.3)--(0.1,0);
    \draw (0.7,0) ++(0:0.6cm and 0.4cm) arc (0:180:0.6cm and 0.4cm) node(n1) {}
             (0.7,0) ++(180:0.6cm and 0.4cm) arc (180:360:0.6cm and 0.4cm) node(n2){};
             \draw (1.3,0)--(1.8,0.3); 
 \draw (1.3,0)--(1.8,-0.3); 
        },\nn
         \btil^{(2)}=&c_2\Scal_6\tikz[baseline=(vert_cent.base)]{
  \node (vert_cent) {\hspace{-13pt}$\phantom{-}$};
 \draw (-0.5,0.3)--(0,0);
\draw (-0.5,-0.3)--(0,0);
 \draw (0,0)--(1.4,0.7);
\draw (0,0)--(1.4,-0.7);
    \draw (1,0) ++(0:0.2cm and 0.5cm) arc (0:180:0.2cm and 0.5cm) node(n1) {}
             (1,0) ++(180:0.2cm and 0.5cm) arc (180:360:0.2cm and 0.5cm) node(n2){};
        }
+c_{2R}\Scal_3 \tikz[baseline=(vert_cent.base)]{
  \node (vert_cent) {\hspace{-13pt}$\phantom{-}$};
 \draw (-0.4,0.3)--(0.1,0);
\draw (-0.4,-0.3)--(0.1,0);
    \draw (0.7,0) ++(0:0.6cm and 0.4cm) arc (0:180:0.6cm and 0.4cm) node(n1) {}
             (0.7,0) ++(180:0.6cm and 0.4cm) arc (180:360:0.6cm and 0.4cm) node(n2){};
 \node (vert_cent) {\hspace{-13pt}$\phantom{-}$};
    \draw (1.9,0) ++(0:0.6cm and 0.4cm) arc (0:180:0.6cm and 0.4cm) node(n1) {}
             (1.9,0) ++(180:0.6cm and 0.4cm) arc (180:360:0.6cm and 0.4cm) node(n2){};
             \draw (2.5,0)--(3.0,0.3); 
 \draw (2.5,0)--(3.0,-0.3); 
        },\nn
     \btil^{(3)}=&
 \Scal_3\left(c_{3a}\tikz[baseline=(vert_cent.base)]{
  \node (vert_cent) {\hspace{-13pt}$\phantom{-}$};
\draw (-0.2,0.3)--(0.1,0);
\draw (-0.2,-0.3)--(0.1,0); 
    \draw (0.7,0) ++(0:0.6cm and 0.4cm) arc (0:180:0.6cm and 0.4cm) node(n1) {}
             (0.7,0) ++(180:0.6cm and 0.4cm) arc (180:360:0.6cm and 0.4cm) node(n2){};
 \draw (0.7,0) ++(0:0.1cm and 0.4cm) arc (0:180:0.1cm and 0.4cm) node(n1) {}
             (0.7,0) ++(180:0.1cm and 0.4cm) arc (180:360:0.1cm and 0.4cm) node(n2){};
    \draw (1.3,0)--(1.5,0.3); 
 \draw (1.3,0)--(1.5,-0.3);          
        }
+c_{3b}\tikz[baseline=(vert_cent.base)]{
  \node (vert_cent) {\hspace{-13pt}$\phantom{-}$};
 \draw (-0.4,0.3)--(0.1,0);
\draw (-0.4,-0.3)--(0.1,0);
    \draw (0.7,0) ++(0:0.6cm and 0.4cm) arc (0:180:0.6cm and 0.4cm) node(n1) {}
         (0.7,0) ++(180:0.6cm) arc (180:360:0.6cm and 0.4cm) node(n2){};
       \draw (1.3,0)--(1.8,0.3); 
 \draw (1.3,0)--(1.8,-0.3);      
\draw (0.7,0.4) circle [radius=0.3cm];
        }\right)
+\Scal_6\left(c_{3c} \tikz[baseline=(vert_cent.base)]{
  \node (vert_cent) {\hspace{-13pt}$\phantom{-}$};
 \draw (-0.5,0.3)--(0,0);
\draw (-0.5,-0.3)--(0,0);
 \draw (0,0)--(1.4,0.7);
\draw (0,0)--(1.4,-0.7);
    \draw (1,0.25) ++(0:0.1cm and 0.25cm) arc (0:180:0.1cm and 0.25cm) node(n1) {}
             (1,0.25) ++(180:0.1cm and 0.25cm) arc (180:360:0.1cm and 0.25cm) node(n2){};
 \draw (1,-0.25) ++(0:0.1cm and 0.5cm) arc (0:180:0.1cm and 0.25cm) node(n1) {}
             (1,-0.25) ++(180:0.1cm and 0.5cm) arc (180:360:0.1cm and 0.25cm) node(n2){};
        }
+c_{3d}\tikz[baseline=(vert_cent.base)]{
  \node (vert_cent) {\hspace{-13pt}$\phantom{-}$};
 \draw (-0.2,0.5)--(0.1,0.5);
\draw (-0.2,-0.5)--(0.1,-0.5);
    \draw (0.7,0.5) ++(0:0.6cm and 0.2cm) arc (0:180:0.6cm and 0.2cm) node(n1) {}
             (0.7,0.5) ++(180:0.6cm and 0.2cm) arc (180:360:0.6cm and 0.2cm) node(n2){};
 \draw (0.7,-0.5) ++(0:0.6cm and 0.2cm) arc (0:180:0.6cm and 0.2cm) node(n1) {}
             (0.7,-0.5) ++(180:0.6cm and 0.2cm) arc (180:360:0.6cm and 0.2cm) node(n2){};
\draw (0.1,0.5)--(0.1,-0.5);
\draw (1.3,0.5)--(1.3,-0.5);
 \draw (1.3,0.5)--(1.5,0.5);
\draw (1.3,-0.5)--(1.5,-0.5);
        }\,\right)\nn
&+c_{3e}\Scal_{12}\tikz[baseline=(vert_cent.base)]{
  \node (vert_cent) {\hspace{-13pt}$\phantom{-}$};
\draw (-0.4,0.3)--(0.1,0);
\draw (-0.4,-0.3)--(0.1,0); 
    \draw (0.7,0) ++(0:0.6cm and 0.4cm) arc (0:50:0.6cm and 0.4cm) node(n1) {}
     (0.7,0) ++(50:0.6cm and 0.4cm) arc (50:130:0.6cm and 0.4cm)  node(n2){}
             (0.7,0) ++(130:0.6cm and 0.4cm) arc (130:250:0.6cm and 0.4cm) 
     (0.7,0) ++(310:0.6cm and 0.4cm) arc (310:250:0.6cm and 0.4cm)  node(n3){}
 (0.7,0) ++(360:0.6cm and 0.4cm) arc (360:310:0.6cm and 0.4cm) node(n4) {}
(n1.base) to [out= 215, in = 325] (n2.base);
\draw (n2.base)--(n4.base);
\draw (n1.base)--(1.6,0.3);
\draw (n4.base)--(1.6,-0.3);
}
+c_{3f}\tikz[baseline=(vert_cent.base)]{
\node (vert_cent) {\hspace{-13pt}$\phantom{-}$};
\draw (-0.7,-0.7) -- (-0.05,-0.05);
\draw (0.05,0.05) -- (0.7,0.7);
\draw (-0.7,0.7) -- (0.7,-0.7);
\draw (-0.5,-0.5) -- (0.5,-0.5);
\draw (-0.5,-0.5) -- (-0.5,0.5);
\draw (-0.5,0.5) -- (0.5,0.5);
\draw (0.5,-0.5) -- (0.5,0.5);
}
+c_{3aR} \Scal_3    \tikz[baseline=(vert_cent.base)]{
  \node (vert_cent) {\hspace{-13pt}$\phantom{-}$};
 \draw (-0.2,0.3)--(0.1,0);
\draw (-0.2,-0.3)--(0.1,0);
    \draw (0.5,0) ++(0:0.4cm and 0.4cm) arc (0:180:0.4cm and 0.4cm) node(n1) {}
             (0.5,0) ++(180:0.4cm and 0.4cm) arc (180:360:0.4cm and 0.4cm) node(n2){};
     \draw (1.3,0) ++(0:0.4cm and 0.4cm) arc (0:180:0.4cm and 0.4cm) node(n1) {}
             (1.3,0) ++(180:0.4cm and 0.4cm) arc (180:360:0.4cm and 0.4cm) node(n2){};
\draw (2.1,0) ++(0:0.4cm and 0.4cm) arc (0:180:0.4cm and 0.4cm) node(n1) {}
             (2.1,0) ++(180:0.4cm and 0.4cm) arc (180:360:0.4cm and 0.4cm) node(n2){};
             \draw (2.5,0)--(2.7,0.3); 
 \draw (2.5,0)--(2.7,-0.3); 
        }\nn
&    +c_{3bR}\Scal_6 \tikz[baseline=(vert_cent.base)]{
  \node (vert_cent) {\hspace{-13pt}$\phantom{-}$};
 \draw (-1.2,0.3)--(-1,0);
\draw (-1.2,-0.3)--(-1,0);
 \draw (-0.5,0) circle [radius=0.5cm];
 \draw (0,0)--(1.3,0.65);
\draw (0,0)--(1.3,-0.65);
    \draw (1,0) ++(0:0.2cm and 0.5cm) arc (0:180:0.2cm and 0.5cm) node(n1) {}
             (1,0) ++(180:0.2cm and 1cm) arc (180:360:0.2cm and 0.5cm) node(n2){};
        }.
\label{twobet}
\end{align}
For later convenience we introduce the notation that $g_{3a}^{\lambda}$ is the graph corresponding to $c_{3a}$, and $g_{2}^{\gamma}$ is the graph corresponding to $d_2$, etc. We note that in Eq.~\eqref{twobet} the graph $g_{3f}^{\lambda}$ is primitive in that it has no divergent subgraph.

Changes of renormalisation scheme are well-known to be equivalent to redefinitions of the coupling, which may be parametrised as\cite{jo1}
\be
g'_{ijkl}=(g+f(g))_{mnpq}C_{mi}C_{nj}C_{pk}C_{ql}
\label{gprime}
\ee
where 
\be
C(g)=(1-2c(g))^{-\tfrac12}.
\label{Cdef}
\ee
 After a scheme change the $\beta$-function and anomalous dimension are represented by a similar diagrammatic series, but with modified coefficients given by 
\be
c_X\go c_X'=c_X+\delta c_X, \quad d_X\go d_X'=d_X+\delta d_X,
\ee 
 where $c_X$ and $d_X$ represent coefficients of generic diagrams in series such as Eqs.~\eqref{twobet}, \eqref{twogam} respectively. As explained in the Appendix (which in turn is a summary of the discussion in Ref.~\cite{jo}), it is useful to parametrise the scheme change by $v$ defined implicitly by Eq.~\eqref{gpdef}. We assume that $v$ is parametrised in a similar way to Eqs.~\eqref{gprime}, \eqref{Cdef}, with analogues of $f(g)$, $c(g)$ given by similar diagrammatic series to those for the $\beta$-function and anomalous dimension, but with $c_X\go \delta_X$ and $d_X\go \epsilon_X$.

At one and two loops we have
\be
\delta c_1=\delta d_1=\delta c_2=\delta c_{2R}=\delta d_2=0.
\ee
 At three loops we find using Eqs.~\eqref{delba}, \eqref{delbb}
\begin{align}
 \delta c_{3a}=2X^{\lambda\lambda}_{2,1}+2X^{\lambda\lambda}_{1,2R},\quad
\delta c_{3b}&=2X^{\gamma\lambda}_{2,1}, \quad
\delta c_{3c}=2X^{\lambda\lambda}_{1,2}+2X^{\lambda\lambda}_{2R,1},\nn
\delta c_{3d}=2X^{\lambda\lambda}_{1,2},\quad
\delta c_{3e}&=0,\quad
\delta c_{3f}=0,\nn
\delta c_{3aR}=X^{\lambda\lambda}_{1,2R},\quad
\delta c_{3bR}&=2X^{\lambda\lambda}_{1,2R},\quad \delta d_3=6X^{\lambda\gamma}_{1,2}.
\label{deldef}
\end{align}
Here 
\be
X^{\lambda\lambda}_{X,Y}=c_X\delta_Y-\delta_Xc_Y,\quad X^{\gamma\lambda}_{X,Y}=d_X\delta_{Y}-\epsilon_{X}c_Y,
\label{deldefa}
\ee
with corresponding definitions for $X^{\lambda\gamma}_{X,Y}$, $X^{\gamma\gamma}_{X,Y}$ when needed. We see from Eq.~\eqref{delba} that the coefficients appearing in $X^{\lambda\lambda}_{2,1}$ etc should in principle be ``hatted'' quantities defined according to Eq.~\eqref{delbb}; but at this level there is no distinction between the two, i.e. $\chat_1=c_1$, $\chat_2=c_2$, $\dhat_2=d_2$.
Note that $c_{3e}$ and $c_{3f}$ are individually invariant--which in the case of $c_{3f}$ follows immediately from the fact that it corresponds to a primitive graph. In deriving invariant combinations of coefficients it is important to note that
\be 
X^{\lambda\lambda}_{X,Y}=-X^{\lambda\lambda}_{Y,X},\quad
X^{\lambda\gamma}_{X,Y}=-X^{\gamma\lambda}_{Y,X},\quad
X^{\gamma\gamma}_{X,Y}=-X^{\gamma\gamma}_{Y,X}.
\ee
We now start the search for these invariant combinations of coefficients at lowest (three-loop) order. {\it A priori} since at this order there are nine three-loop coefficients and five variations $\delta_1$, $\delta_1^2$, $\delta_2$, $\epsilon_2$, $\delta_{2R}$, one's naive expectation would be $9-5=4$ invariants. However, the variations on the right-hand side of Eq.~\eqref{deldef} are expressed in terms of only three independent quantities, $X^{\lambda\lambda}_{1,2}$, $X^{\gamma\lambda}_{2,1}$ and $X^{\lambda\lambda}_{1,2R}$, and so in fact we should have $9-3=6$ independent invariant combinations of three-loop coefficients. Indeed, we easily find from Eqs.~\eqref{deldef} that 
\begin{align}
I^{(3)}_1=c_{3a}+c_{3d}-2c_{3aR},&\quad I^{(3)}_2=2c_{3aR}-c_{3bR},\nn
I^{(3)}_3=c_{3a}+c_{3c},&\quad I^{(3)}_4=3c_{3b}+d_3,
\label{invthree}
\end{align}
are four independent invariant combinations (making a total of six invariants with the individually invariant $c_{3e}$ and $c_{3f}$). 

\section {The four and five loop calculations}
In this section we examine the issue of scheme invariants comprehensively at four loops and partially (due to increased calculational complexity) at five loops. 
The full list of four loop diagrams was presented in Ref.~\cite{kaz}. The anomalous dimension is given at this order by
\be
2\gamma^{(4)}=d_{4a}\tikz[scale=0.6,baseline=(vert_cent.base),scale=1]{
	\node (vert_cent) {\hspace{-13pt}$\phantom{-}$};
	\draw (-1.5,0)--(1.5,0);
	\draw (0,0) circle[radius=1cm];
	\draw (0,0) circle[radius=0.5cm];
}
+d_{4b}\tikz[scale=0.6,baseline=(vert_cent.base),scale=1]{
	\node (vert_cent) {\hspace{-13pt}$\phantom{-}$};
	\draw (-1.5,0)--(-1,0) (1,0)--(1.5,0);
	\draw (0,0) circle[radius=1cm];
	\vertex at \coord{-30} (A) {};
	\vertex at \coord{30} (B) {};
	\draw [bend right=60] (-1,0) to (A) [bend right=60] (A) to (B) [bend right=60] (B) to (1,0);
}
+d_{4c}\tikz[scale=0.6,baseline=(vert_cent.base),scale=1]{
	\node (vert_cent) {\hspace{-13pt}$\phantom{-}$};
	\draw (-1.5,-0.25) to node[fill,circle,inner sep=0pt,minimum size=0pt] (A) {} (1.5,-0.25);
	\draw (0,0) circle[radius=1cm];
	\draw [bend right=50] (0,1) to (A) [bend left=50] (0,1) to (A);
}
+d_{4d}\tikz[scale=0.6,baseline=(vert_cent.base),scale=1]{
	\node (vert_cent) {\hspace{-13pt}$\phantom{-}$};
	\draw (-1.5,0)--(-1,0) (1,0)--(1.5,0);
	\draw (0,0) circle[radius=1cm];
	\draw [bend right=40] (-1,0) to (0,1)--(0,-1) [bend left=40] (0,-1) to (1,0);
},
\ee
while the 1PI part of the $\beta$-function will be parametrised as
\begin{align}
\btil^{(4)}=&\Scal_3\left(c_{4a}\tikz[scale=0.6, baseline=(vert_cent.base)]{
	\node (vert_cent) {\hspace{-13pt}$\phantom{-}$};
	\draw circle[radius=1cm](0,0);
	\node[fill,circle,inner sep=0pt,minimum size=0pt] at \coord{-50} (A) {};
	\vertex at \coord{50} (B) {};
	\vertex at (0,-1) (C) {};
	\draw[bend right] (A) to (B);
	\draw (A)--(C)--(B);
	\draw (-1.5,-1)--(-1,0)--(-1.5,1);
	\draw (1.5,-1)--(1,0)--(1.5,1);
	}
+c_{4b}\tikz[baseline=(vert_cent.base)]{
  \node (vert_cent) {\hspace{-13pt}$\phantom{-}$};
\draw (-0.4,0.3)--(0.1,0);
\draw (-0.4,-0.3)--(0.1,0); 
    \draw (0.7,0) ++(0:0.6cm and 0.4cm) arc (0:180:0.6cm and 0.4cm) node(n1) {}
             (0.7,0) ++(180:0.6cm and 0.4cm) arc (180:360:0.6cm and 0.4cm) node(n2){};
 \draw (0.7,.2) ++(0:0.1cm and 0.2cm) arc (0:180:0.1cm and 0.2cm) node(n1) {}
             (0.7,.2) ++(180:0.1cm and 0.2cm) arc (180:360:0.1cm and 0.2cm) node(n2){};
 \draw (0.7,-.2) ++(0:0.1cm and 0.2cm) arc (0:180:0.1cm and 0.2cm) node(n1) {}
             (0.7,-.2) ++(180:0.1cm and 0.2cm) arc (180:360:0.1cm and 0.2cm) node(n2){};
    \draw (1.3,0)--(1.8,0.3); 
 \draw (1.3,0)--(1.8,-0.3);          
        }
+c_{4c}\tikz[scale=0.6, baseline=(vert_cent.base)]{
	\node (vert_cent) {\hspace{-13pt}$\phantom{-}$};
	\draw circle[radius=1cm](0,0);
	\node[fill,circle,inner sep=0pt,minimum size=0pt] at \coord{-70} (A) {};
	\vertex at \coord{70} (B) {};
	\vertex at (0,1) (C) {};
	\draw [bend right] (A) to (B) [bend left=50] (B) to (C) [bend left=50] (C) to (A);
	\draw (-1.5,-1)--(-1,0)--(-1.5,1);
	\draw (1.5,-1)--(1,0)--(1.5,1);
}\,\,\right)\nn
&+\Scal_6\Biggl(c_{4d}\tikz[scale=0.6, baseline=(vert_cent.base)]{
	\node (vert_cent) {\hspace{-13pt}$\phantom{-}$};
	\draw circle[radius=1cm](0,0);
	\vertex at \coord{40} (A) {};
	\vertex at \coord{140} (B) {};
	\draw [bend right=30] (1.25,1) to (A) [bend right] (A) to (B) [bend right=30] (B) to (1.25,-1);
	\vertex at \coord{-10} (C) {};
	\vertex at \coord{-170} (D) {};
	\draw [bend left] (C) to (D) [bend right] (C) to (D);
	\draw (-1.5,1)--(-1,0)--(-1.5,-1);
}
+c_{4e}\tikz[scale=0.6, baseline=(vert_cent.base)]{
	\node (vert_cent) {\hspace{-13pt}$\phantom{-}$};
	\draw circle[radius=1cm](0,0);
	\draw (-0.1,0) circle[radius=0.35cm];
	\vertex at \coord{20} (A) {};
	\vertex at \coord{160} (B) {};
	\draw [bend right=20] (1.25,1) to (A) [bend right=60] (A) to (B) [bend right=20] (B) to (1.25,-1);
	\draw (-1.5,1)--(-1,0)--(-1.5,-1);
}
+c_{4f} \tikz[baseline=(vert_cent.base)]{
  \node (vert_cent) {\hspace{-13pt}$\phantom{-}$};
 \draw (-0.5,0.3)--(0,0);
\draw (-0.5,-0.3)--(0,0);
 \draw (0,0)--(1.4,0.7);
\draw (0,0)--(1.4,-0.7);
    \draw (1,0) ++(0:0.2cm and 0.5cm) arc (0:180:0.2cm and 0.5cm) node(n1) {}
             (1,0) ++(180:0.2cm and 0.5cm) arc (180:360:0.2cm and 0.5cm) node(n2){};
\draw (1,0) ++(0:0.2cm and 0.1cm) arc (0:180:0.2cm and 0.1cm) node(n1) {}
             (1,0) ++(180:0.2cm and 0.1cm) arc (180:360:0.2cm and 0.1cm) node(n2){};
        }
+c_{4g} \tikz[baseline=(vert_cent.base)]{
  \node (vert_cent) {\hspace{-13pt}$\phantom{-}$};
 \draw (-0.5,0.3)--(0,0);
\draw (-0.5,-0.3)--(0,0);
 \draw (0,0)--(1.4,0.7);
\draw (0,0)--(1.4,-0.7);
    \draw (1.2,0.4) ++(0:0.1cm and 0.2cm) arc (0:180:0.1cm and 0.2cm) node(n1) {}
             (1.2,0.4) ++(180:0.1cm and 0.2cm) arc (180:360:0.1cm and 0.2cm) node(n2){};
 \draw (1.2,0) ++(0:0.1cm and 0.5cm) arc (0:180:0.1cm and 0.2cm) node(n1) {}
             (1.2,0) ++(180:0.1cm and 0.5cm) arc (180:360:0.1cm and 0.2cm) node(n2){};
 \draw (1.2,-0.4) ++(0:0.1cm and 0.5cm) arc (0:180:0.1cm and 0.2cm) node(n1) {}
             (1.2,-0.4) ++(180:0.1cm and 0.5cm) arc (180:360:0.1cm and 0.2cm) node(n2){};
        }\nn
&+c_{4h}\tikz[baseline=(vert_cent.base)]{
\node (vert_cent) {\hspace{-13pt}$\phantom{-}$};
 \draw (-0.5,0.3)--(0,0);
\draw (-0.5,-0.3)--(0,0);
\draw[name path=path1] [rotate=30] (.8,0) ellipse [x radius=.3, y radius=.07];
 \draw [name path=path5][rotate=30] (0,0)--(0.51,0);
\draw[name path=path2][rotate=30] (1.09,0)--(1.4,0);
\draw[name path=path3] [rotate=-30] (.8,0) ellipse [x radius=.3, y radius=.07];
 \draw [name path=path6][rotate=-30] (0,0)--(0.51,0);
\draw[name path=path4][rotate=-30] (1.09,0)--(1.4,0);
\draw [name intersections={of=path1 and path2, by=x}]
[name intersections={of=path3 and path4, by=y}] (x)-- (y);
\draw [name intersections={of=path1 and path5, by=x}]
[name intersections={of=path3 and path6, by=y}] (x)-- (y);}
+c_{4i}\tikz[baseline=(vert_cent.base)]{
\node (vert_cent) {\hspace{-13pt}$\phantom{-}$};
 \draw (-0.5,0.3)--(0,0);
\draw (-0.5,-0.3)--(0,0);
\draw[name path=path1] [rotate=30] (.8,0) ellipse [x radius=.3, y radius=.07];
 \draw [name path=path5][rotate=30] (0,0)--(0.51,0);
\draw[name path=path2][rotate=30] (1.09,0)--(1.4,0);
\draw[name path=path3] [rotate=-30] (.8,0) ellipse [x radius=.3, y radius=.07];
 \draw [name path=path6][rotate=-30] (0,0)--(0.51,0);
\draw[name path=path4][rotate=-30] (1.09,0)--(1.4,0);
\path [name path=path8][name intersections={of=path1 and path2, by=x1}]
[name intersections={of=path3 and path6, by=y1}] (x1)-- (y1);
\draw [name path=path7] [name intersections={of=path1 and path5, by=x2}]
[name intersections={of=path3 and path4, by=y2}] (x2)-- (y2);
\draw [name intersections={of=path7 and path8, by=y0}] let  \p1=(y0) in (x1)--({\x1+1.5},{\y1+1.5});
\draw [name intersections={of=path7 and path8, by=y0}] let  \p1=(y0) in (y1)--({\x1-1.5},{\y1-1.5});
}\,\Biggr)+\Scal_{12}\Biggl(c_{4j} \tikz[baseline=(vert_cent.base)]{
  \node (vert_cent) {\hspace{-13pt}$\phantom{-}$};
 \draw (-0.5,0.3)--(0,0);
\draw (-0.5,-0.3)--(0,0);
 \draw (0,0)--(1.4,0.7);
\draw (0,0)--(1.4,-0.7);
    \draw (1,0) ++(0:0.1cm and 0.5cm) arc (0:180:0.1cm and 0.5cm) node(n1) {}
             (1,0) ++(180:0.1cm and 0.5cm) arc (180:360:0.1cm and 0.5cm) node(n2){};
        \draw (0.5,-0.25) circle [radius=0.25cm];
}
+c_{4k}\tikz[baseline=(vert_cent.base)]{
\node (vert_cent) {\hspace{-13pt}$\phantom{-}$};
 \draw (-0.5,0.3)--(0,0);
\draw (-0.5,-0.3)--(0,0);
\draw[name path=path1] [rotate=30] (.8,0) ellipse [x radius=.3, y radius=.05];
 \draw [name path=path5][rotate=30] (0,0)--(0.51,0);
\draw[name path=path2][rotate=30] (1.09,0)--(1.4,0);
\draw (0,0)--(-30:1.4);
\draw ({0.8*cos(30)},-0.2) ellipse [x radius=.05, y radius=.2];
\draw [name intersections={of=path1 and path5, by=x}] (x)--({0.8*cos(30)},0);
\draw [name intersections={of=path2 and path1, by=y}] (y)--({0.8*cos(30)},0);
}\nn&
+c_{4l}\tikz[baseline=(vert_cent.base)]{
\node (vert_cent) {\hspace{-13pt}$\phantom{-}$};
 \draw (-0.5,0.3)--(0,0);
\draw (-0.5,-0.3)--(0,0);
\draw[name path=path1] [rotate=30] (.5,0) ellipse [x radius=.2, y radius=.05];
\draw[name path=path3] [rotate=30] (.9,0) ellipse [x radius=.2, y radius=.05];
 \draw [name path=path5][rotate=30] (0,0)--(0.31,0);
\draw[name path=path2][rotate=30] (1.09,0)--(1.4,0);
\draw (0,0)--(-30:1.4);
\draw [name intersections={of=path1 and path5, by=x}] (x)--({0.8*cos(30)},-0.4);
\draw [name intersections={of=path2 and path3, by=y}] (y)--({0.8*cos(30)},-0.4);
}+c_{4m}\tikz[baseline=(vert_cent.base)]{
\node (vert_cent) {\hspace{-13pt}$\phantom{-}$};
 \draw (-0.5,0.3)--(0,0);
\draw (-0.5,-0.3)--(0,0);
\draw[name path=path1] [rotate around={-30:({0.6*cos(30)},0)}] ({0.6*cos(30)-0.15},0) ellipse [x radius=.15, y radius=.05];
\draw (0,0) -- ({1.2*cos(30)},{-1.2*sin(30)});
\draw (0,0) -- ({1.2*cos(30)},{1.2*sin(30)});
\draw ({0.9*cos(30)},{-0.9*sin(30)}) -- ({0.6*cos(30)},0);
\draw ({0.9*cos(30)},{0.9*sin(30)}) -- ({0.6*cos(30)},0);
\draw  ({0.9*cos(30)},{-0.9*sin(30)}) --  ({0.9*cos(30)},{0.9*sin(30)})
}+c_{4n}\tikz[baseline=(vert_cent.base)]{
\node (vert_cent) {\hspace{-13pt}$\phantom{-}$};
 \draw (-0.5,0.3)--(0,0);
\draw (-0.5,-0.3)--(0,0);
\draw[name path=path2][rotate=30] (0,0)--(1.4,0);
\draw[name path=path2][rotate=-30] (0,0)--(1.4,0);
\draw ({cos(30)},{sin(30)})--({cos(30)},0);
\draw ({0.5*cos(30)},{-0.5*sin(30)})--({cos(30)},0);
\draw ({0.5*cos(30)},{-0.5*sin(30)})--({cos(30)},sin(30);
\draw ({cos(30)},-0.25) ellipse [x radius=.05, y radius=.25];
}
+c_{4o}\tikz[scale=0.6, baseline=(vert_cent.base)]{
	\node (vert_cent) {\hspace{-13pt}$\phantom{-}$};
	\draw circle[radius=1cm](0,0);
	\vertex at \coord{-40} (A) {};
	\vertex at \coord{-140} (B) {};
	\draw (-0.9,1.15)--(A) [bend left] (A) to (B) (B)--(-0.9,-1.15);
	\vertex at \coord{30} (C) {};
	\vertex at \coord{90} (D) {};
	\vertex at \coord{150} (E) {};
	\draw [bend right=10] (1,1.25) to (C) [bend right=60] (C) to (D) [bend right=60] (D) to (E) [bend right=10] (E) to (1,-1.25);
}
+c_{4p}\tikz[baseline=(vert_cent.base)]{
\node (vert_cent) {\hspace{-13pt}$\phantom{-}$};
\draw (-0.7,-0.7) -- (-0.5,-0.5);
\path [name path=path1] (-0.5,-0.5) -- (0.5,0);
\draw[name path=path2] (-0.7,0.7) -- (0.7,-0.7);
\draw [name intersections={of=path1 and path2, by=y0}] let  \p1=(y0) in (-0.5,-0.5)--({\x1-1.5},{\y1-1});
\draw [name intersections={of=path1 and path2, by=y0}] let  \p1=(y0) in (0.5,0)--({\x1+1.5},{\y1+1});
\draw (0.5,0.5)--(0.7,0.7);
\draw (-0.5,-0.5) -- (0.5,-0.5);
\draw (-0.5,-0.5) -- (-0.5,0.5);
\draw (-0.5,0.5) -- (0.5,0.5);
\draw (0.5,-0.5) -- (0.5,0);
 \draw (0.5,0.25) ++(0:0.1cm and 0.25cm) arc (0:180:0.1cm and 0.25cm) node(n1) {}
             (0.5,0.25) ++(180:0.1cm and 0.25cm) arc (180:360:0.1cm and 0.25cm) node(n2){};
}\,\Biggr)\nn
&
+c_{4q}\Scal_6\tikz[baseline=(vert_cent.base)]{
\node (vert_cent) {\hspace{-13pt}$\phantom{-}$};
\path [name path=path1] (-0.7,-0.7) -- (0.5,0.5);
\draw [name path=path2] (-0.7,0.7) -- (0.5,-0.5);
\draw [name intersections={of=path1 and path2, by=y0}] let  \p1=(y0) in (-0.7,-0.7)--({\x1-1.5},{\y1-1});
\draw [name intersections={of=path1 and path2, by=y0}] let  \p1=(y0) in (0.5,0.5)--({\x1+1.5},{\y1+1});
 \draw (0.5,0) ++(0:0.1cm and 0.5cm) arc (0:180:0.1cm and 0.5cm) node(n1) {}
             (0.5,0) ++(180:0.1cm and 0.5cm) arc (180:360:0.1cm and 0.5cm) node(n2){};
\draw (-0.5,-0.5) -- (0.5,-0.5);
\draw (-0.5,-0.5) -- (-0.5,0.5);
\draw (-0.5,0.5) -- (0.5,0.5);
\draw (0.6,0) -- (0.8,0.2);
\draw (0.6,0) -- (0.8,-0.2);
}
+c_{4r}\Scal_{24}\tikz[scale=0.6, baseline=(vert_cent.base)]{
	\node (vert_cent) {\hspace{-13pt}$\phantom{-}$};
	\draw circle[radius=1cm](0,0);
	\vertex at \coord{-40} (A) {};
	\vertex at \coord{-140} (B) {};
	\draw (-0.9,1.15)--(A) [bend left] (A) to (B) (B)--(-0.9,-1.15);
	\vertex at \coord{-20} (C) {};
	\vertex at \coord{60} (D) {};
	\draw [bend left] (1.25,1) to (D) [bend left] (D) to (C) (C)--(1,-1.25);
	}+c_{4s}\tikz[baseline=(vert_cent.base)]{
\node (vert_cent) {\hspace{-13pt}$\phantom{-}$};
\draw (-0.7,-0.7) -- (0.7,0.7);
\draw (-0.7,0.7) -- (0.7,-0.7);
\draw (-0.5,-0.5) -- (0.5,-0.5);
\draw (-0.5,-0.5) -- (-0.5,0.5);
\draw (-0.5,0.5) -- (0.5,0.5);
\draw (0.5,-0.5) -- (0.5,0.5);
}\nn
&+\Scal_3\left(c_{4aR}\tikz[scale=0.6, baseline=(vert_cent.base)]{
	\node (vert_cent) {\hspace{-13pt}$\phantom{-}$};
	\draw (-1.5,0) circle[radius=0.5cm];
	\draw (-0.5,0) circle[radius=0.5cm];
	\draw (0.5,0) circle[radius=0.5cm];
	\draw (1.5,0) circle[radius=0.5cm];
	\draw (-2.25,0.5)--(-2,0)--(-2.25,-0.5);
	\draw (2.25,0.5)--(2,0)--(2.25,-0.5);
}
+c_{4bR}\tikz[scale=0.6, baseline=(vert_cent.base)]{
	\node (vert_cent) {\hspace{-13pt}$\phantom{-}$};
	\draw (-0.75,0) circle[radius=0.75cm];
	\draw (0.75,0) circle[radius=0.75cm];
	\vertex at ({0.75+0.75*sin(40)},{0.75*cos(40)}) (A) {};
	\vertex at ({0.75+0.75*sin(140)},{0.75*cos(140)}) (B) {};
	\draw [bend right=20] (1.75,0.75) to (A) [bend right=40] (A) to (B) [bend right=20] (B) to (1.75,-0.75);
	\vertex at ({-0.75+0.75*sin(-40)},{0.75*cos(-40)}) (C) {};
	\vertex at ({-0.75+0.75*sin(-140)},{0.75*cos(-140)}) (D) {};
	\draw [bend left=20] (-1.75,0.75) to (C) [bend left=40] (C) to (D) [bend left=20] (D) to (-1.75,-0.75);
}\,\,\right)\nn&+\Scal_6\Biggl(c_{4cR}\tikz[scale=0.6, baseline=(vert_cent.base)]{
	\node (vert_cent) {\hspace{-13pt}$\phantom{-}$};
	\draw (-1.5,0) circle[radius=0.5cm];
	\draw (-0.5,0) circle[radius=0.5cm];
	\draw (0.75,0) circle[radius=0.75cm];
	\draw (-2.25,0.5)--(-2,0)--(-2.25,-0.5);
	\vertex at ({0.75+0.75*sin(40)},{0.75*cos(40)}) (A) {};
	\vertex at ({0.75+0.75*sin(140)},{0.75*cos(140)}) (B) {};
	\draw [bend right=20] (1.75,0.75) to (A) [bend right=40] (A) to (B) [bend right=20] (B) to (1.75,-0.75);
}
+c_{4dR}\tikz[scale=0.6, baseline=(vert_cent.base)]{
	\node (vert_cent) {\hspace{-13pt}$\phantom{-}$};
	\draw (-0.5,0) circle[radius=0.5cm];
	\draw (0.75,0) circle[radius=0.75cm];
	\draw (-1.25,0.5)--(-1,0)--(-1.25,-0.5);
	\vertex at ({0.75+0.75*sin(30)},{0.75*cos(30)}) (A) {};
	\vertex at (1.5,0) (B) {};
	\vertex at ({0.75+0.75*sin(150)},{0.75*cos(150)}) (C) {};
	\draw [bend right=30] (1.5,0.75) to (A) [bend right=60] (A) to (B) [bend right=60] (B) to (C) [bend right=30] (C) to (1.5,-0.75);
}
+c_{4eR}\tikz[scale=0.6, baseline=(vert_cent.base)]{
	\node (vert_cent) {\hspace{-13pt}$\phantom{-}$};
	\draw (-0.75,0) circle[radius=0.75cm];
	\draw (0.75,0) circle[radius=0.75cm];
	\vertex at ({0.75+0.75*sin(-50)},{0.75*cos(-50)}) (A) {};
	\vertex at ({0.75+0.75*sin(50)},{0.75*cos(50)}) (B) {};
	\draw [bend left=0] (A) to (B) [bend left=60] (B) to (A);
	\draw (-2,0.75)--(-1.5,0)--(-2,-0.75);
	\draw (2,0.75)--(1.5,0)--(2,-0.75);
}\nn&
+c_{4fR}\tikz[scale=0.6, baseline=(vert_cent.base)]{
	\node (vert_cent) {\hspace{-13pt}$\phantom{-}$};
	\draw (-0.75,0) circle[radius=0.75cm];
	\draw (0.75,0) circle[radius=0.75cm];
	\vertex at ({0.75+0.75*sin(0)},{0.75*cos(0)}) (A) {};
	\vertex at ({0.75+0.75*sin(180)},{0.75*cos(180)}) (B) {};
	\draw [bend right=40] (A) to (B) [bend left=40] (A) to (B);
	\draw (-2,0.75)--(-1.5,0)--(-2,-0.75);
	\draw (2,0.75)--(1.5,0)--(2,-0.75);
}\,\,\Biggr)
+c_{4gR}\Scal_{12}\tikz[scale=0.6, baseline=(vert_cent.base)]{
	\node (vert_cent) {\hspace{-13pt}$\phantom{-}$};
	\draw (-0.75,0) circle[radius=0.75cm];
	\draw (0.75,0) circle[radius=0.75cm];
	\vertex at ({0.75+0.75*sin(-50)},{0.75*cos(-50)}) (A) {};
	\vertex at ({0.75+0.75*sin(60)},{0.75*cos(60)}) (B) {};
	\draw [bend left=10] (1.75,0.75) to (B) [bend left=30] (B) to (A) [bend right=10] (A) to (1.75,-0.75);
	\draw (-2,0.75)--(-1.5,0)--(-2,-0.75);
}.
\label{fourbet}
\end{align}
In Eq.~\eqref{fourbet} the graph $g_{4s}^{\lambda}$ is the only primitive one.

We find (again using Eqs.~\eqref{delba}, \eqref{delbb}) variations of the four-loop coefficients given by 
\begin{align}
\delta c_{4a}=&4\hX^{\lambda\lambda}_{1,3a}+4\hX^{\lambda\lambda}_{3e,1}+4X^{\lambda\lambda}_{2,2R},\nn
\delta c_{4b}=&-\delta c_{4f}=2\hX^{\lambda\lambda}_{1,3a}+2\hX^{\lambda\lambda}_{3c,1},\nn
\delta c_{4c}=&6\hX^{\lambda\lambda}_{1,3b}+2\hX^{\gamma\lambda}_{3,1},\nn
\delta c_{4d}=&2\hX^{\lambda\lambda}_{1,3a}+2\hX^{\lambda\lambda}_{1,3bR}+2\hX^{\lambda\lambda}_{3d,1},\nn
\delta c_{4e}=&2\hX^{\lambda\lambda}_{3b,1}+2X^{\gamma\lambda}_{2,2},\nn
\delta c_{4g}=&3\hX^{\lambda\lambda}_{1,3c}+2\hX^{\lambda\lambda}_{3aR,1}+2X^{\lambda\lambda}_{2R,2},\nn
\delta c_{4h}=&\delta c_{4i}=2\hX^{\lambda\lambda}_{3d,1}+2\hX^{\lambda\lambda}_{1,3e},\nn
\delta c_{4j}=&\hX^{\lambda\lambda}_{1,3b}+X^{\gamma\lambda}_{2,2},\nn
\delta c_{4k}=&2\hX^{\lambda\lambda}_{1,3c}+\hX^{\lambda\lambda}_{1,3e}+2\hX^{\lambda\lambda}_{3bR,1},\nn
\delta c_{4l}=&2\hX^{\lambda\lambda}_{1,3e}+2\hX^{\lambda\lambda}_{3c,1}+2X^{\lambda\lambda}_{2R,2},\nn
\delta c_{4m}=&\delta c_{4n}=\delta c_{4s}=0,\nn
\delta c_{4o}=&\hX^{\lambda\lambda}_{1,3c}+2\hX^{\lambda\lambda}_{1,3d}+X^{\lambda\lambda}_{2R,2},\nn
\delta c_{4p}=&-\delta c_{4q}=\hX^{\lambda\lambda}_{1,3f},\nn
\delta c_{4r}=&2\hX^{\lambda\lambda}_{1,3d}+\hX^{\lambda\lambda}_{1,3e},
\label{delfour}
\end{align}
for the one-vertex irreducible coefficients, 
\begin{align}
\delta c_{4aR}=&2\hX^{\lambda\lambda}_{1,3aR},\nn
\delta c_{4bR}=&4\hX^{\lambda\lambda}_{1,3bR},\nn
\delta c_{4cR}=&2\hX^{\lambda\lambda}_{1,3aR}+\hX^{\lambda\lambda}_{1,3bR},\nn
\delta c_{4dR}=&2\hX^{\lambda\lambda}_{1,3bR},\nn
\delta c_{4eR}=&2X^{\gamma\lambda}_{2,2R},\nn
\delta c_{4fR}=&2\hX^{\lambda\lambda}_{1,3aR}+\hX^{\lambda\lambda}_{3bR,1}+2X^{\lambda\lambda}_{2,2R},\nn
\delta c_{4gR}=&2\hX^{\lambda\lambda}_{1,3bR}+2X^{\lambda\lambda}_{2,2R}\nn
\label{delfourR}
\end{align}
for the 1VR coefficients and
\begin{align}
\delta d_{4a}=&0,\nn
\delta d_{4b}=&3\hX^{\lambda\gamma}_{1,3}+6X^{\lambda\gamma}_{2R,2},\nn
\delta d_{4c}=&2\hX^{\lambda\gamma}_{1,3}+6X^{\lambda\gamma}_{2,2},\nn
\delta d_{4d}=&4\hX^{\lambda\gamma}_{1,3}+6X^{\lambda\gamma}_{2,2}
\label{delfoura}
\end{align}
for the anomalous dimension coefficients. At this level, in contrast to the earlier three-loop calculation, we do need to distinguish ``hatted'' from ``unhatted'' quantities. The $\hX^{\lambda\lambda}$ quantities are defined by 
\be
\hX^{\lambda\lambda}_{X,Y}=\chat_X\delta_Y-\delta_X\chat_Y,
\label{deldefp}
\ee 
in other words as for $X^{\lambda\lambda}$ in Eq.~\eqref{deldefa} but with the $\beta$-function quantities $c_{X,Y}$
replaced by hatted quantities $\chat_{X,Y}$. Similar definitions apply to $X^{\lambda\gamma}$, etc, but with $d_{X,Y}$ replaced by hatted quantities $\dhat_{X,Y}$ where relevant.
 Here again $\chat_1=c_1$, $\chat_2=c_2$, $\dhat_2=d_2$, while the quantities $\chat_{3a}$ etc are defined by 
\be
\chat_{3a}=c_{3a}+\tfrac12\delta c_{3a}
\label{hatdef}
\ee
with $\delta c_{3a}$ as defined as  in Eq.~\eqref{deldef}, and similar expressions for $\chat_{3b}$ etc, and also $\dhat_3$.  The additional terms in the hatted quantities derive from the first Lie derivative term on the right-hand side of Eq.~\eqref{delba}. 

Now again we look for invariants at this order. Note that $c_{4m}$, $c_{4n}$, $c_{4s}$, $d_{4a}$ are individually invariant--which again in the case of $c_{4s}$ follows immediately from the fact that it corresponds to a primitive graph. There are thirty four-loop coefficients whose variations are given in Eqs.~\eqref{delfour}, \eqref{delfourR}, \eqref{delfoura}; and eighteen variations up to the three-loop level, namely
$\delta_{3a-3f,3aR,3bR}$, $\epsilon_3$, $\delta_1^3$, $\delta_1\delta_2$, $\delta_1\delta_{2R}$, $\delta_1\epsilon_2$,  $\delta_2$, $\epsilon_2$, $\delta_{2R}$, $\delta_1^2$, $\delta_1$. We would therefore naively expect $30-18=12$ invariants. However, the variations on the right-hand sides of  Eqs.~\eqref{delfour}, \eqref{delfourR}, \eqref{delfoura} are expressed in terms of only twelve independent $X$/$\hX$combinations and therefore the correct expectation is $30-12=18$ invariants. Indeed, together with the four individually invariant coefficients $c_{4m}$, $c_{4n}$, $c_{4s}$, $d_{4a}$ we find the following fourteen linear invariant combinations:
\begin{align}
I^{(4)L}_1=&c_{4h}-c_{4i},\nn
I^{(4)L}_2=&c_{4b}+c_{4f},\nn
I^{(4)L}_3=&c_{4a}+2c_{4f}+2c_{4l},\nn
I^{(4)L}_4=&c_{4l}+2c_{4o}-2c_{4r}-c_{4bR}+2c_{4gR},\nn
I^{(4)L}_5=&c_{4c}+3c_{4e}+d_{4c},\nn
I^{(4)L}_6=&c_{4d}+c_{4f}-c_{4k}+c_{4r}-c_{4bR},\nn
I^{(4)L}_7=&c_{4b}-c_{4d}+c_{4g}-c_{4o}+c_{4cR}+\tfrac12c_{4gR},\nn
I^{(4)L}_8=&c_{4h}-c_{4k}+2c_{4o}-c_{4r}-c_{4bR}+c_{4gR},\nn
I^{(4)L}_9=&3c_{4e}+6c_{4j}+4d_{4c}-2d_{4d},\nn
I^{(4)L}_{10}=&c_{4p}+c_{4q},\nn
I^{(4)L}_{11}=&2d_{4b}+3d_{4c}-3d_{4d}+6c_{4eR},\nn
I^{(4)L}_{12}=&4c_{4aR}-4c_{4cR}+c_{4bR},\nn
I^{(4)L}_{13}=&c_{4bR}-2c_{4dR},\nn
I^{(4)L}_{14}=&c_{4cR}-c_{4bR}+c_{4gR}-c_{4fR}.
\label{invfour}
\end{align}
We call these 18 invariants ``linear''. 
We also find three ``quadratic'' invariants
\begin{align}
I^{(4)Q}_{1}=&c_1(2d_{4c}-d_{4d})+3c_2c_{3b}+3d_2c_{3d},\nn
I^{(4)Q}_{2}=&c_1c_{4eR}-c_{2R}c_{3b}-d_2c_{3bR},\nn
I^{(4)Q}_{3}=&c_1(c_{4dR}-c_{4gR})+c_2c_{3aR}-c_{2R}c_{3d},
\label{qinv}
\end{align}
which are a consequence of the relations
\begin{align}
c_2X^{\lambda\gamma}_{1,2}-d_2X^{\lambda\lambda}_{1,2}=&c_1X^{\lambda\gamma}_{2,2},\nn
c_{2R}X^{\lambda\gamma}_{1,2}-d_2X^{\lambda\lambda}_{1,2R}=&c_1X^{\lambda\gamma}_{2R,2},\nn
c_2X^{\lambda\lambda}_{1,2R}-c_{2R}X^{\lambda\lambda}_{1,2}=&c_1X^{\lambda\lambda}_{2,2R},
\label{quadrels}
\end{align}
respectively. Altogether we have found twenty-one invariants, considerably more than (in fact almost double) the twelve which might naively have been expected.  

We note that one may derive a fourth identity 
\be
c_2X^{\lambda\gamma}_{2R,2}+d_2X^{\lambda\lambda}_{2,2R}=c_{2R}X^{\lambda\gamma}_{2,2}
\label{idext}
\ee
which leads to an invariant 
\be
I^{(4)Q}_{4}=d_2(c_{4bR}-2c_{4gR})+2c_2c_{4eR}+\tfrac23c_{2R}(2d_{4c}-d_{4d});
\label{invext}
\ee
but in fact Eq.~\eqref{idext} may be derived from linear combinations of the identities in Eq.~\eqref{quadrels} and correspondingly $I^{(4)Q}_{4}$ is a linear combination of invariants already found in Eqs.~\eqref{invfour}, \eqref{qinv}.

We now proceed to a very partial five-loop calculation. The number of diagrams at five loops is dauntingly high, so we have not undertaken a complete calculation of all the invariants. A natural place to start is with the five-loop anomalous dimension which has only eleven terms:
\begin{align}
2\gamma^{(5)}=&
d_{5a} \tikz[scale=0.6, baseline=(vert_cent.base)]{
	\node (vert_cent) {\hspace{-13pt}$\phantom{-}$};
	\draw (-2,0)--(2,0);
	\draw (0,0) circle[radius=1.5cm];
	\draw (0,1.5)--(0,-1.5);
	\vertex at (0.75,0.2) (A) {};
	\vertex at (0.75,-0.2) (B) {};
	\draw [bend left=25] (0,1.5) to (A) [bend left=25] (B) to (0,-1.5);
}
+d_{5b}\tikz[scale=0.6, baseline=(vert_cent.base)]{
	\node (vert_cent) {\hspace{-13pt}$\phantom{-}$};
	\draw (-2,-0.5)--(2,-0.5) node[fill,circle,inner sep=0pt,minimum size=0pt,pos=0.5] (C) {};
	\draw (0,0) circle[radius=1.5cm];
	\vertex at ({1.5*sin(40)},{1.5*cos(40)}) (A) {};
	\vertex at ({1.5*sin(-40)},{1.5*cos(-40)}) (B) {};
	\draw [bend left=50] (A) to (B) (B)--(C)--(A);
}
+d_{5c}\tikz[scale=0.6, baseline=(vert_cent.base)]{
	\node (vert_cent) {\hspace{-13pt}$\phantom{-}$};
	\draw (-2,0)--(-1.5,0) (1.5,0)--(2,0);
	\draw (0,0) circle[radius=1.5cm];
	\vertex at ({1.5*sin(-30)},{1.5*cos(-30)}) (A) {};
	\vertex at (0,-1.5) (B) {};
	\vertex at ({1.5*sin(30)},{1.5*cos(30)}) (C) {};
	\draw [bend right=50] (-1.5,0) to (A) [bend left=10] (A) to (B) [bend left=10] (B) to (C) [bend right=50] (C) to (1.5,0);
}
+d_{5d}\tikz[scale=0.6, baseline=(vert_cent.base)]{
	\node (vert_cent) {\hspace{-13pt}$\phantom{-}$};
	\draw (-2,-0.5)--(2,-0.5);
	\draw (0,0) circle[radius=1.5cm];
	\vertex at ({1.5*sin(-80)},{1.5*cos(-80)}) (A) {};
	\vertex at ({1.5*sin(80)},{1.5*cos(80)}) (B) {};
	\vertex at (0,1.5) (C) {};
	\draw [bend right=20] (A) to (B) [bend left=50] (B) to (C) [bend left=50] (C) to (A);
}\nn
&
+d_{5e}\tikz[scale=0.6, baseline=(vert_cent.base)]{
	\node (vert_cent) {\hspace{-13pt}$\phantom{-}$};
	\draw (-2,-0.50)--(2,-0.50) node[fill,circle,inner sep=0pt,minimum size=0pt,pos=0.5] (C) {};
	\draw (0,0) circle[radius=1.5cm];
	\vertex at (0,1.5) (A) {};
	\vertex at ($(A)!0.5!(C)$) (B) {};
	\draw [bend left=60] (A) to (B) [bend right=60] (A) to (B) [bend left=60] (B) to (C) [bend right=60] (B) to (C);
}
+d_{5f}\tikz[scale=0.6, baseline=(vert_cent.base)]{
	\node (vert_cent) {\hspace{-13pt}$\phantom{-}$};
	\draw (-2,0)--(-1.5,0) (1.5,0)--(2,0);
	\draw (0,0) circle[radius=1.5cm];
	\vertex at ({1.5*sin(-45)},{1.5*cos(-45)}) (A) {};
	\vertex at (0,1.5) (B) {};
	\vertex at ({1.5*sin(45)},{1.5*cos(45)}) (C) {};
	\draw [bend right=60] (-1.5,0) to (A) [bend right=60] (A) to (B) [bend right=60] (B) to (C) [bend right=60] (C) to (1.5,0) 
}
+d_{5g}\tikz[scale=0.6, baseline=(vert_cent.base)]{
	\node (vert_cent) {\hspace{-13pt}$\phantom{-}$};
	\draw (-2,0)--(-1.5,0) (1.5,0)--(2,0);
	\draw (0,0) circle[radius=1.5cm];
	\vertex at (0,1.5) (A) {};
	\draw [bend right=45] (-1.5,0) to (A) [bend right=45] (A) to (1.5,0);
	\vertex at ({1.5*sin(-135)},{1.5*cos(-135)}) (B) {};
	\vertex at ({1.5*sin(135)},{1.5*cos(135)}) (C) {};
	\draw [bend left=60] (B) to (C) (B)--(C);
}
+\Scal_2\Biggl(d_{5h}\tikz[scale=0.6, baseline=(vert_cent.base)]{
	\node (vert_cent) {\hspace{-13pt}$\phantom{-}$};
	\draw (-2,0)--(-1.5,0) (0,0)--(2,0);
	\draw (0,0) circle[radius=1.5cm];
	\draw [bend right=45] (-1.5,0) to (0,1.5)--(0,-1.5) [bend right=70] (0,-1.5) to (0,0);
}\nn
&+d_{5i}\tikz[scale=0.6, baseline=(vert_cent.base)]{
	\node (vert_cent) {\hspace{-13pt}$\phantom{-}$};
	\draw (-2,0)--(-1.5,0) (1.5,0)--(2,0);
	\draw (0,0) circle[radius=1.5cm];
	\vertex at (0,1.5) (A) {};
	\draw [bend right=45] (-1.5,0) to node[fill,circle,inner sep=0pt,minimum size=0pt,pos=0.5] (C) {} (A);
	\draw [bend right=45] (A) to (1.5,0);
	\vertex at ({1.5*sin(-45)},{1.5*cos(-45)}) (B) {};
	\draw [bend left=60] (B) to (C) [bend right=60] (B) to (C)
}
+d_{5j}\tikz[scale=0.6, baseline=(vert_cent.base)]{
	\node (vert_cent) {\hspace{-13pt}$\phantom{-}$};
	\draw (-2,0)--(-1.5,0) (1.5,0)--(2,0);
	\draw (0,0) circle[radius=1.5cm];
	\vertex at ({1.5*sin(-45)},{1.5*cos(-45)}) (A) {};
	\draw [bend right=70] (-1.5,0) to (A) [bend right=70] (A) to (0,1.5)--(0,-1.5) [bend left=45] (0,-1.5) to (1.5,0);
}
+d_{5k}\tikz[scale=0.6, baseline=(vert_cent.base)]{
	\node (vert_cent) {\hspace{-13pt}$\phantom{-}$};
	\draw (-2,0)--(-1.5,0) (1.5,0)--(2,0);
	\draw (0,0) circle[radius=1.5cm];
	\vertex at (0,1.5) (A) {};
	\draw [bend right=45] (-1.5,0) to node[fill,circle,inner sep=0pt,minimum size=0pt,pos=0.5] (C) {} (A);
	\draw [bend right=45] (A) to (1.5,0);
	\draw (C) circle[radius=0.4];
}\Biggr).
\label{anomfive}
\end{align}
We find from Eqs.~\eqref{delba}, \eqref{delbb} that the variations of the coefficients in Eq.~\eqref{anomfive} are given by
\begin{align}
\delta d_{5a}=&2\hX^{\lambda\gamma}_{3f,2},\nn
\delta d_{ 5b}=&12\hX^{\lambda\gamma}_{3e,2}+4\hX^{\lambda\gamma}_{1,4c}+4\hX^{\lambda\gamma}_{2,3},\nn
\delta d_{5c}=&6\hX^{\lambda\gamma}_{3e,2}+4\hX^{\lambda\gamma}_{1,4d}+4\hX^{\lambda\gamma}_{2,3},\nn
\delta d_{5d}=&6\hX^{\lambda\gamma}_{1,4a}+3\hX^{\gamma\gamma}_{3,2},\nn
\delta d_{5e}=&6\hX^{\lambda\gamma}_{3c,2}+2\hX^{\lambda\gamma}_{1,4c}+2\hX^{\lambda\gamma}_{2R,3},\nn
\delta d_{5f}=&6\hX^{\lambda\gamma}_{3aR,2}+4\hX^{\lambda\gamma}_{1,4b}+3\hX^{\lambda\gamma}_{2R,3},\nn
\delta d_{5g}=&2\hX^{\lambda\gamma}_{1,4a}+\hX^{\gamma\gamma}_{2,3},\nn
\delta d_{5h}=&6\hX^{\lambda\gamma}_{3d,2}+3\hX^{\lambda\gamma}_{3e,2}+2\hX^{\lambda\gamma}_{1,4c}+2\hX^{\lambda\gamma}_{1,4d}+2X^{\lambda\gamma}_{2,3},\nn
\delta d_{5i}=&3\hX^{\lambda\gamma}_{3a,2}+3\hX^{\lambda\gamma}_{3bR,2}+2\hX^{\lambda\gamma}_{1,4b}+\hX^{\lambda\gamma}_{1,4c}+2\hX^{\lambda\gamma}_{2,3},\nn
\delta d_{5j}=&3\hX^{\lambda\gamma}_{3c,2}+3\hX^{\lambda\gamma}_{3bR,2}+2\hX^{\lambda\gamma}_{1,4b}+2\hX^{\lambda\gamma}_{1,4d}+\hX^{\lambda\gamma}_{2,3}+2\hX^{\lambda\gamma}_{2R,3},\nn
\delta d_{5k}=&3\hX^{\lambda\gamma}_{3b,2}+2\hX^{\lambda\gamma}_{1,4a}+2\hX^{\gamma\gamma}_{2,3}.
\label{delfive}
\end{align}
The  hatted $X$-type terms are defined in a similar manner to Eq.~\eqref{deldefp}, i.e. by replacing $\beta$-function quantities $c_{X,Y}$ and $d_{X,Y}$, in Eq.~\eqref{deldefa} by hatted quantities $\chat_{X,Y}$, and $\dhat_{X,Y}$. The hatted coefficients are in turn defined in terms of the corresponding unhatted quantities in a manner similar to Eq.~\eqref{hatdef}. However, in the case  of four-loop anomalous dimension coefficients, we need to define
\be
\dhat_{4b}=d_{4b}+\tfrac12\delta'd_{4b},
\ee
where $\delta'd_{4b}$ (and similarly$\delta'd_{4c,d}$) are defined as in Eq.~\eqref{delfoura}, but with hatted replaced by tilded quantities, namely
\be
\delta d_{4b}=3\Xtil^{\lambda\gamma}_{1,3}+6X^{\lambda\gamma}_{2R,2}.
\ee
$\Xtil^{\lambda\gamma}_{1,3}$ is defined as for $X^{\lambda\gamma}_{1,3}$  but with $d_3$
replaced by $\dtill_3$. This in turn is defined by a similar equation to Eq.~\eqref{hatdef}, but with $\tfrac12\go\tfrac13$, i.e.
\be
\dtill_3=d_3+\tfrac13\delta d_3,
\ee
with $\delta d_3$ as in Eq.~\eqref{deldef}.
 This appears rather complicated, but simply reflects the nested structure of Eq.~\eqref{delbb}. This feature has not been apparent in our calculations until now, because there the terms quadratic in $\cal L$ have not hitherto contributed. 

However it proves impossible to construct an invariant combination purely of anomalous dimension coefficients and in fact we need to include some 1VR four-point contributions, depicted below:
\be
\tikz[baseline=(vert_cent.base)]{
  \node (vert_cent) {\hspace{-13pt}$\phantom{-}$};
 \draw (-0.4,0.3)--(0.1,0);
\draw (-0.4,-0.3)--(0.1,0);
    \draw (0.7,0) ++(0:0.6cm and 0.4cm) arc (0:180:0.6cm and 0.4cm) node(n1) {}
         (0.7,0) ++(180:0.6cm) arc (180:360:0.6cm and 0.4cm) node(n2){};
\draw (0.7,0.4) circle [radius=0.3cm];
   \draw (1.6,0) circle [radius=0.3cm];
\draw (2.2,0) circle [radius=0.3cm];
\draw (2.5,0)--(2.7,0.3); 
 \draw (2.5,0)--(2.7,-0.3);   
\draw (1,-0.7) node {\small{$5aR$}};
}
\quad
\tikz[baseline=(vert_cent.base)]{
  \node (vert_cent) {\hspace{-13pt}$\phantom{-}$};
 \draw (-0.4,0.3)--(0.1,0);
\draw (-0.4,-0.3)--(0.1,0);
    \draw (0.7,0) ++(0:0.6cm and 0.4cm) arc (0:180:0.6cm and 0.4cm) node(n1) {}
         (0.7,0) ++(180:0.6cm) arc (180:360:0.6cm and 0.4cm) node(n2){};
\draw (0.7,0.4) circle [radius=0.3cm];
     \draw (1.3,0)--(2.0,0.35);
\draw (1.3,0)--(2.0,-0.35);
    \draw (1.8,0) ++(0:0.1cm and 0.5cm) arc (0:180:0.1cm and 0.25cm) node(n1) {}
             (1.8,0) ++(180:0.1cm and 0.5cm) arc (180:360:0.1cm and 0.25cm) node(n2){};
 \draw (0.5,-0.7) node {\small{$5bR$}};    
}\quad
\tikz[scale=0.6,baseline=(vert_cent.base)]{
  \node (vert_cent) {\hspace{-13pt}$\phantom{-}$};
\draw (2.5,1)--(2,0);
\draw (2.5,-1)--(2,0); 
\draw (1.5,0) circle [radius=0.5cm];
	\node (vert_cent) {\hspace{-13pt}$\phantom{-}$};
	\draw circle[radius=1cm](0,0);
	\node[fill,circle,inner sep=0pt,minimum size=0pt] at \coord{-70} (A) {};
	\vertex at \coord{70} (B) {};
	\vertex at (0,1) (C) {};
	\draw [bend right] (A) to (B) [bend left=50] (B) to (C) [bend left=50] (C) to (A);
	\draw (-1.5,-1)--(-1,0)--(-1.5,1); 
\draw (0,-1.5) node {\small{$5cR$}}; 
        }.
\ee
The variations of the corresponding coefficients are given by 
\begin{align}
\delta c_{5aR}=&\hX^{\lambda\lambda}_{1,4eR}+2\hX^{\gamma\lambda}_{2,3aR},\nn
\delta c_{5bR}=&2\hX^{\lambda\lambda}_{1,4eR}+2\hX^{\gamma\lambda}_{2,3bR},\nn
\delta c_{5cR}=&6\hX^{\lambda\lambda}_{1,4eR}+2\hX^{\gamma\lambda}_{3,2R},
\label{bubfive}
\end{align}
where the hatted quantities are again defined in a similar way to Eq.~\eqref{deldefp}. Note that (as we see in Eq.~\eqref{delfourR}) the variation $\delta c_{4eR}$ is expressed in terms of unhatted quantities, so there is no need to invoke the modified $\delta'$ here. Naively, no linear invariant constructed purely from the coefficients in Eqs.~\eqref{anomfive}, \eqref{bubfive} would be expected--there are 16 independent variations in Eq.~\eqref{delfive} and only 14 coefficients. However, it turns out that there are three unexpected relations among the invariance conditions, resulting in 
 just one five-loop linear invariant formed using only anomalous dimension and 1VR coefficients, namely
\be
I^{(5)L}_1=d_{5b}-2d_{5c}-2d_{5e}-2d_{5f}+4d_{5j}-6c_{5aR}+6c_{5bR}-c_{5cR}.
\label{invfive}
\ee
In addition,  we also find several quadratic invariants, namely
\begin{align}
I^{(5)Q}_1=&c_1d_{5a}+2d_2c_{4p}+c_{3b}c_{3f},\nn
 I^{(5)Q}_2=&2c_1(d_{5g}-d_{5k})-d_2c_{4c}+d_3c_{3b},\nn
I^{(5)Q}_3=&c_1(2c_{5aR}-c_{5bR})-2d_2(2c_{4aR}-c_{4cR})-c_{3b}(2c_{3aR}-c_{3bR}),\nn
I^{(5)Q}_{4}=&c_1(d_{5d}-3d_{5g})-\tfrac32d_2J-\tfrac12d_3^2,\nn
I^{(5)Q}_5=&c_1(3c_{5bR}-c_{5cR})+\tfrac12c_{2R}J+6d_2(c_{4aR}-c_{4cR})+d_3c_{3bR},\nn
I^{(5)Q}_{6}=&c_1(d_{5c}+2d_{5e}-2d_{5h})-6d_2(c_{4b}-c_{4d}-2c_{4aR}+2c_{4cR})+c_{2R}J\nn
&-2d_3(c_{3c}-c_{3d}),\nn
I^{(5)Q}_{7}=&c_1(d_{5b}-2d_{5e})+3d_2(c_{4l}-2c_{4o}+2c_{4r})\nn
&+(c_2-c_{2R})J+6c_{3b}c_{3e}+2c_{3c}d_3,\nn
I^{(5)Q}_{8}=&c_1(d_{5e}+d_{5f}-2d_{5i})
-3d_2(c_{4b}-3c_{4aR}+2c_{4cR})\nn
&+\tfrac14(5c_{2R}-4c_2)J
+d_3(c_{3a}-c_{3c}-c_{3aR}+c_{3bR}),
\label{quadfive}
\end{align}
where $J$ denotes the frequently occurring combination defined by
\be
J=2c_{4c}+3c_{4e}-6c_{4j}.
\label{Jdef}
\ee
These owe their existence to relations like
\be
c_1\hX^{\lambda\gamma}_{3a,2}+d_2\hX^{\lambda\lambda}_{1,3a}+\chat_{3a}\hX^{\gamma\lambda}_{2,1}=0
\label{newrel}
\ee
together with similar relations for $3b$--$3f$, $3aR$, $3bR$; together with
\begin{align}
c_1\hX^{\lambda\gamma}_{2,3}+\dhat_3\hX^{\lambda\lambda}_{1,2}+c_{2}\hX^{\gamma\lambda}_{3,1}=&0,\nn
c_1\hX^{\lambda\gamma}_{2R,3}+\dhat_3\hX^{\lambda\lambda}_{1,2R}+c_{2R}\hX^{\gamma\lambda}_{3,1}=&0,\nn
c_1\hX^{\gamma\gamma}_{2,3}+\dhat_3\hX^{\lambda\gamma}_{1,2}+d_{2}\hX^{\gamma\lambda}_{3,1}=&0.
\label{newrela}
\end{align}
The number of invariants is as expected, since the eleven relations of the form Eqs.~\eqref{newrel}, \eqref{newrela} reduce the effective number of independent variations from 16 to 5, yielding 14-5=9 invariants (both quadratic and linear).
  
In the absence of a complete calculation, one may estimate the total number of invariants which will be found at five loops.
The five-loop $\beta$-function was calculated in Ref.~\cite{klein}, and contained contributions from 124 1PI 5-loop 4-point diagrams and 11 5-loop 2-point anomalous dimension diagrams, making 135 coefficients in total\footnote{The six-loop $\beta$-function was recently computed in Ref.~\cite{kompan}}. There are 67 independent variations at 5 loops, implying a naive expectation of 135-67=68 linear invariants.  On the other hand there are 57 5-loop $X$-type terms (some of which of course appear in Eq.~\eqref{delfive}), which following the argument explained at four loops implies an actual total of 135-57=78 linear invariants. But furthermore there are altogether 27 identities of the form Eqs.~\eqref{newrel}, \eqref{newrela},  constructed from the one one-loop quantity, the three two-loop quantities and the nine three-loop quantities. This implies an additional 27 quadratic invariants making 105 invariants in total. As at four loops, there are considerably more invariants than might have been expected.
One may also speculate on the possible existence of higher-order invariants based on higher-order Jacobi-style identities.

\section {One-vertex reducible graphs}
In this section we briefly discuss the issue of  $\beta$-function contributions from one-particle reducible (1VR) graphs. It is well-known that no such contributions arise using minimal subtraction within dimensional regularisation ($\msbar$), as may easily be established by  consideration of the diagram-by-diagram subtraction process. It would be convenient if when considering scheme redefinitions one could restrict attention to schemes which have the same feature. In fact, if we start from a scheme such as $\msbar$  in which the $\beta$-function coefficients corresponding to 4-point 1VR graphs $G_R$ are zero, i.e. $c_{G_R}=0$, it is clear from Eqs. \eqref{delba}, \eqref{delbb} that the simple conditions 
\be
\delta_{G_R}=0
\label{vdef}
\ee
 will ensure that 
the redefined coefficients will also satisfy $c'_{G_R}=0$.\footnote{There are no 1VR 2-point graphs and therefore there is no need to impose $\epsilon_{G_R}=0$.} This relies on the fact that for $L$, $L'$ loop graphs $G$, $G'$, with $L+L'\ge3$,  if (in the notation of the appendix) ${\cal L}_GG'$ contains 1VR graphs, then at least one of $G$ or $G'$ must itself be 1VR. We therefore have a simple all-orders prescription given by Eq.~\eqref{vdef} for defining schemes with no 1VR contributions.

The redefined coupling as given by Eqs.~\eqref{gprime}, \eqref{Cdef} turns out to adopt a simple form when $c_{G_R}=\delta_{G_R}=0$. We assume that $f(g)$, $c(g)$ in Eqs.~\eqref{gprime}, \eqref{Cdef} are given by similar diagrammatic series to those for the $\beta$-function and anomalous dimension, but with $c_X\go \tdelta_X$ and $d_X\go \tepsilon_X$.
At one loop we simply find $\tdelta_1=\delta_1$. At two loops we find 
\begin{align}
\tdelta_2=&\delta_2+\delta_1^2,\nn
\tdelta_{2R}=&\delta_{2R}+\tdelta_1^2,
\label{vtwo}
\end{align}
so that the condition for 1VI graphs is 
\be
c_{2R}=0,\quad \tdelta_{2R}=\tdelta_1^2, 
\label{nobub}
\ee
At three loops 
\begin{align}
\tdelta_{3a}=&\delta_{3a}+\delta_1(\delta_2+\delta_{2R})+\tfrac23\delta_1^3,\nn
\tdelta_{3b}=&\delta_{3b}+\delta_1\epsilon_2,\nn
\tdelta_{3c}=&\delta_{3c}+\delta_1(\delta_2+\delta_{2R})+\tfrac23\delta_1^3,\nn
\tdelta_{3d}=&\delta_{3d}+\delta_1\delta_2+\tfrac23\delta_1^3,\nn
\tdelta_{3e}=&\delta_{3e}+2\delta_1\delta_2+\tfrac23\delta_1^3,\nn
\tdelta_{3aR}=&\delta_{3aR}+\tfrac52\delta_1\delta_{2R}+\delta_1^3,\nn
\tdelta_{3bR}=&\delta_{3bR}+\delta_1(\delta_2+\delta_{2R})+\delta_1^3,\nn
\tepsilon_3=&\epsilon_3+3\delta_1\epsilon_2.
\label{vthree}
\end{align}
It is easy to confirm using Eq.~\eqref{vtwo} that $\delta_{2R}=\delta_{3aR}=\delta_{3bR}=0$ corresponds to
\be
\tdelta_{3aR}=\tdelta_1^3,\quad \tdelta_{3bR}=\tdelta_1\tdelta_2.
\label{nobuba}
\ee
The emerging pattern is clear; the value for $\tdelta_{G_R}$ is the product of the $\tdelta$s for its 1VI subgraphs.
At four loops we find
\begin{align}
\tdelta_{4aR}=&\delta_{4aR}+3\delta_1\delta_{3aR}+\tfrac32\delta_{2R}^2+\tfrac{13}{3}\delta_1^2\delta_{2R}+\delta_1^4,\nn
\tdelta_{4bR}=&\delta_{4bR}+2\delta_1\delta_{3bR}+\delta_2^2+\tfrac43\delta_1^2\delta_{2R}+2\delta_1^2\delta_2+\delta_1^4,\nn
\tdelta_{4cR}=&\delta_{4cR}+\delta_1\delta_{3aR}+\tfrac32\delta_1\delta_{3bR}+\delta_1^2\delta_2+\tfrac83\delta_1^2\delta_{2R}+\delta_1^4,\nn
\tdelta_{4dR}=&\delta_{4dR}+\delta_1\delta_{3c}+\delta_1\delta_{3bR}+\delta_{2R}^2+\delta_1^2\delta_2+\tfrac53\delta_1^2\delta_{2R}+\tfrac23\delta_1^4,\nn
\tdelta_{4eR}=&\delta_1\delta_{3b}+\epsilon_2\delta_{2R}+\delta_1^2\epsilon_2,\nn
\tdelta_{4fR}=&\delta_{4fR}+\delta_1\delta_{3a}+\delta_1\delta_{3aR}+\tfrac12\delta_1\delta_{3bR}+\delta_1^2\delta_2+2\delta_1^2\delta_{2R}+\tfrac23\delta_1^4,\nn
\tdelta_{4gR}=&\delta_{4gR}+\delta_1\delta_{3e}+\delta_1\delta_{3bR}+2\delta_1^2\delta_2+\tfrac23\delta_1^2\delta_{2R}+\tfrac23\delta_1^4,
\end{align}
Using Eqs.~\eqref{vthree}, \eqref{vtwo} we find that $\delta_{G_R}=0$ up to this level corresponds to taking
\begin{align}
\tdelta_{4aR}=\tdelta_1^4,\quad \tdelta_{4bR}=\tdelta_2^2,&\quad \tdelta_{4cR}=\tdelta_1^2\tdelta_2,\quad
\tdelta_{4dR}=\tdelta_1\tdelta_{3c},\nn\tdelta_{4eR}=\tdelta_1\tdelta_{3b},\quad \tdelta_{4fR}=&\tdelta_1\tdelta_{3a},\quad \tdelta_{4gR}=\tdelta_1\tdelta_{3e},
\label{nobubb}
\end{align}
so that each four-loop 1VR $\delta$ is the product of the $\delta$s for its 1VI subgraphs, as expected. It seems highly likely that this simple pattern persists to all orders, but we have not been able to construct a proof.

When considering the scheme invariants, we can therefore restrict ourselves to those schemes with $c_{G_R}=0$. The counting of invariants is then slightly different.  Upon setting $c_{3aR}=c_{3bR}=0$ in Eq.~\eqref{invthree}, there are then just three invariant combinations, namely $I_1^{(3\prime)}=c_{3a}+c_{3d}$, 
$I_3^{(3)}$ and $I_4^{(3)}$. We have lost two coefficients ($c_{3aR}$ and $c_{3bR}$) and one independent variation ($X^{\lambda\lambda}_{1,2R}$) and so we expect to lose $2-1=1$ invariants.

The pattern is similar at four loops; if we impose Eq.~\eqref{nobuba}, then we have $\delta c_{4aR-4gR}=0$ and so we can can consistently set $c_{4aR-4gR}=0$ in Eq.~\eqref{invfour}. We now have 23 coefficients and the 14 variations 
$\tdelta_{3a-3f}$, $\tepsilon_3$, $\tdelta_1^3$, $\tdelta_1\tdelta_2$,  $\tdelta_1\tepsilon_2$,  $\tdelta_2$, $\tepsilon_2$, $\tdelta_1^2$, $\tdelta_1$, leading to a naive expectation of 23-14=9 invariants.  On the other hand, out of the original eighteen linear invariants in Eq.~\eqref{invfour} we are left with eleven invariant linear combinations, plus the four individual invariants, making 15. Again this is as anticipated, since we have lost the seven coefficients  $c_{4aR-4gR}$ and the four independent variations $X^{\lambda\lambda}_{1, 3aR}$, $X^{\lambda\lambda}_{1, 3bR}$,
$X^{\lambda\lambda}_{2, 2R}$ and $X^{\lambda\gamma}_{2R,2}$ so we lose $7-4=3$ invariant linear combinations. Furthermore it is clear that in the 1VI case only one of the identities in Eq.~\eqref{quadrels} remains, and consequently only one of the quadratic invariants in Eq.~\eqref{qinv} survives. The total number of invariants is therefore 16; once again, almost double the naively expected number.

Finally we can consistently set $c_{5aR}=c_{5bR}=c_{5cR}=0$ in Eq.~\eqref{invfive}, to obtain a invariant constructed solely from anomalous dimension coefficients
\be
I^{(5)\prime}_1=d_{5b}-2d_{5c}-2d_{5e}-2d_{5f}+4d_{5j}.
\ee

\section {Relation with Hopf algebra}
Scheme invariants may be described graphically by adopting and extending rules described by Panzer\cite{Panzer} using the Hopf algebra coproduct $\Delta:\G\go\G\otimes\G$, where $\G$ is the vector space  spanned by the set of connected 1PI superficially divergent graphs and the disconnected products of such graphs. The action of the coproduct $\Delta$ on a Feynman graph $g\in\G$ is defined by 
\be
\Delta g=\sum_ig_i\otimes g/g_i\quad\forall\quad \hbox{subgraphs}\quad g_i\subset g, g_i,g\in\G,\quad g_i\ne 1,g,\quad\hbox{otherwise}\quad \Delta g=\emptyset.
\ee
Here $g/g_i$ denotes the graph obtained from $g$ by contracting each connected 1PI graph in the subgraph to a single vertex, or a single line if the connected 1PI graph has two external lines. Further details and a general discussion will be presented in Ref.~\cite{jo}, but this brief overview is sufficient for our present purposes.
The invariants of Eqs.~\eqref{invfour}, \eqref{qinv} and \eqref{invfive} should correspond to combinations of graphs with a symmetric, or cocommutative, coproduct, following the general results of Ref.\cite{jo}. In this section we verify this by explicit calculation.
Firstly, we readily derive the following useful results:
At three loops
\begin{align}
\Delta(g_{\lambda}\!{}^{3a})=&g_{\lambda}\!{}^1\otimes g_{\lambda}\!{}^{2R}+2g_{\lambda}\!{}^2\otimes g_{\lambda}\!{}^1,\nn
\Delta(g_{\lambda}\!{}^{3b})=&g_{\gamma}\!{}^{2}\otimes g_{\lambda}\!{}^1,\nn
\Delta(g_{\lambda}\!{}^{3c})=&2g_{\lambda}\!{}^1\otimes g_{\lambda}\!{}^2+g_{\lambda}\!{}^{2R}\otimes g_{\lambda}\!{}^1,\nn
\Delta(g_{\lambda}\!{}^{3d})=&2g_{\lambda}\!{}^1\otimes g_{\lambda}\!{}^2+(g_{\lambda}^1)^2\otimes g_{\lambda}^1,\nn
\Delta(g_{\lambda}\!{}^{3e})=&g_{\lambda}\!{}^1\otimes g_{\lambda}\!{}^2+g_{\lambda}\!{}^2\otimes g_{\lambda}\!{}^1,\nn
\Delta(g_{\lambda}\!{}^{3f})=&0,\nn
\Delta(g_{\gamma}\!{}^{3})=&2g_{\lambda}\!{}^1\otimes g_{\gamma}\!{}^{2},\nn
\Delta(g_{\lambda}\!{}^{3aR})=&3g_{\lambda}\!{}^1\otimes g_{\lambda}\!{}^{2R}+2g_{\lambda}\!{}^{2R}\otimes g_{\lambda}\!{}^{1}+(g_{\lambda}^1)^2\otimes g_{\lambda}^1,\nn
\Delta(g_{\lambda}\!{}^{3bR})=&g_{\lambda}\!{}^1\otimes g_{\lambda}\!{}^{2R}+g_{\lambda}\!{}^1\otimes g_{\lambda}\!{}^2+g_{\lambda}\!{}^2\otimes g_{\lambda}\!{}^1+(g_{\lambda}^1)^2\otimes g_{\lambda}^1,
\end{align}
and at four loops we have for the 4-point graphs 
\begin{align}
\Delta(g_{\lambda}\!{}^{4a})=&g_{\lambda}\!{}^1\otimes g_{\lambda}\!{}^{3a}+2g_{\lambda}\!{}^{3e}\otimes g_{\lambda}\!{}^1 +g_{\lambda}\!{}^2\otimes g_{\lambda}\!{}^{2R},\nn
\Delta(g_{\lambda}\!{}^{4b})=&2g_{\lambda}\!{}^1\otimes g_{\lambda}\!{}^{3a}+2g_{\lambda}\!{}^{3c}\otimes g_{\lambda}\!{}^1+g_{\lambda}\!{}^{2R}\otimes g_{\lambda}\!{}^{2R},\nn
\Delta(g_{\lambda}\!{}^{4c})=&2g_{\lambda}\!{}^1\otimes g_{\lambda}\!{}^{3b}+g_{\gamma}\!{}^{3}\otimes g_{\lambda}\!{}^1,\nn
\Delta(g_{\lambda}\!{}^{4d})=&g_{\lambda}\!{}^1\otimes g_{\lambda}\!{}^{3a}+g_{\lambda}\!{}^1\otimes g_{\lambda}\!{}^{3bR}+g_{\lambda}\!{}^{3d}\otimes g_{\lambda}\!{}^1+g_{\lambda}\!{}^2\otimes g_{\lambda}\!{}^2\nn
&+(g_{\lambda}\!{}^1)^2\otimes g_{\lambda}\!{}^{2R}+g_{\lambda}\!{}^1g_{\lambda}\!{}^2\otimes g_{\lambda}\!{}^1,\nn
\Delta(g_{\lambda}\!{}^{4e})=&g_{\gamma}\!{}^{2}\otimes g_{\lambda}\!{}^2+g_{\lambda}\!{}^{3b}\otimes g_{\lambda}\!{}^1,\nn
\Delta(g_{\lambda}\!{}^{4f})=&g_{\lambda}\!{}^{3a}\otimes g_{\lambda}\!{}^1+g_{\lambda}\!{}^1\otimes g_{\lambda}\!{}^{3c}+2g_{\lambda}\!{}^2\otimes g_{\lambda}\!{}^2,\nn
\Delta(g_{\lambda}\!{}^{4g})=&3g_{\lambda}\!{}^1\otimes g_{\lambda}\!{}^{3c}+g_{\lambda}\!{}^{3aR}\otimes g_{\lambda}\!{}^1+2g_{\lambda}\!{}^{2R}\otimes g_{\lambda}\!{}^2+(g_{\lambda}\!{}^1)^2\otimes g_{\lambda}\!{}^2,\nn
\Delta(g_{\lambda}\!{}^{4h})=&\Delta(g_{\lambda}\!{}^{4i})= 2g_{\lambda}\!{}^1\otimes g_{\lambda}\!{}^{3e}+g_{\lambda}\!{}^{3d}\otimes g_{\lambda}\!{}^1+(g_{\lambda}\!{}^1)^2\otimes g_{\lambda}\!{}^2,\nn\Delta(g_{\lambda}\!{}^{4j})=&g_{\lambda}\!{}^1\otimes g_{\lambda}\!{}^{3b}+g_{\gamma}\!{}^{2}\otimes g_{\lambda}\!{}^2+g_{\lambda}\!{}^1g_{\gamma}\!{}^{2}\otimes g_{\lambda}\!{}^1,\nn
\Delta(g_{\lambda}\!{}^{4k})=&g_{\lambda}\!{}^1\otimes g_{\lambda}\!{}^{3c}+g_{\lambda}\!{}^1\otimes g_{\lambda}\!{}^{3e}+g_{\lambda}\!{}^{3bR}\otimes g_{\lambda}\!{}^1+g_{\lambda}\!{}^2\otimes g_{\lambda}\!{}^2+(g_{\lambda}\!{}^1)^2\otimes g_{\lambda}\!{}^2,\nn
\Delta(g_{\lambda}\!{}^{4l})=&2g_{\lambda}\!{}^1\otimes g_{\lambda}\!{}^{3e}+g_{\lambda}\!{}^{2R}\otimes g_{\lambda}\!{}^2+g_{\lambda}\!{}^{3c}\otimes g_{\lambda}\!{}^1,\nn
\Delta(g_{\lambda}\!{}^{4m})=&\Delta(g_{\lambda}\!{}^{4n})=g_{\lambda}\!{}^1\otimes g_{\lambda}\!{}^{3e}
+g_{\lambda}\!{}^{3e}\otimes g_{\lambda}\!{}^{1} +g_{\lambda}\!{}^2\otimes g_{\lambda}\!{}^2,\nn
\Delta(g_{\lambda}\!{}^{4o})=&g_{\lambda}\!{}^1\otimes g_{\lambda}\!{}^{3c}+2g_{\lambda}\!{}^1\otimes g_{\lambda}\!{}^{3d}+g_{\lambda}\!{}^{2R}\otimes g_{\lambda}\!{}^2+2(g_{\lambda}\!{}^1)^2\otimes g_{\lambda}\!{}^2+g_{\lambda}\!{}^1g_{\lambda}\!{}^{2R}\otimes g_{\lambda}\!{}^1,\nn
\Delta(g_{\lambda}\!{}^{4p})=&g_{\lambda}\!{}^1\otimes g_{\lambda}\!{}^{3f},\nn
\Delta(g_{\lambda}\!{}^{4q})=&g_{\lambda}\!{}^{3f}\otimes g_{\lambda}\!{}^1,\nn
\Delta(g_{\lambda}\!{}^{4r})=&g_{\lambda}\!{}^1\otimes g_{\lambda}\!{}^{3d}+g_{\lambda}\!{}^1\otimes g_{\lambda}\!{}^{3e}+g_{\lambda}\!{}^2\otimes g_{\lambda}\!{}^2+(g_{\lambda}\!{}^1)^2\otimes g_{\lambda}\!{}^2+g_{\lambda}\!{}^1g_{\lambda}\!{}^2\otimes g_{\lambda}\!{}^1,\nn
\Delta(g_{\lambda}\!{}^{4s})=&0,\nn
\Delta(g_{\lambda}\!{}^{4aR})=&4g_{\lambda}\!{}^1\otimes g_{\lambda}\!{}^{3aR}+2g_{\lambda}\!{}^{3aR}\otimes g_{\lambda}\!{}^1+3g_{\lambda}\!{}^{2R}\otimes g_{\lambda}\!{}^{2R}\nn
&+3(g_{\lambda}\!{}^1)^2\otimes g_{\lambda}\!{}^{2R}+2g_{\lambda}\!{}^1g_{\lambda}\!{}^{2R}\otimes g_{\lambda}\!{}^1,\nn
\Delta(g_{\lambda}\!{}^{4bR})=&2g_{\lambda}\!{}^1\otimes g_{\lambda}\!{}^{3bR}+2g_{\lambda}\!{}^2\otimes g_{\lambda}\!{}^2+(g_{\lambda}\!{}^1)^2\otimes g_{\lambda}\!{}^{2R}+2g_{\lambda}\!{}^1g_{\lambda}\!{}^2\otimes g_{\lambda}\!{}^1,\nn
\Delta(g_{\lambda}\!{}^{4cR})=&g_{\lambda}\!{}^1\otimes g_{\lambda}\!{}^{3aR}+2g_{\lambda}\!{}^1\otimes g_{\lambda}\!{}^{3bR}+g_{\lambda}\!{}^{3bR}\otimes g_{\lambda}\!{}^1+g_{\lambda}\!{}^2\otimes g_{\lambda}\!{}^{2R}+g_{\lambda}\!{}^{2R}\otimes g_{\lambda}\!{}^2\nn
&+2(g_{\lambda}\!{}^1)^2\otimes g_{\lambda}\!{}^{2R}
+g_{\lambda}\!{}^1g_{\lambda}\!{}^2\otimes g_{\lambda}\!{}^1+g_{\lambda}\!{}^1g_{\lambda}\!{}^{2R}\otimes g_{\lambda}\!{}^1,\nn
\Delta(g_{\lambda}\!{}^{4dR})=&2g_{\lambda}\!{}^1\otimes g_{\lambda}\!{}^{3bR}+g_{\lambda}\!{}^1\otimes g_{\lambda}\!{}^{3c}+g_{\lambda}\!{}^{3c}\otimes g_{\lambda}\!{}^{1}+g_{\lambda}\!{}^{2R}\otimes g_{\lambda}\!{}^{2R}\nn
&+2(g_{\lambda}\!{}^1)^2\otimes g_{\lambda}\!{}^2+g_{\lambda}\!{}^1g_{\lambda}\!{}^{2R}\otimes g_{\lambda}\!{}^1,\nn
\Delta(g_{\lambda}\!{}^{4eR})=&g_{\gamma}\!{}^{2}\otimes g_{\lambda}\!{}^{2R}+g_{\lambda}\!{}^1\otimes g_{\lambda}\!{}^{3b}+g_{\lambda}\!{}^{3b}\otimes g_{\lambda}\!{}^{1}+g_{\lambda}\!{}^1g_{\gamma}\!{}^{2}\otimes g_{\lambda}\!{}^1,\nn
\Delta(g_{\lambda}\!{}^{4fR})=&g_{\lambda}\!{}^1\otimes g_{\lambda}\!{}^{3aR}+g_{\lambda}\!{}^{3bR}\otimes g_{\lambda}\!{}^1+g_{\lambda}\!{}^1\otimes g_{\lambda}\!{}^{3a}+g_{\lambda}\!{}^{3a}\otimes g_{\lambda}\!{}^{1}+2g_{\lambda}\!{}^2\otimes g_{\lambda}\!{}^{2R}\nn
&+(g_{\lambda}\!{}^1)^2\otimes g_{\lambda}\!{}^{2R}+g_{\lambda}\!{}^1g_{\lambda}\!{}^2\otimes g_{\lambda}\!{}^1,\nn
\Delta(g_{\lambda}\!{}^{4gR})=&g_{\lambda}\!{}^1\otimes g_{\lambda}\!{}^{3bR}+g_{\lambda}\!{}^2\otimes g_{\lambda}\!{}^{2R}+g_{\lambda}\!{}^1\otimes g_{\lambda}\!{}^{3e}+g_{\lambda}\!{}^{3e}\otimes g_{\lambda}\!{}^{1}\nn
&+(g_{\lambda}\!{}^1)^2\otimes g_{\lambda}\!{}^2+g_{\lambda}\!{}^1g_{\lambda}\!{}^2\otimes g_{\lambda}\!{}^1,
\end{align}
and for the 2-point graphs
\begin{align}\Delta(g_{\gamma}\!{}^{4a})=&g_{\gamma}\!{}^{2}\otimes g_{\gamma}\!{}^{2},\nn
\Delta(g_{\gamma}\!{}^{4b})=&3g_{\lambda}\!{}^1\otimes g_{\gamma}\!{}^{3}+2g_{\lambda}\!{}^{2R}\otimes g_{\gamma}\!{}^{2}+(g_{\lambda}\!{}^1)^2\otimes g_{\gamma}\!{}^{2},\nn
\Delta(g_{\gamma}\!{}^{4c})=&g_{\lambda}\!{}^1\otimes g_{\gamma}\!{}^{3}+2g_{\lambda}\!{}^2\otimes g_{\gamma}\!{}^{2},\nn
\Delta(g_{\gamma}\!{}^{4d})=&2g_{\lambda}\!{}^1\otimes g_{\gamma}\!{}^{3}+2g_{\lambda}\!{}^2\otimes g_{\gamma}\!{}^{2}+(g_{\lambda}\!{}^1)^2\otimes g_{\gamma}\!{}^{2}.
\end{align}
At five loops, the basic co-products are
\begin{align}
\Delta(g_{\gamma}^{5a})=&g_{\lambda}^{3f}\otimes g_{\gamma}^2,\nn
\Delta(g_{\gamma}^{5b})=&2g_{\lambda}^{3e}\otimes g_{\gamma}^2+g_{\lambda}^2\otimes g_{\gamma}^3
+g_{\lambda}^1\otimes g_{\gamma}^{4c},\nn
\Delta(g_{\gamma}^{5c})=&2g_{\lambda}^{3e}\otimes g_{\gamma}^2+2g_{\lambda}^2\otimes g_{\gamma}^3
+2g_{\lambda}^1\otimes g_{\gamma}^{4d}+(g_{\lambda}^1)^2\otimes g_{\gamma}^3
+2g_{\lambda}^1g_{\lambda}^2\otimes g_{\gamma}^2,\nn
\Delta(g_{\gamma}^{5d})=&g_{\gamma}^{3}\otimes g_{\gamma}^2
+2g_{\lambda}^1\otimes g_{\gamma}^{4a},\nn
\Delta(g_{\gamma}^{5e})=&2g_{\lambda}^{3c}\otimes g_{\gamma}^2+g_{\lambda}^{2R}\otimes g_{\gamma}^3
+2g_{\lambda}^1\otimes g_{\gamma}^{4c},\nn
\Delta(g_{\gamma}^{5f})=&2g_{\lambda}^{3aR}\otimes g_{\gamma}^2+3g_{\lambda}^{2R}\otimes g_{\gamma}^3
+4g_{\lambda}^1\otimes g_{\gamma}^{4b}+3(g_{\lambda}^1)^2\otimes g_{\gamma}^3
+2g_{\lambda}^1g_{\lambda}^{2R}\otimes g_{\gamma}^2,\nn
\Delta(g_{\gamma}^{5g})=&g_{\gamma}^{2}\otimes g_{\gamma}^3
+2g_{\lambda}^1\otimes g_{\gamma}^{4a}+2g_{\lambda}^1g_{\gamma}^{2}\otimes g_{\gamma}^2,\nn
\Delta(g_{\gamma}^{5h})=&g_{\lambda}^{3d}\otimes g_{\gamma}^2+g_{\lambda}^{3e}\otimes g_{\gamma}^2+g_{\lambda}^{2}\otimes g_{\gamma}^3
+g_{\lambda}^1\otimes g_{\gamma}^{4c}+g_{\lambda}^1\otimes g_{\gamma}^{4d}\nn
&+(g_{\lambda}^1)^2\otimes g_{\gamma}^3
+g_{\lambda}^1g_{\lambda}^{2}\otimes g_{\gamma}^2,\nn
\Delta(g_{\gamma}^{5i})=&g_{\lambda}^{3a}\otimes g_{\gamma}^2+g_{\lambda}^{3bR}\otimes g_{\gamma}^2+2g_{\lambda}^{2}\otimes g_{\gamma}^3
+g_{\lambda}^1\otimes g_{\gamma}^{4b}+g_{\lambda}^1\otimes g_{\gamma}^{4c}\nn
&+(g_{\lambda}^1)^2\otimes g_{\gamma}^3
+g_{\lambda}^1g_{\lambda}^{2}\otimes g_{\gamma}^2,\nn
\Delta(g_{\gamma}^{5j})=&g_{\lambda}^{3c}\otimes g_{\gamma}^2+g_{\lambda}^{3bR}\otimes g_{\gamma}^2+g_{\lambda}^{2}\otimes g_{\gamma}^3+g_{\lambda}^{2R}\otimes g_{\gamma}^3
+g_{\lambda}^1\otimes g_{\gamma}^{4b}+2g_{\lambda}^1\otimes g_{\gamma}^{4d}\nn
&+2(g_{\lambda}^1)^2\otimes g_{\gamma}^3
+g_{\lambda}^1g_{\lambda}^{2}\otimes g_{\gamma}^2+g_{\lambda}^1g_{\lambda}^{2R}\otimes g_{\gamma}^2,\nn
\Delta(g_{\gamma}^{5k})=&g_{\lambda}^{3b}\otimes g_{\gamma}^2+g_{\gamma}^2\otimes g_{\gamma}^3
+g_{\lambda}^1\otimes g_{\gamma}^{4a}
+g_{\lambda}^1g_{\gamma}^2\otimes g_{\gamma}^2,\nn
\Delta(g_{\lambda}^{5aR})=&2g_{\lambda}^1\otimes g_{\lambda}^{4eR}+g_{\lambda}^{4eR}\otimes g_{\lambda}^{1}+g_{\gamma}^2\otimes g_{\lambda}^{3aR}+g_{\lambda}^{3b}\otimes g_{\lambda}^{2R}+ g_{\lambda}^{2R}\otimes g_{\lambda}^{3b}\nn
&+g_{\lambda}^1g_{\lambda}^{3b}\otimes g_{\lambda}^1+2g_{\lambda}^1g_{\gamma}^2\otimes g_{\lambda}^{2R}+g_{\lambda}^{2R}g_{\gamma}^2\otimes g_{\lambda}^1,\nn
\Delta(g_{\lambda}^{5bR})=&g_{\lambda}^1\otimes g_{\lambda}^{4eR}+g_{\gamma}^2\otimes g_{\lambda}^{3bR}+g_{\lambda}^{3b}\otimes g_{\lambda}^{2}+ g_{\lambda}^{2}\otimes g_{\lambda}^{3b}\nn
&+g_{\lambda}^1g_{\lambda}^{3b}\otimes g_{\lambda}^1+g_{\lambda}^1g_{\gamma}^2\otimes g_{\lambda}^{2R}+g_{\lambda}^{2}g_{\gamma}^2\otimes g_{\lambda}^1,\nn
\Delta(g_{\lambda}^{5cR})=&2g_{\lambda}^1\otimes g_{\lambda}^{4eR}+g_{\gamma}^3\otimes g_{\lambda}^{2R}+g_{\lambda}^{4c}\otimes g_{\lambda}^1+ g_{\lambda}^1\otimes g_{\lambda}^{4c}\nn
&+2(g_{\lambda}^1)^2\otimes g_{\lambda}^{3b}+g_{\lambda}^1g_{\gamma}^3\otimes g_{\lambda}^1.
\end{align}

At three loops, the coproducts for $g_{\lambda}\!{}^{3e}$ and $g_{\lambda}\!{}^{3f}$ are cocommutative and zero respectively, corresponding to the individual invariance of $c_{3e}$, $c_{3f}$. Corresponding to the invariants in Eq.~\eqref{invthree} we have the following combinations with cocommutative coproducts:
\begin{align}
\Delta(g_{\lambda}\!{}^{3a}+g_{\lambda}\!{}^{3d}-g_{\lambda}\!{}^{3aR})=&2g_{\lambda}\!{}^1\otimes_s g_{\lambda}\!{}^2-2g_{\lambda}\!{}^1\otimes_s g_{\lambda}\!{}^{2R},\nn
\Delta(g_{\lambda}\!{}^{3aR}-g_{\lambda}\!{}^{3bR})=&2g_{\lambda}\!{}^1\otimes_s g_{\lambda}\!{}^{2R}-g_{\lambda}\!{}^1\otimes_s g_{\lambda}\!{}^2,\nn
\Delta(g_{\lambda}\!{}^{3a}+g_{\lambda}\!{}^{3c})=&2g_{\lambda}\!{}^1\otimes_s g_{\lambda}\!{}^2+g_{\lambda}\!{}^1\otimes_s g_{\lambda}\!{}^{2R},\nn
\Delta(2g_{\lambda}\!{}^{3b}+g_{\gamma}\!{}^{3})=&2g_{\lambda}\!{}^1\otimes_s g_{\gamma}\!{}^{2},
\label{hopfthree}
\end{align}
where
\be
G_1\otimes_s G_2=G_1\otimes G_2+G_2\otimes G_1.
\ee
The scheme-invariant combination of RG coefficients corresponding to a combination of graphs 
$\sum_i\alpha_ig^i_{\lambda}+\sum_j\tilde\alpha_jg^j_{\gamma}$ with a cocommutative coproduct is\cite{jo}
$\sum_i\alpha_iS_ic_i+\sum_j\tilde\alpha_jS'_jd_j$
where $S_i$ are the symmetry factors for the 4-point graphs, and $S'_i$ those for the 2-point graphs. The relevant symmetry factors at this loop order are given by
\be
S_{3f}=1,\quad S_{3e}=2,\quad
S'_3=S_{3a}=S_{3c}=S_{3d}=S_{3bR}=4,\quad
S_{3b}=6,\quad S_{3aR}=8.
\label{symthree}
\ee
So for instance
\be
g_{\lambda}\!{}^{3a}+g_{\lambda}\!{}^{3d}-g_{\lambda}\!{}^{3aR}\go
4c_{3a}+4c_{3d}-8c_{3aR}
\ee
which agrees with $I^{(3)}_1$ in Eq.~\eqref{invthree} up to an overall factor.

At four loops, the coproducts for $g_{\lambda}\!{}^{4m}$, $g_{\lambda}\!{}^{4n}$ and $g_{\gamma}\!{}^{4a}$ are cocommutative and that for $g_{\lambda}\!{}^{4s}$ is zero, corresponding to the individual invariance of $c_{4m}$, $c_{4n}$, $c_{4s}$ and $d_{4a}$. Corresponding to the invariants in Eq.~\eqref{invfour} we have the following combinations with cocommutative coproducts:
\begin{align}
\Delta(g_{\lambda}\!{}^{4h}-g_{\lambda}\!{}^{4i})&=C_1^{(4)L},\nn
\Delta(g_{\lambda}\!{}^{4b}+2g_{\lambda}\!{}^{4f})&=C_2^{(4)L},\nn
\Delta(g_{\lambda}\!{}^{4a}+g_{\lambda}\!{}^{4f}+g_{\lambda}\!{}^{4l})&=C_3^{(4)L},\nn
\Delta(g_{\lambda}\!{}^{4l}+g_{\lambda}\!{}^{4o}-2g_{\lambda}\!{}^{4r}-g_{\lambda}\!{}^{4bR}+2g_{\lambda}\!{}^{4gR}-g_{\lambda}\!{}^1g_{\lambda}\!{}^{3c})&=C_4^{(4)L},\nn
\Delta(g_{\lambda}\!{}^{4c}+2g_{\lambda}\!{}^{4e}+g_{\gamma}\!{}^{4c})&=C_5^{(4)L},\nn
\Delta(g_{\lambda}\!{}^{4d}+g_{\lambda}\!{}^{4f}-g_{\lambda}\!{}^{4k}+g_{\lambda}\!{}^{4r}-g_{\lambda}\!{}^{4bR})&=C_6^{(4)L},\nn
\Delta(g_{\lambda}\!{}^{4b}-2g_{\lambda}\!{}^{4d}+g_{\lambda}\!{}^{4g}-g_{\lambda}\!{}^{4o}+g_{\lambda}\!{}^{4cR}+g_{\lambda}\!{}^{4gR})&=C_7^{(4)L},\nn
\Delta(g_{\lambda}\!{}^{4h}-g_{\lambda}\!{}^{4k}+g_{\lambda}\!{}^{4o}-g_{\lambda}\!{}^{4r}-g_{\lambda}\!{}^{4bR}+g_{\lambda}\!{}^{4gR}\nn
-g_{\lambda}\!{}^1g_{\lambda}\!{}^{3d}-g_{\lambda}\!{}^1g_{\lambda}\!{}^{3e}+g_{\lambda}\!{}^1g_{\lambda}\!{}^{3bR})
&=C_8^{(4)L},\nn
\Delta(g_{\lambda}\!{}^{4e}+g_{\lambda}\!{}^{4j}+2g_{\gamma}\!{}^{4c}-g_{\gamma}\!{}^{4d}+\tfrac12g_{\lambda}\!{}^1g_{\gamma}\!{}^{3})&=C_9^{(4)L}
,\nn
\Delta(g_{\lambda}\!{}^{4q}+g_{\lambda}\!{}^{4p})&=C_{10}^{(4)L},\nn
\Delta(g_{\gamma}\!{}^{4b}+3g_{\gamma}\!{}^{4c}-3g_{\gamma}\!{}^{4d}+2g_{\lambda}\!{}^{4eR}+g_{\lambda}\!{}^1g_{\gamma}\!{}^{3})
&=C_{11}^{(4)L},\nn
\Delta(g_{\lambda}\!{}^{4aR}+g_{\lambda}\!{}^{4bR}-2g_{\lambda}\!{}^{4cR})&=C_{12}^{(4)L},\nn
\Delta(g_{\lambda}\!{}^{4bR}-g_{\lambda}\!{}^{4dR}+g_{\lambda}\!{}^1g_{\lambda}\!{}^{3c})&=C_{13}^{(4)L},\nn
\Delta(2g_{\lambda}\!{}^{4bR}-g_{\lambda}\!{}^{4cR}+g_{\lambda}\!{}^{4fR}-2g_{\lambda}\!{}^{4gR}
-g_{\lambda}\!{}^1g_{\lambda}\!{}^{3a})
&=C_{14}^{(4)L}.
\label{hopffour}
\end{align}
Here, rather than give explicit expressions on the right-hand side, we use $C_i^{(l)L}\in\Gcal\otimes_s\Gcal$ to denote $l$-loop cocommutative coproducts corresponding to linear invariants. Since their exact form is not especially significant, we relegate the full expressions to Appendix B. The noteworthy new feature here is the necessity sometimes to add quadratic terms, of course with no counterpart in the original linear invariants of Eq.~\eqref{invfour}, on the left-hand side in order to obtain co-commutative results. The need for this is explained in general in Ref.~\cite{jo}.

Corresponding to the quadratic invariants in Eq.~\eqref{qinv} we have
\begin{align}
\Delta(2g_{\lambda}\!{}^1g_{\gamma}\!{}^{4c}-g_{\lambda}\!{}^1g_{\gamma}\!{}^{4d}+g_{\gamma}\!{}^{2}g_{\lambda}\!{}^{3d}+2g_{\lambda}\!{}^2g_{\lambda}\!{}^{3b}
-(g_{\lambda}\!{}^1)^2g_{\lambda}\!{}^{3b})&=C_1^{(4)Q},\nn
\Delta(g_{\lambda}\!{}^1g_{\lambda}\!{}^{4eR}-g_{\lambda}\!{}^{2R}g_{\lambda}\!{}^{3b}-g_{\gamma}\!{}^{2}g_{\lambda}\!{}^{3bR})&=C_2^{(4)Q},\nn
\Delta[g_{\lambda}\!{}^1(g_{\lambda}\!{}^{4dR}-2g_{\lambda}\!{}^{4gR})+2g_{\lambda}\!{}^2
g_{\lambda}\!{}^{3aR}-g_{\lambda}\!{}^{2R}g_{\lambda}\!{}^{3d}]&=C_3^{(4)Q}.
\label{hopffourq}
\end{align}
Here we see the need for additional cubic terms on the left-hand side, in addition to the quadratic terms corresponding to those in the invariant. The relevant graph combination corresponding to the additional invariant in Eq.~\eqref{invext} may be derived from those already given and hence is not displayed here.
Here we use $C_i^{(l)Q}\in\Gcal\otimes_s\Gcal$ to denote $l$-loop cocommutative coproducts corresponding to quadratic  invariants. The coefficients of the linear invariants in Eq.~\eqref{invfour} may be obtained from the linear terms on the left-hand side of Eq.~\eqref{hopffour}
by substitutions similar to those described at three loops after Eq.~\eqref{hopfthree}. Likewise, the coefficients of the quadratic invariants in Eq.~\eqref{qinv} may be obtained from the quadratic  terms on the left-hand side of Eq.~\eqref{hopffourq} by similar substitutions. Here the relevant symmetry factors are given by
\begin{align}
S_{4s}=1,\quad S_1=S_2=S_{4a}=&S_{4m}=S_{4n}=S_{4p}=S_{4q}=2,\nn
S_{2R}=S_{4bR}=&S_{4gR}=S_{4c}=S_{4d}=S_{4f}=S_{4h}=S_{4i}\nn
=S_{4k}=S_{4l}=S_{4r}=&S'_{4c}=S'_{4d}=4,\nn
S'_2=S_{4e}=6,&\quad S_{4eR}=S_{4j}=S'_{4a}=12,\nn
S_{4cR}=S_{4dR}=S_{4fR}=S_{4b}=&S_{4g}=S_{4o}=S'_{4b}=8,\quad S_{4aR}=16,
\label{symfour}
\end{align}
together with those in Eq.~\eqref{symthree}.
We also find corresponding to Eq.~\eqref{invfive}
\begin{align}
\Delta(4g_{\gamma}\!{}^{5b}-4g_{\gamma}\!{}^{5c}-2g_{\gamma}\!{}^{5e}-g_{\gamma}\!{}^{5f}
+4g_{\gamma}\!{}^{5j}-2g_{\lambda}\!{}^{5aR}+4g_{\lambda}\!{}^{5bR}&\nn
-g_{\lambda}\!{}^{5cR}
+g_{\lambda}\!{}^1g_{\lambda}\!{}^{4c}-4g_{\lambda}\!{}^2g_{\lambda}\!{}^{3b}+2g_{\lambda}\!{}^{2R}g_{\lambda}\!{}^{3b})
&=C_1^{(5)L}.
\label{hopffive}
\end{align}
Corresponding to the quadratic invariants in Eq.~\eqref{quadfive}, we find
\begin{align}
\Delta[g_{\lambda}\!{}^1g_{\lambda}\!{}^{5a}+g_{\gamma}\!{}^2g_{\lambda}\!{}^{4p}+g_{\lambda}\!{}^{3b}
g_{\lambda}\!{}^{3f}]&=C_1^{(5)Q},\nn
\Delta[g_{\lambda}\!{}^1(g_{\gamma}\!{}^{5g}-2g_{\gamma}\!{}^{5k})-g_{\gamma}\!{}^2g_{\lambda}\!{}^{4c}+g_{\gamma}\!{}^3g_{\lambda}\!{}^{3b}]
&=C_2^{(5)Q}\nn
\Delta[g_{\lambda}\!{}^1(g_{\lambda}\!{}^{5aR}-g_{\lambda}\!{}^{5bR})+g_{\gamma}\!{}^2(-g_{\lambda}\!{}^{4aR}+g_{\lambda}\!{}^{4cR})
+g_{\lambda}\!{}^{3b}(-g_{\lambda}\!{}^{3aR}+g_{\lambda}\!{}^{3bR})]
&=C_3^{(5)Q},\nn
\Delta[g_{\lambda}\!{}^1(g_{\gamma}\!{}^{5d}-g_{\gamma}\!{}^{5g})-2g_{\gamma}\!{}^2G_J
-\tfrac12(g_{\gamma}\!{}^{3})^2+g_{\lambda}\!{}^1g_{\gamma}\!{}^2g_{\gamma}\!{}^3]
&=C_{4}^{(5)Q},\nn
\Delta[g_{\lambda}\!{}^1(2g_{\lambda}\!{}^{5bR}-g_{\lambda}\!{}^{5cR})+g_{\lambda}\!{}^{2R}G_J
+g_{\gamma}\!{}^2(g_{\lambda}\!{}^{4aR}-2g_{\lambda}\!{}^{4cR})\nn+g_{\gamma}\!{}^3g_{\lambda}\!{}^{3bR}
+(g_{\lambda}\!{}^1)^2(g_{\lambda}\!{}^{4j}-g_{\lambda}\!{}^{4e}-g_{\lambda}\!{}^{4eR})+g_{\lambda}\!{}^1g_{\lambda}\!{}^{2R}
g_{\lambda}\!{}^{3b}]&=C_5^{(5)Q},\nn
\Delta[g_{\lambda}\!{}^1(g_{\gamma}\!{}^{5c}+g_{\gamma}\!{}^{5e}
-2g_{\gamma}\!{}^{5h})
-g_{\gamma}\!{}^2(g_{\lambda}\!{}^{4b}-2g_{\lambda}\!{}^{4d}-g_{\lambda}\!{}^{4aR}+2g_{\lambda}\!{}^{4cR})+g_{\lambda}\!{}^{2R}G_J&\nn
-g_{\gamma}\!{}^3(g_{\lambda}\!{}^{3c}-g_{\lambda}\!{}^{3d})
+(g_{\lambda}\!{}^1)^2(g_{\lambda}\!{}^{4j}-g_{\lambda}\!{}^{4c}-g_{\lambda}\!{}^{4e}-g_{\lambda}\!{}^{4eR})+g_{\lambda}\!{}^1g_{\lambda}\!{}^{2R}
g_{\lambda}\!{}^{3b}]&=C_{6}^{(5)Q},\nn
\Delta[g_{\lambda}\!{}^1(2g_{\gamma}\!{}^{5b}-g_{\gamma}\!{}^{5e})+g_{\gamma}\!{}^2(
g_{\lambda}\!{}^{4l}-g_{\lambda}\!{}^{4o}+2g_{\lambda}\!{}^{4r})&\nn
+(2g_{\lambda}\!{}^2-g_{\lambda}\!{}^{2R})G_J
+4g_{\lambda}\!{}^{3b}g_{\lambda}\!{}^{3e}
+g_{\lambda}\!{}^{3c}g_{\gamma}\!{}^3]&=C_{7}^{(5)Q},\nn
\Delta[g_{\lambda}\!{}^1(2g_{\gamma}\!{}^{5e}+g_{\gamma}\!{}^{5f}
-4g_{\gamma}\!{}^{5i})
-g_{\gamma}\!{}^2(2g_{\lambda}\!{}^{4b}-3g_{\lambda}\!{}^{4aR}+4g_{\lambda}\!{}^{4cR})&\nn
+(5g_{\lambda}\!{}^{2R}-8g_{\lambda}\!{}^2)G_J
+g_{\gamma}\!{}^{3}(2g_{\lambda}\!{}^{3a}-2g_{\lambda}\!{}^{3c}
-g_{\lambda}\!{}^{3aR}+2g_{\lambda}\!{}^{3bR})&\nn
+(g_{\lambda}\!{}^{2}-g_{\lambda}\!{}^{2R})g_{\lambda}\!{}^1g_{\gamma}\!{}^{3}-g_{\lambda}\!{}^1g_{\gamma}\!{}^2g_{\lambda}\!{}^{3a}+(g_{\lambda}\!{}^1)^2g_{\gamma}\!{}^{4c}
]&=C_{8}^{(5)Q},
\label{hopffiveq}
\end{align}
where
\be
G_J=g_{\lambda}\!{}^{4c}+g_{\lambda}\!{}^{4e}-g_{\lambda}\!{}^{4j}
\ee
corresponds to $J$ defined in Eq.~\eqref{Jdef}.
The invariants of Eqs.~\eqref{invfive}, \eqref{quadfive} may be recovered from Eqs.~\eqref{hopffive}, \eqref{hopffiveq} as before. Here the relevant symmetry factors (in addition to those in Eqs.~\eqref{symthree}, \eqref{symfour}) are
\begin{align}
S'_{5a}=1,\quad S'_{5b}=2,\quad S'_{5c}=S'_{5h}=4,\quad& S_{5cR}=S'_{5d}=S'_{5e}=S'_{5i}=S'_{5j}
=8,\nn S_{5bR}=S'_{5k}=12,\quad& S'_{5f}=16,\quad S_{5aR}=S'_{5g}=24.
\end{align}

\section{$a$-function considerations}
A good deal of effort has been invested in recent years\cite{Cardy,KS,KS1,Luty} in the search for an $a$-theorem, a generalisation of Zamolodchikov's two-dimensional $c$-theorem\cite{Zam} to four dimensions (or indeed to other dimensions higher than two\cite{ElvangST, GrinsteinCKA, GrinsteinCKA1,GrinsteinCKA2, GrinsteinCKA3}). From our point of view, as mentioned in the introduction, the crucial development is the demonstration that the $\beta$-functions in theories in four and six dimensions obey a gradient flow equation similar to one which plays a critical role in the derivation of the $c$-theorem\cite{Analog,Analog1,OsbJacnew,Weyl}. These gradient flow equations often place constraints relating the $\beta$-function coefficients, as has been shown for four-dimensional gauge theories \cite{jp2} and six-dimensional $\phi^3$ theories\cite{gap} (similar gradient flows have been demonstrated in three dimensions\cite{JJP,asusy,jp1} though here the theoretical underpinning has not yet been provided). Our purpose in this section is to apply the same considerations to our four-dimensional $\phi^4$ theory where we are able to confirm our results using the explicit calculations available to a high loop order. We start by presenting the basic theoretical background  in general notation in the interests of clarity and brevity. For a theory with couplings $g^I$, the corresponding $\beta$-functions are defined by
\be
\beta^I=\mu\frac{d}{d\mu}g^I
\ee
where $\mu$ is a mass scale (in practice usually the standard dimensional regularisation mass scale).
The essential conclusion of Refs.~\cite{Analog1}, \cite{OsbJacnew} is the existence of a function $A$ such that
\be
\pa_IA=T_{IJ}\beta^J
\label{grad}
\ee
where $\pa_I\equiv\frac{\pa}{\pa g^I}$ and
\be
T_{IJ}=G_{IJ}+\pa_IW_J-\pa_JW_I
\label{Tdef}
\ee
with $G_{IJ}$ symmetric\footnote{In general for a theory with a symmetry, the $\beta$-function should be replaced by a ``generalised'' $\beta$-function\cite{OsbJacnew}. It was shown by explicit calculation in Ref.~\cite{grin} that the difference between the two becomes non-trivial at three loops for a fermion-scalar theory in four dimensions. However, for a pure scalar theory we do not expect any distinction until five loops which is beyond our interests in this section.}. The function $A$ is invariant up to 
\be
A\go A+g_{IJ}\beta^I\beta^J,
\label{Ainv}
\ee
where $g_{IJ}$ is an arbitrary symmetric matrix. At lowest order we have an $a$-function given by 
\be
A^{(4)}=A^{(4)}_1\tikz[baseline=(vert_cent.base),scale=1]{
	\node (vert_cent) {\hspace{-13pt}$\phantom{-}$};
	\draw (0,0) circle[radius=1cm];
	\vertex at \coord{0} (A) {};
	\vertex at \coord{-120} (B) {};
	\vertex at \coord{120} (C) {};
	\draw [bend left=25] (A) to (B) [bend left=25] (B) to node[fill,circle,inner sep=0pt,minimum size=0pt,pos=0.5] (D) {} (C) [bend left=25] (C) to (A);
}
\ee
and Eq.~\eqref{grad} simply implies 
\be
3A^{(4)}_1=3c_1\implies A^{(4)}_1=c_1
\ee
(the factor of 3 on the right-hand side derives from the multiplicity factor of $\Scal_3$ for the corresponding term in the $\beta$-function).
At the next order we have
\be
A^{(5)}=A^{(5)}_1\tikz[baseline=(vert_cent.base),scale=1]{
	\node (vert_cent) {\hspace{-13pt}$\phantom{-}$};
	\draw (0,0) circle[radius=1cm];
	\vertex at \coord{-40} (A) {};
	\vertex at \coord{-140} (B) {};
	\vertex at \coord{40} (C) {};
	\vertex at \coord{140} (D) {};
	\draw [bend left=60] (A) to (B) [bend right=60] (C) to (D);
	\draw (A)--(B) (C)--(D);
}
+A^{(5)}_2\tikz[baseline=(vert_cent.base),scale=1]{
	\node (vert_cent) {\hspace{-13pt}$\phantom{-}$};
	\draw (0,0) circle[radius=1cm];
	\vertex at \coord{-45} (A) {};
	\vertex at \coord{-135} (B) {};
	\vertex at \coord{135} (C) {};
	\vertex at \coord{45} (D) {};
	\draw [bend left=40] (A) to (B) [bend left=40] (B) to (C) [bend left=40] (C) to (D) [bend left=40] (D) to (A); 
}
+A^{(5)}_3\tikz[baseline=(vert_cent.base),scale=1]{
	\node (vert_cent) {\hspace{-13pt}$\phantom{-}$};
	\draw (0,0) circle[radius=1cm];
	\vertex at (0,1) (A) {};
	\vertex at (0,-1) (B) {};
	\draw [bend right=50] (A) to node[fill,circle,inner sep=0pt,minimum size=0pt,pos=0.5] (C) {} (B) [bend right=50] (B) to node[fill,circle,inner sep=0pt,minimum size=0pt,pos=0.5] (D) {} (A);
	\draw [bend right=50] (C) to (D) [bend left=50] (C) to (D);
}
\ee
and now Eq.~\eqref{grad} entails
\begin{align}
4A^{(5)}_1=&2d_2,\nn
4A^{(5)}_2=&3c_{2R}+c_1T^{(4)},\nn
4A^{(5)}_3=&6c_2+2c_1T^{(4)},
\label{Aone}
\end{align}
Here $T^{(4)}$ represents the coefficient of the single fourth-order metric term. The figure below displays this structure by showing its contraction with a $dg$ (represented by a cross) and a $\beta^{(1)}$ (represented by a diamond).
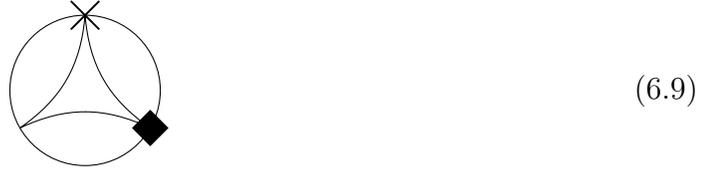
\be
\tikz[baseline=(vert_cent.base),scale=1]{
	\node (vert_cent) {\hspace{-13pt}$\phantom{-}$};
	\draw (0,0) circle[radius=1cm];
	\vertex at \coord{0} (A) {};
	\vertex at \coord{-120} (B) {};
	\vertex at \coord{120} (C) {};
	\draw [bend left=25] (A) to (B) [bend left=25] (B) to node[fill,circle,inner sep=0pt,minimum size=0pt,pos=0.5] (D) {} (C) [bend left=25] (C) to (A);
\vcross at (A) {};
	\vbeta at (C) {};
}
\label{Tfour}
\ee
In Eq.~\eqref{Aone} there are two equations and three unknowns resulting in one residual free parameter. This corresponds to the invariance under 
\be
 A^{(5)}_2\go A^{(5)}_2+3g^{(3)}c_1^2,\quad A^{(5)}_3\go A^{(5)}_3+6g^{(3)}c_1^2,\quad  T^{(4)}\go T^{(4)}+12g^{(3)}c_1
\label{freeone}
\ee
reflecting the freedom described by Eq.~\eqref{Ainv} at lowest order (with $g_{IJ}=g^{(3)}\tdelta_{IJ}$, $g^{(3)}$ arbitrary).
The six-loop $a$-function is given by
\begin{align}
A^{(6)}=&A^{(6)}_1\tikz[baseline=(vert_cent.base),scale=1]{
	\node (vert_cent) {\hspace{-13pt}$\phantom{-}$};
	\draw (0,0) circle[radius=1cm];
	\vertex at \coord{0} (A) {};
	\vertex at \coord{-120} (B) {};
	\vertex at \coord{120} (C) {};
	\draw [bend left=25] (A) to (B) [bend left=25] (B) to node[fill,circle,inner sep=0pt,minimum size=0pt,pos=0.5] (D) {} (C) [bend left=25] (C) to (A);
	\draw (D) circle[radius=0.25cm];
}
+A^{(6)}_2\tikz[baseline=(vert_cent.base),scale=1]{
	\node (vert_cent) {\hspace{-13pt}$\phantom{-}$};
	\draw (0,0) circle[radius=1cm];
	\vertex at (1,0) (A) {};
	\vertex at (-1,0) (B) {};
	\vertex at (0,1) (C) {};
	\vertex at (0,-1) (D) {};
	\draw (A)--(B) (C)--(D);
	\vertex at (0.3,0.1) (E) {};
	\vertex at (0.3,-0.1) (F) {};
	\draw [bend left=15] (C) to (E) [bend right=15] (D) to (F);
	\vertex at (0.4,0.3) (G) {};
	\vertex at (-0.15,0.3) (H) {};
	\draw [bend right=20] (A) to (G) [bend left=15] (B) to (H);
}
+A^{(6)}_3\tikz[baseline=(vert_cent.base),scale=1]{
	\node (vert_cent) {\hspace{-13pt}$\phantom{-}$};
	\draw (0,0) circle[radius=1cm];
	\vertex at \coord{-45} (A) {};
	\vertex at \coord{-135} (B) {};
	\vertex at \coord{135} (C) {};
	\vertex at \coord{45} (D) {};
	\draw (A)--(B)--(D)--(C)--(A); 
}
+A^{(6)}_4\tikz[baseline=(vert_cent.base),scale=1]{
	\node (vert_cent) {\hspace{-13pt}$\phantom{-}$};
	\draw (0,0) circle[radius=1cm];
	\vertex at \coord{0} (A) {};
	\vertex at \coord{-72} (B) {};
	\vertex at \coord{-144} (C) {};
	\vertex at \coord{144} (D) {};
	\vertex at \coord{72} (E) {};
	\draw [bend left=40] (A) to (B) [bend left=40] (B) to (C) [bend left=40] (C) to (D) [bend left=40] (D) to (E) [bend left=40] (E) to (A);
}\nn
&+A^{(6)}_5\tikz[baseline=(vert_cent.base),scale=1]{
	\node (vert_cent) {\hspace{-13pt}$\phantom{-}$};
	\draw (0,0) circle[radius=1cm];
	\vertex at \coord{0} (A) {};
	\vertex at \coord{-120} (B) {};
	\vertex at \coord{120} (C) {};
	\draw [bend left=25] (A) to (B) [bend left=25] (B) to node[fill,circle,inner sep=0pt,minimum size=0pt,pos=0.5] (D) {} (C) [bend left=25] (C) to (A);
	\vertex at \coord{180} (E) {};
	\draw [bend right=40] (D) to (E) [bend left=40] (D) to (E);
}
\label{Asix}
\end{align}
and the seven associated five-loop metric contributions are depicted below, with the same conventions as for $T^{(4)}$ earlier.
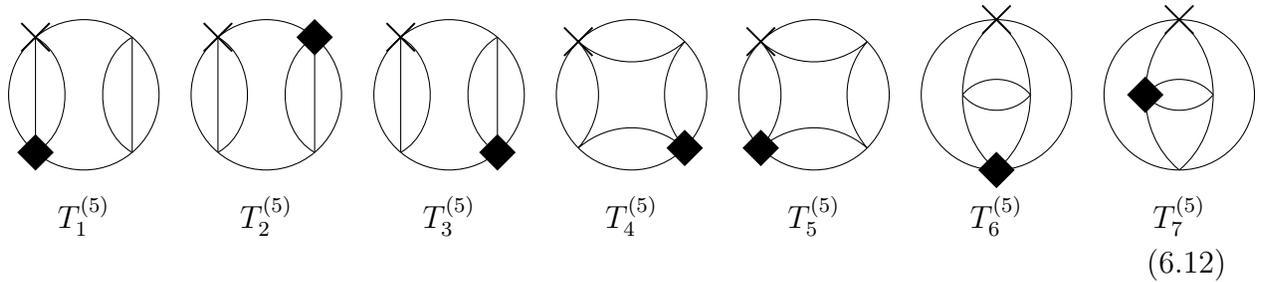
\be
\tikz[baseline=(vert_cent.base),scale=1]{
	\node (vert_cent) {\hspace{-13pt}$\phantom{-}$};
	\draw (0,0) circle[radius=1cm];
	\vertex at \coord{-40} (A) {};
	\vertex at \coord{-140} (B) {};
	\vertex at \coord{40} (C) {};
	\vertex at \coord{140} (D) {};
	\draw [bend left=60] (A) to (B) [bend right=60] (C) to (D);
	\draw (A)--(B) (C)--(D);
	\vcross at (A) {};
	\vbeta at (B) {};
\draw (0,-1.2) node [below]{$T^{(5)}_1$}
}\quad
\tikz[baseline=(vert_cent.base),scale=1]{
	\node (vert_cent) {\hspace{-13pt}$\phantom{-}$};
	\draw (0,0) circle[radius=1cm];
	\vertex at \coord{-40} (A) {};
	\vertex at \coord{-140} (B) {};
	\vertex at \coord{40} (C) {};
	\vertex at \coord{140} (D) {};
	\draw [bend left=60] (A) to (B) [bend right=60] (C) to (D);
	\draw (A)--(B) (C)--(D);
	\vcross at (A) {};
	\vbeta at (C) {};
\draw (0,-1.2) node [below]{$T^{(5)}_2$}
}\quad
\tikz[baseline=(vert_cent.base),scale=1]{
	\node (vert_cent) {\hspace{-13pt}$\phantom{-}$};
	\draw (0,0) circle[radius=1cm];
	\vertex at \coord{-40} (A) {};
	\vertex at \coord{-140} (B) {};
	\vertex at \coord{40} (C) {};
	\vertex at \coord{140} (D) {};
	\draw [bend left=60] (A) to (B) [bend right=60] (C) to (D);
	\draw (A)--(B) (C)--(D);
	\vcross at (A) {};
	\vbeta at (D) {};
\draw (0,-1.2) node [below]{$T^{(5)}_3$}
}
\quad\tikz[baseline=(vert_cent.base),scale=1]{
	\node (vert_cent) {\hspace{-13pt}$\phantom{-}$};
	\draw (0,0) circle[radius=1cm];
	\vertex at \coord{-45} (A) {};
	\vertex at \coord{-135} (B) {};
	\vertex at \coord{135} (C) {};
	\vertex at \coord{45} (D) {};
	\draw [bend left=40] (A) to (B) [bend left=40] (B) to (C) [bend left=40] (C) to (D) [bend left=40] (D) to (A); 
	\vcross at (A) {};
	\vbeta at (C) {};
\draw (0,-1.2) node [below]{$T^{(5)}_4$}
}\quad
\tikz[baseline=(vert_cent.base),scale=1]{
	\node (vert_cent) {\hspace{-13pt}$\phantom{-}$};
	\draw (0,0) circle[radius=1cm];
	\vertex at \coord{-45} (A) {};
	\vertex at \coord{-135} (B) {};
	\vertex at \coord{135} (C) {};
	\vertex at \coord{45} (D) {};
	\draw [bend left=40] (A) to (B) [bend left=40] (B) to (C) [bend left=40] (C) to (D) [bend left=40] (D) to (A); 
	\vcross at (A) {};
	\vbeta at (B) {};
\draw (0,-1.2) node [below]{$T^{(5)}_5$}
}\quad
\tikz[baseline=(vert_cent.base),scale=1]{
	\node (vert_cent) {\hspace{-13pt}$\phantom{-}$};
	\draw (0,0) circle[radius=1cm];
	\vertex at (0,1) (A) {};
	\vertex at (0,-1) (B) {};
	\draw [bend right=50] (A) to node[fill,circle,inner sep=0pt,minimum size=0pt,pos=0.5] (C) {} (B) [bend right=50] (B) to node[fill,circle,inner sep=0pt,minimum size=0pt,pos=0.5] (D) {} (A);
	\draw [bend right=50] (C) to (D) [bend left=50] (C) to (D);
	\vcross at (A) {};
	\vbeta at (B) {};
\draw (0,-1.2) node [below]{$T^{(5)}_6$}
}\quad
\tikz[baseline=(vert_cent.base),scale=1]{
	\node (vert_cent) {\hspace{-13pt}$\phantom{-}$};
	\draw (0,0) circle[radius=1cm];
	\vertex at (0,1) (A) {};
	\vertex at (0,-1) (B) {};
	\draw [bend right=50] (A) to node[fill,circle,inner sep=0pt,minimum size=0pt,pos=0.5] (C) {} (B) [bend right=50] (B) to node[fill,circle,inner sep=0pt,minimum size=0pt,pos=0.5] (D) {} (A);
	\draw [bend right=50] (C) to (D) [bend left=50] (C) to (D);
	\vcross at (A) {};
	\vbeta at (C) {};
\draw (0,-1.2) node [below]{$T^{(5)}_7$}
}
\label{Tfive}
\ee
We now find from Eq.~\eqref{grad}
\begin{align}
A^{(6)}_1=&3c_{3b}+d_2T^{(4)},\nn
2A^{(6)}_1=&2d_{3}+3c_1(T^{(5)}_2+T^{(5)}_3),\nn
2A^{(6)}_1=&3c_1T^{(5)}_1+d_2T^{(4)},\nn
5A^{(6)}_2=&c_{3f},\nn
A^{(6)}_3=&6c_{3d}+2c_1T^{(5)}_6,\nn
4A^{(6)}_3=&12c_{3e}+2c_1T^{(5)}_7+4c_2T^{(4)},\nn
5A^{(6)}_4=&3c_{3aR}+c_1(T^{(5)}_4+T^{(5)}_5)+c_{2R}T^{(4)},\nn
2A^{(6)}_5=&6c_{3c}+c_1T^{(5)}_7+2c_{2R}T^{(4)},\nn
2A^{(6)}_5=&6c_{3bR}+c_1(2T^{(5)}_5+T^{(5)}_6)+c_2T^{(4)},\nn
A^{(6)}_5=&3c_{3a}+2c_1T^{(5)}_4+c_2T^{(4)}.
\label{eqthree}
\end{align}
The values of the coefficients may be extracted from Ref.~\cite{kaz} and are given at one and two loops by
\be
c_1=1,\quad c_2=-1,\quad c_{2R}=0,\quad d_2=\tfrac{1}{6}
\label{cotwo}
\ee
and at three loops by
\begin{align}
 c_{3a}=\tfrac12,\quad c_{3b}=-\tfrac38,\quad c_{3c}=c_{3d}=&-\tfrac12, \quad c_{3e}=2,\quad c_{3f}=12\zeta_3,\nn
d_3=-\tfrac{1}{8},\quad c_{3aR}=&c_{3bR}=0.
\label{cothree}
\end{align}
In Eq.~\eqref{eqthree} we have ten equations for twelve unknowns, namely the five $A$ parameters in Eq.~\eqref{Asix}, the six $T$ parameters in Eq.~\eqref{Tfive} (note that $T^{(5)}_2$ and $T^{(5)}_3$ only appear in the combination $T^{(5)}_2+T^{(5)}_3$), plus $T^{(4)}$. This results in two free parameters which are taken to be $T^{(4)}$ and  $A^{(6)}_4$. 
The solution of Eq.~\eqref{eqthree} is then
\begin{align}
A^{(6)}_1=&-\tfrac{9}{8}+\tfrac{1}{6}T^{(4)},\nn
A^{(6)}_2=&\tfrac{12}{5}\zeta_3,\nn
A^{(6)}_3=&\tfrac{51}{5}-2T^{(4)}+4A^{(6)}_4,\nn
A^{(6)}_5=&\tfrac{27}{10}-T^{(4)}+4A^{(6)}_4,\nn
T^{(5)}_1=&-\tfrac34+\tfrac{1}{18}T^{(4)},\nn
T^{(5)}_2+T^{(5)}_3=&-\tfrac23+\tfrac19T^{(4)},\nn
T^{(5)}_4=&\tfrac35+2A^{(6)}_4,\nn
T^{(5)}_5=&-\tfrac35+3A^{(6)}_4,\nn
T^{(5)}_6=&\tfrac{33}{5}-T^{(4)}+2A^{(6)}_4,\nn
T^{(5)}_7=&\tfrac{42}{5}-2T^{(4)}+8A^{(6)}_4.
\label{solthree}
\end{align}
The first free parameter corresponds to the invariance already noted in Eq.~\eqref{freeone}, while the second corresponds to the invariance under 
\begin{align}
A^{(6)}_3\go A^{(6)}_3+4g^{(4)},\quad A^{(6)}_4&\go 
A^{(6)}_4+g^{(4)},\quad A^{(6)}_5\go A^{(6)}_5+4g^{(4)},\nn
T^{(5)}_4\go T^{(5)}_4+2g^{(4)},\quad T^{(5)}_5 &\go T^{(5)}_5+3g^{(4)},\quad  T^{(5)}_6\go T^{(5)}_6+2g^{(4)},\nn
T^{(5)}_7&\go T^{(5)}_7+8 g^{(4)},
\end{align}
reflecting the freedom under 
\be
A\go A+g^{(4)} \beta_{ijkl}\beta_{ijmn}g_{klmn},
\label{freetwo}
\ee
with $g^{(4)}$ arbitrary.
Finally, the seven-loop $a$-function is parametrised as
\begin{align}
A^{(7)}=&A^{(7)}_1\tikz[baseline=(vert_cent.base)]{
\node (vert_cent) {\hspace{-13pt}$\phantom{-}$};
\draw[name path=path1] [rotate=-60] (0,0)--(1,0);
\draw (1,{-sin(60)}) ellipse [x radius=0.5, y radius=.1];
\draw [rotate around={60:(2,0)}] (1,0)--(2,0);
\draw[name path=path1] [rotate=60] (.5,0) ellipse [x radius=.5, y radius=.1];
\draw (0,0)--(0.5,{sin(60)});
\draw [rotate around={-60:(2,0)}](1.5,0) ellipse [x radius=.5, y radius=.1];
\draw (1.5,{sin(60)})--(2,0);
\draw (0.5,{sin(60)})--(1.5,{-sin(60)});
\draw (1.5,{sin(60)})--(0.5,{-sin(60)});
}
+
A^{(7)}_2\tikz[baseline=(vert_cent.base)]{
\node (vert_cent) {\hspace{-13pt}$\phantom{-}$};
\draw[name path=path1] [rotate=60] (0,0)--(1,0);
\draw (1,{sin(60)}) ellipse [x radius=0.5, y radius=.1];
\draw (0.5,{sin(60)})--(1.5,{sin(60});
\draw [rotate around={-60:(2,0)}] (1,0)--(2,0);
\draw[name path=path1] [rotate=-60] (.5,0) ellipse [x radius=.5, y radius=.1];
\draw (0,0)--(0.5,{-sin(60)});
\draw (0.5,{-sin(60)})--(1.5,{-sin(60});
\draw [rotate around={60:(2,0)}](1.5,0) ellipse [x radius=.5, y radius=.1];
\draw (1.5,{-sin(60)})--(2,0);
}
+
A^{(7)}_3\tikz[baseline=(vert_cent.base)]{
\node (vert_cent) {\hspace{-13pt}$\phantom{-}$};
\draw (0,0)--(2,0);
\draw[name path=path1] [rotate=60] (0,0)--(1,0);
\draw (1,{sin(60)}) ellipse [x radius=0.5, y radius=.1];
\draw (0.5,{sin(60)})--(1.5,{sin(60});
\draw [rotate around={-60:(2,0)}] (1,0)--(2,0);
\draw[name path=path1] [rotate=-60] (.5,0) ellipse [x radius=.5, y radius=.05];
\draw (1,{-sin(60)}) ellipse [x radius=.5, y radius=.05];
\draw [rotate around={60:(2,0)}](1.5,0) ellipse [x radius=.5, y radius=.05];
}+
A^{(7)}_4\tikz[baseline=(vert_cent.base)]{
\node (vert_cent) {\hspace{-13pt}$\phantom{-}$};
\draw[name path=path1] [rotate=60] (0,0)--(1,0);
\draw (1,{sin(60)}) ellipse [x radius=0.5, y radius=.1];
\draw [rotate around={-60:(2,0)}] (1,0)--(2,0);
\draw[name path=path1] [rotate=-60] (.5,0) ellipse [x radius=.5, y radius=.1];
\draw [rotate around={60:(2,0)}](1.5,0) ellipse [x radius=.5, y radius=.1];
\draw (0.5,{-sin(60)})--(1.5,{sin(60)});
\draw (1.5,{-sin(60)})--(0.5,{sin(60)});
\draw (0,0)--(1.5,{-sin(60)});
\draw (2,0)--(0.5,{-sin(60)});
}\nn&+
A^{(7)}_5\tikz[baseline=(vert_cent.base)]{
\node (vert_cent) {\hspace{-13pt}$\phantom{-}$};
\draw (0,0)--(0.5,{sin(60)});
\draw (2,0)--(1.5,{sin(60)});
\draw (0.5,{sin(60)})--(1.5,{sin(60}) ;
\draw (0,0)--(1.5,{sin(60});
\draw (2,0)--(0.5,{sin(60});
\draw (0,0)--(0.5,{-sin(60)});
\draw (2,0)--(1.5,{-sin(60)});
\draw (0.5,{-sin(60)})--(1.5,{-sin(60}) ;
\draw (0,0)--(1.5,{-sin(60});
\draw (2,0)--(0.5,{-sin(60});
\draw (0.5,{sin(60)})--(.5,{-sin(60}) ;
\draw (1.5,{sin(60)})--(1.5,{-sin(60}) ;
}
+A^{(7)}_6\tikz[baseline=(vert_cent.base)]{
\node (vert_cent) {\hspace{-13pt}$\phantom{-}$};
\draw[name path=path1] [rotate=60] (0,0)--(1,0);
\draw (1,{sin(60)}) ellipse [x radius=0.5, y radius=.1];
\draw (0.5,{sin(60)})--(1.5,{sin(60});
\draw [rotate around={-60:(2,0)}] (1,0)--(2,0);
\draw[name path=path1] [rotate=-60] (.5,0) ellipse [x radius=.5, y radius=.1];
\draw (2,0)--(0.5,{-sin(60)});
\draw (0.5,{-sin(60)})--(1.5,{-sin(60});
\draw [rotate around={60:(2,0)}](1.5,0) ellipse [x radius=.5, y radius=.1];
\draw (1.5,{-sin(60)})--(0,0);
}+
A^{(7)}_7\tikz[baseline=(vert_cent.base)]{
\node (vert_cent) {\hspace{-13pt}$\phantom{-}$};
\draw[name path=path1] [rotate=-60] (0,0)--(1,0);
\draw (1,{-sin(60)}) ellipse [x radius=0.5, y radius=.1];
\draw [rotate around={60:(2,0)}] (1,0)--(2,0);
\draw (0,0)--(0.5,{sin(60)});
\draw (2,0)--(1.5,{sin(60)});
\draw (0.5,{sin(60)})--(1.5,{sin(60)});
\draw (0,0)--(2,0);
\draw (1,{sin(60)}) ellipse [x radius=.5, y radius=.1];
\draw (0,0)--(1.5,{-sin(60)});
\draw (2,0)--(0.5,{-sin(60)});
}+
A^{(7)}_8\tikz[baseline=(vert_cent.base)]{
\node (vert_cent) {\hspace{-13pt}$\phantom{-}$};
\draw[name path=path1] [rotate=60] (.5,0) ellipse [x radius=.5, y radius=.05];
\draw (0.5,{sin(60)})--(1.5,{sin(60}) ;
\draw [rotate around={-60:(2,0)}](1.5,0) ellipse [x radius=.5, y radius=.05];
\draw[name path=path1] [rotate=-60] (.5,0) ellipse [x radius=.5, y radius=.05];
\draw (0.5,{-sin(60)})--(1.5,{-sin(60}) ;
\draw [rotate around={60:(2,0)}](1.5,0) ellipse [x radius=.5, y radius=.05];
\draw (0.5,{sin(60)})--(.5,{-sin(60}) ;
\draw (1.5,{sin(60)})--(1.5,{-sin(60}) ;
}\nn&+
A^{(7)}_9\tikz[baseline=(vert_cent.base)]{
\node (vert_cent) {\hspace{-13pt}$\phantom{-}$};
\draw[name path=path1] [rotate=60] (0,0)--(1,0);
\draw (0.5, {sin(60)}) -- (1.5,{sin(60)});
\draw [rotate around={-60:(2,0)}] (1,0)--(2,0);
\draw (0,0)--(0.5,{-sin(60)});
\draw (1,{-sin(60)}) ellipse [x radius=.5, y radius=.05];
\draw (2,0)--(1.5,{-sin(60)});
\draw (0,0)--(1.5,{sin(60)});
\draw (2,0)--(0.5,{sin(60)});
\draw (0,0)--(2,0);
\draw (0.5,{sin(60)})--(0.5,{-sin(60)});
\draw (1.5,{sin(60)})--(1.5,{-sin(60)});
}+
A^{(7)}_{10}\tikz[baseline=(vert_cent.base)]{
\node (vert_cent) {\hspace{-13pt}$\phantom{-}$};
\draw[name path=path1] [rotate=60] (.5,0) ellipse [x radius=.5, y radius=.05];
\draw (1,{sin(60)}) ellipse [x radius=.5, y radius=.05];
\draw [rotate around={-60:(2,0)}](1.5,0) ellipse [x radius=.5, y radius=.05];
\draw[name path=path1] [rotate=-60] (.5,0) ellipse [x radius=.5, y radius=.05];
\draw (1,{-sin(60)}) ellipse [x radius=.5, y radius=.05];
\draw [rotate around={60:(2,0)}](1.5,0) ellipse [x radius=.5, y radius=.05];
}
+A^{(7)}_{11}\tikz[baseline=(vert_cent.base)]{
\node (vert_cent) {\hspace{-13pt}$\phantom{-}$};
\draw[name path=path1] [rotate=60] (.5,0) ellipse [x radius=.5, y radius=.05];
\draw (0.5,{sin(60)})--(1.5,{sin(60}) ;
\draw [rotate around={-60:(2,0)}](1.5,0) ellipse [x radius=.5, y radius=.05];
\draw[name path=path1] [rotate=-60] (.5,0) ellipse [x radius=.5, y radius=.05];
\draw (0.5,{-sin(60)})--(1.5,{-sin(60}) ;
\draw [rotate around={60:(2,0)}](1.5,0) ellipse [x radius=.5, y radius=.05];
\draw (0.5,{sin(60)})--(1.5,{-sin(60}) ;
\draw (1.5,{sin(60)})--(0.5,{-sin(60}) ;
}
+A^{(7)}_{12}\tikz[baseline=(vert_cent.base)]{
\node (vert_cent) {\hspace{-13pt}$\phantom{-}$};
\draw[name path=path1] [rotate=-60] (0,0)--(1,0);
\draw (1,{-sin(60)}) ellipse [x radius=0.5, y radius=.1];
\draw [rotate around={60:(2,0)}] (1,0)--(2,0);
\draw[name path=path1] [rotate=60] (.5,0) ellipse [x radius=.5, y radius=.1];
\draw (0.5,{sin(60)})--(1.5,{sin(60});
\draw [rotate around={-60:(2,0)}](1.5,0) ellipse [x radius=.5, y radius=.1];
\draw (0.5,{sin(60)})--(0.5,{-sin(60)});
\draw (1.5,{sin(60)})--(1.5,{-sin(60)});
\draw (0,0)--(2,0);
}\nn&+
A^{(7)}_{13}\tikz[baseline=(vert_cent.base)]{
\node (vert_cent) {\hspace{-13pt}$\phantom{-}$};
\draw[name path=path1] [rotate=-60] (0,0)--(1,0);
\draw (1,{-sin(60)}) ellipse [x radius=0.5, y radius=.1];
\draw [rotate around={60:(2,0)}] (1,0)--(2,0);
\draw[name path=path1] [rotate=60] (.5,0) ellipse [x radius=.5, y radius=.1];
\draw (0.5,{sin(60)})--(1.5,{sin(60});
\draw [rotate around={-60:(2,0)}](1.5,0) ellipse [x radius=.5, y radius=.1];
\draw (0.5,{sin(60)})--(1.5,{-sin(60)});
\draw (1.5,{sin(60)})--(0.5,{-sin(60)});
\draw (0,0)--(2,0);
}+
A^{(7)}_{14}\tikz[baseline=(vert_cent.base)]{
\node (vert_cent) {\hspace{-13pt}$\phantom{-}$};
\draw[name path=path1] [rotate=60] (0,0)--(1,0);
\draw (1,{sin(60)}) ellipse [x radius=0.5, y radius=.1];
\draw [rotate around={-60:(2,0)}] (1,0)--(2,0);
\draw (0,0)--(1.5,{sin(60)});
\draw (2,0)--(0.5,{sin(60)});
\draw[name path=path1] [rotate=-60] (0,0)--(1,0);
\draw (1,{-sin(60)}) ellipse [x radius=0.5, y radius=.1];
\draw [rotate around={60:(2,0)}] (1,0)--(2,0);
\draw (0,0)--(1.5,{-sin(60)});
\draw (2,0)--(0.5,{-sin(60)});
}+
A^{(7)}_{15}\tikz[baseline=(vert_cent.base)]{
\node (vert_cent) {\hspace{-13pt}$\phantom{-}$};
\draw[name path=path1] [rotate=60] (0,0)--(1,0);
\draw (1,{sin(60)}) ellipse [x radius=0.5, y radius=.1];
\draw [rotate around={-60:(2,0)}] (1,0)--(2,0);
\draw[name path=path1] [rotate=-60] (.5,0) ellipse [x radius=.5, y radius=.05];
\draw (1,{-sin(60)}) ellipse [x radius=.5, y radius=.05];
\draw [rotate around={60:(2,0)}](1.5,0) ellipse [x radius=.5, y radius=.05];
\draw (0,0)--(1.5,{sin(60)});
\draw (2,0)--(0.5,{sin(60)});
}
+A^{(7)}_{16}\tikz[baseline=(vert_cent.base)]{
\node (vert_cent) {\hspace{-13pt}$\phantom{-}$};
\draw (0,0)--(0.5,{sin(60)});
\draw (0.5,{sin(60)})--(1.5,{sin(60)});
\draw [rotate around={-60:(2,0)}](1.5,0) ellipse [x radius=.5, y radius=.05];
\draw[name path=path1] [rotate=-60] (.5,0) ellipse [x radius=.5, y radius=.05];
\draw (1,{-sin(60)}) ellipse [x radius=.5, y radius=.05];
\draw (2,0)--(1.5,{-sin(60)});
\draw (0.5,{sin(60)})--(2,0);
\draw (0.5,{sin(60)})--(1.5,{-sin(60)});
\draw (0,0)--(1.5,{sin(60)});
}\nn&+
A^{(7)}_{17}\tikz[baseline=(vert_cent.base)]{
\node (vert_cent) {\hspace{-13pt}$\phantom{-}$};
\draw (0,0)--(0.5,{sin(60)});
\draw (2,0)--(1.5,{sin(60)});
\draw (0.5,{sin(60)})--(1.5,{sin(60}) ;
\draw[name path=path1] [rotate=-60] (.5,0) ellipse [x radius=.5, y radius=.05];
\draw (0.5,{-sin(60)})--(1.5,{-sin(60}) ;
\draw [rotate around={60:(2,0)}](1.5,0) ellipse [x radius=.5, y radius=.05];
\draw (0.5,{sin(60)})--(.5,{-sin(60}) ;
\draw (1.5,{sin(60)})--(1.5,{-sin(60}) ;
\draw (0,0)--(1.5,{sin(60});
\draw (2,0)--(0.5,{sin(60});
}.
\label{Aseven}
\end{align}
These seven-loop vacuum diagrams were given in Fig.~6 of Ref.~\cite{tarn} and we have retained their ordering (similarly, the five and six loop vacuum diagrams were depicted in their Figs. 4 and 5 respectively).
Since there are 24  6-loop metric contributions, we have introduced a compact notation to avoid depicting them all individually. Eq.~\eqref{Tsix} shows the six-loop vacuum diagrams; seen already in Eq.~\eqref{Asix}, but now with some vertices labelled. We introduce the notation $T^{(6)}_{nxy}$ to denote a metric contribution where the vertices $x$, $y$ in diagram $n$ correspond to the $I$, $J$ indices respectively of a contribution to $T_{IJ}$. The labellings shown are sufficient to cover all the independent possibilities.
\be
\tikz[baseline=(vert_cent.base),scale=1]{
	\node (vert_cent) {\hspace{-13pt}$\phantom{-}$};
	\draw (0,0) circle[radius=1cm];
	\vertex at \coord{0} (A) {};
	\vertex at \coord{-120} (B) {};
	\vertex at \coord{120} (C) {};
	\draw [bend left=25] (A) to (B) [bend left=25] (B) to node[fill,circle,inner sep=0pt,minimum size=0pt,pos=0.5] (D) {} (C) [bend left=25] (C) to (A);
	\draw (D) circle[radius=0.25cm];
\draw \coord{120} node[above]{$b$};
\draw (0,1) node[above]{$a$};
\draw (0.3,-.75) node[above]{$c$};
\draw (-0.3,-.75) node[above]{$d$};
\draw \coord{-120} node[below]{$e$};
\draw (0,-1.4) node {1};
}\quad 
\tikz[baseline=(vert_cent.base),scale=1]{
	\node (vert_cent) {\hspace{-13pt}$\phantom{-}$};
	\draw (0,0) circle[radius=1cm];
	\vertex at (1,0) (A) {};
	\vertex at (-1,0) (B) {};
	\vertex at (0,1) (C) {};
	\vertex at (0,-1) (D) {};
	\draw (A)--(B) (C)--(D);
	\vertex at (0.3,0.1) (E) {};
	\vertex at (0.3,-0.1) (F) {};
	\draw [bend left=15] (C) to (E) [bend right=15] (D) to (F);
	\vertex at (0.4,0.3) (G) {};
	\vertex at (-0.15,0.3) (H) {};
	\draw [bend right=20] (A) to (G) [bend left=15] (B) to (H);
\draw (0,1) node[above]{$a$};
\draw (1,0) node[right]{$b$};
\draw (0,-1.4) node {2};
}
\quad \tikz[baseline=(vert_cent.base),scale=1]{
	\node (vert_cent) {\hspace{-13pt}$\phantom{-}$};
	\draw (0,0) circle[radius=1cm];
	\vertex at \coord{-45} (A) {};
	\vertex at \coord{-135} (B) {};
	\vertex at \coord{135} (C) {};
	\vertex at \coord{45} (D) {};
	\draw (A)--(B)--(D)--(C)--(A); 
\draw (A) node[above] {$a$};
\draw (D) node[above] {$b$};
\draw (C) node[below] {$c$};
\draw (0,0) node[below] {$d$};
\draw (0,-1.4) node {3};
}
\quad\tikz[baseline=(vert_cent.base),scale=1]{
	\node (vert_cent) {\hspace{-13pt}$\phantom{-}$};
	\draw (0,0) circle[radius=1cm];
	\vertex at \coord{0} (A) {};
	\vertex at \coord{-72} (B) {};
	\vertex at \coord{-144} (C) {};
	\vertex at \coord{144} (D) {};
	\vertex at \coord{72} (E) {};
	\draw [bend left=40] (A) to (B) [bend left=40] (B) to (C) [bend left=40] (C) to (D) [bend left=40] (D) to (E) [bend left=40] (E) to (A);
\draw (A) node[above] {$a$};
\draw (E) node[above] {$b$};
\draw (D) node[below] {$c$};
\draw (0,-1.4) node {4};
}\quad \tikz[baseline=(vert_cent.base),scale=1]{
	\node (vert_cent) {\hspace{-13pt}$\phantom{-}$};
	\draw (0,0) circle[radius=1cm];
	\vertex at \coord{0} (A) {};
	\vertex at \coord{-120} (B) {};
	\vertex at \coord{120} (C) {};
	\draw [bend left=25] (A) to (B) [bend left=25] (B) to node[fill,circle,inner sep=0pt,minimum size=0pt,pos=0.5] (D) {} (C) [bend left=25] (C) to (A);
	\vertex at \coord{180} (E) {};
	\draw [bend right=40] (D) to (E) [bend left=40] (D) to (E);
\draw (A) node[above] {$a$};
\draw (E) node[above] {$c$};
\draw (C) node[below] {$b$};
\draw (D) node[above] {$d$};
\draw (B) node[below] {$e$};
\draw (0,-1.4) node {5};
}
\label{Tsix}
\ee
The number of $T$-type contributions is the number of distinct ways of selecting an ordered pair of vertices from the diagrams shown in \eqref{Tsix}, namely 24. At this order Eq.~\eqref{grad} implies
 \begin{align}
4A^{(7)}_1=&\tfrac32d_2(T^{(5)}_2+T^{(5)}_3)+2d_{4a},\nn
2A^{(7)}_1=&\tfrac32d_2T^{(5)}_1,\nn
6A^{(7)}_2=&\tfrac12d_2(T^{(5)}_1+T^{(5)}_2+T^{(5)}_3),\nn
2A^{(7)}_3=&c_1T^{(6)}_{1ab}+c_{3b}T^{(4)}+d_2(T^{(5)}_5+2T^{(5)}_4)+6c_{4eR},\nn
2A^{(7)}_3=&c_1(T^{(6)}_{1ca}+T^{(6)}_{1cb}+T^{(6)}_{1ce})+3c_{2R}(T^{(5)}_2+T^{(5)}_3)+2d_{4b},\nn
2A^{(7)}_3=&c_1T^{(6)}_{1be}+c_1T^{(6)}_{1ba}+3c_{2R}T^{(5)}_1+d_2T^{(5)}_5,\nn
6A^{(7)}_4=&c_1T^{(6)}_{3ad}+2c_{3d}T^{(4)}+6c_{4i},\nn
6A^{(7)}_5=&c_{4s},\nn
2A^{(7)}_6=&2c_1T^{(6)}_{1ab}+\tfrac12d_2(2T^{(5)}_6+T^{(5)}_7)+12c_{4j},\nn
2A^{(7)}_6=&2c_1(T^{(6)}_{1cb}+T^{(6)}_{1ce})+3c_2(T^{(5)}_2+T^{(5)}_3)+2d_{4d},\nn
2A^{(7)}_6=&2c_1T^{(6)}_{1be}+3c_2T^{(5)}_1+\tfrac12d_2T^{(5)}_7,\nn
2A^{(7)}_7=&2c_{3b}T^{(4)}+d_2T^{(5)}_7+6c_{4e},\nn
2A^{(7)}_7=&2c_1T^{(6)}_{1ca}+3c_2(T^{(5)}_2+T^{(5)}_3)+2d_{4c},\nn
2A^{(7)}_7=&2c_1T^{(6)}_{1ba}+3c_2T^{(5)}_1+d_2T^{(5)}_6,\nn
4A^{(7)}_8=&3c_1(T^{(6)}_{1bc}+T^{(6)}_{1bd}+T^{(6)}_{1cd})+d_3T^{(4)},\nn
2A^{(7)}_8=&3c_1T^{(6)}_{1ac}+d_3T^{(4)}+3c_{4c},\nn
4A^{(7)}_9=&3c_1T^{(6)}_{2ab}+12c_{4p},\nn
2A^{(7)}_9=&6c_{4q}+c_{3f}T^{(4)},\nn
6A^{(7)}_{10}=&c_1(T^{(6)}_{4ab}+T^{(6)}_{4ac})+c_{2R}(T^{(5)}_5+T^{(5)}_4)+c_{3aR}T^{(4)}+3c_{4aR},\nn
4A^{(7)}_{11}=&c_1(T^{(6)}_{5cd}+T^{(6)}_{5bc})+2c_{2R}T^{(5)}_5+c_{3c}T^{(4)}+6c_{4dR},\nn
2A^{(7)}_{11}=&c_1T^{(6)}_{5ac}+2c_{2R}T^{(5)}_4+c_{3c}T^{(4)}+3c_{4b},\nn
4A^{(7)}_{12}=&c_1(2T^{(6)}_{5cb}+T^{(6)}_{3ad})+c_2T^{(5)}_7+4c_{3bR}T^{(4)}+12c_{4k},\nn
2A^{(7)}_{12}=&2c_1(T^{(6)}_{5be}+T^{(6)}_{5ab})+c_2T^{(5)}_6+2c_{3d}T^{(4)}+6c_{4d},\nn
6A^{(7)}_{13}=&c_1T^{(6)}_{3ad}+2c_{3d}T^{(4)}+6c_{4h},\nn
4A^{(7)}_{14}=&2c_1T^{(6)}_{5ca}+c_2T^{(5)}_7+2c_{3a}T^{(4)}+6c_{4f},\nn
2A^{(7)}_{14}=&2c_1T^{(6)}_{5ba}+c_2T^{(5)}_6+3c_{4bR},\nn
2A^{(7)}_{15}=&c_1(T^{(6)}_{5cb}+T^{(6)}_{5ca})+c_{2R}T^{(5)}_7+2c_{3aR}T^{(4)}+6c_{4g},\nn
2A^{(7)}_{15}=&c_1(T^{(6)}_{5be}+T^{(6)}_{5ba}+2T^{(6)}_{4ab})+c_2T^{(5)}_5+c_{2R}T^{(5)}_6+c_{3bR}T^{(4)}+6c_{4cR},\nn
2A^{(7)}_{15}=&c_1(T^{(6)}_{5ab}+2T^{(6)}_{4ac})+c_2(T^{(5)}_5+2T^{(5)}_4)+c_{3a}T^{(4)}+c_{3bR}T^{(4)}+6c_{4fR},\nn
2A^{(7)}_{16}=&c_1(T^{(6)}_{3ab}+T^{(6)}_{3ac})+2c_{2R}T^{(5)}_7+4c_{3c}T^{(4)}+12c_{4l},\nn
2A^{(7)}_{16}=&c_1(2T^{(6)}_{5bc}+T^{(6)}_{3bc})+4c_2T^{(5)}_5+2c_{3e}T^{(4)}+12c_{4gR},\nn
A^{(7)}_{16}=&c_1(2T^{(6)}_{5cd}+T^{(6)}_{3da})+2c_{2R}T^{(5)}_6+12c_{4o},\nn
A^{(7)}_{16}=&2c_1T^{(6)}_{5ac}+4c_2T^{(5)}_4+2c_{3e}T^{(4)}+3c_{4a},\nn
2A^{(7)}_{17}=&2c_1(T^{(6)}_{3bc}+T^{(6)}_{3da})+4c_2T^{(5)}_6+24c_{4r},\nn
2A^{(7)}_{17}=&2c_1T^{(6)}_{3ac}+2c_2T^{(5)}_7+4c_{3e}T^{(4)}+12c_{4m},\nn
2A^{(7)}_{17}=&2c_1T^{(6)}_{3ab}+2c_2T^{(5)}_7+4c_{3e}T^{(4)}+12c_{4n}.
\label{fourAs}
\end{align}
The counting of unknowns is now slightly more subtle; we shall explain in some detail since the solution of Eqs.~\eqref{fourAs} leads to constraints on the $\beta$-function coefficients, and we would like to be sure that we have obtained the correct number of these.  There are thirty-six four-loop structures (including 1PR structures which cannot contribute to the $\beta$-function and hence must be set to zero) leading to the  thirty-six equations  in Eq.~\eqref{fourAs}; and there are 17 $A$ coefficients (as shown in Eq.~\eqref{Aseven}) and 24 $T$ coefficients at this order . However, $T^{(6)}_{1cb}$ and $T^{(6)}_{1ce}$ only appear in the combination $T^{(6)}_{1cb}+T^{(6)}_{1ce}$, and $T^{(6)}_{1bc}$, $T^{(6)}_{1bd}$,  and $T^{(6)}_{1cd}$ only appear in the combination $T^{(6)}_{1bc}+T^{(6)}_{1bd}+T^{(6)}_{1cd}$; furthermore, there are two invariances, under shifts among
$T^{(6)}_{3bc}$, $T^{(6)}_{3da}$, $T^{(6)}_{5cd}$, $T^{(6)}_{5bc}$, and among $T^{(6)}_{4ab}$, $T^{(6)}_{4ac}$, $T^{(6)}_{5ab}$, $T^{(6)}_{5be}$. Therefore there is a total of $17+26-5=38$ unknowns at this order. The lower-order metric coefficients $T^{(5)}_1$--$T^{(5)}_7$ get determined in Eq.~\eqref{solthree} up to two unknowns (one of which is $T^{(4)}$ which of course already appears as an unknown in Eq.~\eqref{fourAs}), resulting in 40 unknowns in total. There are seven five-loop vacuum diagrams which can contribute to the freedom in Eq.~\eqref{Ainv} (the diagrams appearing in \eqref{Tfive} but with insertions of $\beta^{(1)}$ replacing the diamonds and crosses), but two of these give the same contribution. There is also one four-loop vacuum diagram corresponding to the freedom in Eq.~\eqref{freetwo} (diagrammatically corresponding to that appearing in \eqref{Tfour}, but with insertions of $\beta^{(1)}$,  $\beta^{(2)}$ replacing the diamond and cross respectively). Finally there is a three-loop vacuum diagram corresponding to the freedom in Eq.~\eqref{freeone}, with insertions of $\beta^{(1)}$, $\beta^{(3)}$ or $\beta^{(2)}$, $\beta^{(2)}$. Therefore the number of unknowns which are solved for is only $40-6-1-1=32$. This implies that $36-32=4$ of the 36 equations must remain as constraints. Indeed after solving the equations we find the constraints
\begin{align}
2c_1d_{4a}-d_2I^{(3)}_4=&0,\nn
2c_1(I^{(4)}_{11}-I^{(4)}_{15}-3I^{(4)}_{16})+3d_2(I^{(3)}_2-I^{(3)}_3+\tfrac12c_{3e})+2(c_2-c_{2R})I^{(3)}_4=&0,\nn
2c_1(I^{(4)}_{11}-I^{(4)}_9)+4(c_2-c_{2R})I^{(3)}_4-3d_2(2I^{(3)}_2-c_{3e})=&0,\nn
c_1(2I^{(4)}_2-I^{(4)}_3+I^{(4)}_4+I^{(4)}_{13}+\tfrac12c_{4n}+\tfrac12c_{4o})
+(c_2-c_{2R})(2I^{(3)}_2-c_{3e})=&0.
\label{consist}
\end{align}
We note that as is to be expected, these constraints may be expressed in terms of the invariants defined in Eqs.~\eqref{invthree}, \eqref{invfour} and \eqref{qinv}.
At four loops (again extracted from Ref.~\cite{kaz}) the coefficients are 
\begin{align}
c_{4a}=\tfrac13(6\zeta_3-11),\quad c_{4b}=1-\zeta_3,\quad c_{4c}&=\tfrac{7}{12},\quad\quad c_{4d}=\tfrac12,\quad  c_{4e}=\tfrac{121}{144},\nn c_{4f}=1-2\zeta_3, \quad  c_{4g}=c_{4o}=\tfrac14(2\zeta_3-1),&\quad c_{4h}=c_{4l}=\tfrac{1}{6}(5-6\zeta_3),\nn
 c_{4i}=\tfrac{5}{6},\quad 
  c_{4j}=-\tfrac{37}{288},\quad 
 c_{4k}=c_{4r}=\tfrac23,\quad& c_{4m}=4\zeta_3-5,\quad c_{4n}=-5,\nn
c_{4p}=3(\zeta_4-2\zeta_3),\quad& c_{4q}=-3(2\zeta_3+\zeta_4),\quad
c_{4s}=-40\zeta_5,\nn
d_{4a}=-\tfrac{5}{48},\quad d_{4b}=-\tfrac{5}{32},&\quad d_{4c}=\tfrac{13}{48},\quad d_{4d}=\tfrac23, 
\label{cofour}
\end{align} 
with $c_{4aR}=\ldots=c_{4fR}=0$, and we may easily check that the values in Eqs.~\eqref{cotwo}, \eqref{cothree} and \eqref{cofour} satisfy the constraints in Eq.~\eqref{consist}.

We refrain from giving the values of the $a$-coefficients in the general case. However an interesting special case is that of a symmetric $T_{IJ}$. It turns out that we can impose symmetry on $T_{IJ}$ up to this order without needing to impose any further constraints on the $\beta$-function coefficients. The $a$-function coefficients are then
\begin{align}
A^{(7)}_{1}=&-\tfrac{3}{32}+\tfrac{1}{144}T^{(4)},\nn
A^{(7)}_{2}=&-\tfrac{17}{864}+\tfrac{1}{432}T^{(4)},\nn
A^{(7)}_3=&\tfrac{79}{96}-\tfrac{3}{8}T^{(4)}+\tfrac56A^{(6)}_4,\nn
A^{(7)}_{4}=&\tfrac{7}{10}-3\zeta_3+T^{(4)}-6A^{(6)}_4+4A^{(7)}_{10}-2A^{(7)}_{11},\nn
A^{(7)}_{5}=&-\tfrac{20}{3}\zeta_5,\nn
A^{(7)}_{6}=&\tfrac{67}{40}-\tfrac{13}{24}T^{(4)}+A^{(6)}_4,\nn
A^{(7)}_{7}=&\tfrac{773}{240}-\tfrac{13}{24}T^{(4)}+\tfrac23A^{(6)}_4,\nn
A^{(7)}_8=&\tfrac{19}{5}-\tfrac{5}{8}T^{(4)}+A^{(6)}_4,\nn
A^{(7)}_{9}=&-9(2\zeta_3+\zeta_4)+6\zeta_3T^{(4)},\nn
A^{(7)}_{12}=&\tfrac{18}{5}-\tfrac{21}{2}\zeta_3+3T^{(4)}-18A^{(6)}_4+12A^{(7)}_{10}-5A^{(7)}_{11},\nn
A^{(7)}_{13}=&\tfrac{7}{10}-4\zeta_3+T^{(4)}-6A^{(6)}_4+4A^{(7)}_{10}-2A^{(7)}_{11},\nn
A^{(7)}_{14}=&-\tfrac{21}{10}-\tfrac32\zeta_3+T^{(4)}-2A^{(6)}_4+A^{(7)}_{11},\nn
A^{(7)}_{15}=&\tfrac{33}{20}-3\zeta_3+T^{(4)}-7A^{(6)}_4+6A^{(7)}_{10}-A^{(7)}_{11},\nn
A^{(7)}_{16}=&-\tfrac{97}{5}+12\zeta_3+5T^{(4)}-8A^{(6)}_4+4A^{(7)}_{11},\nn
A^{(7)}_{17}=&-\tfrac{314}{5}+30\zeta_3+12T^{(4)}-16A^{(6)}_4+4A^{(7)}_{11},\nn
\end{align}
and
\begin{align}
T^{(6)}_{1ab}=&\tfrac{371}{240}-\tfrac38T^{(4)}+\tfrac12A^{(6)}_{4},\nn
T^{(6)}_{1ac}=&\tfrac{39}{20}-\tfrac38T^{(4)}+\tfrac23A^{(6)}_4,\nn
T^{(6)}_{1bc}+T^{(6)}_{1bd}=&\tfrac{1}{120}-\tfrac38T^{(4)}+A^{(6)}_4,\nn
T^{(6)}_{1cd}=&\tfrac{607}{120}-\tfrac{5}{12}T^{(4)}+\tfrac13A^{(6)}_4,\nn
T^{(6)}_{1be}=&\tfrac15-\tfrac38T^{(4)}+\tfrac23A^{(6)}_4,\nn
T^{(6)}_{2ab}=&-24\zeta_4+8\zeta_3T^{(4)},\nn
T^{(6)}_{3ab}=&-\tfrac{122}{5}+30\zeta_3+6T^{(4)}-8A^{(6)}_4+4A^{(7)}_{11},\nn
T^{(6)}_{3ac}=&T^{(6)}_{3ab}-24\zeta_3,\nn
T^{(6)}_{3ad}=&-\tfrac{4}{5}-18\zeta_3+7T^{(4)}-36A^{(6)}_4+24A^{(7)}_{10}-12A^{(7)}_{11},\nn
T^{(6)}_{3bc}=&-\tfrac{284}{5}+48\zeta_3+3T^{(4)}+24A^{(6)}_4-24A^{(7)}_{10}+16A^{(7)}_{11},\nn
T^{(6)}_{4ab}=&-\tfrac{27}{20}+\tfrac94\zeta_3-\tfrac12T^{(4)}+3A^{(6)}_4+\tfrac32A^{(7)}_{11},\nn
T^{(6)}_{4ac}=&-T^{(6)}_{4ab}+6A^{(7)}_{10},\nn
T^{(6)}_{5ab}=&\tfrac{6}{5}-\tfrac32\zeta_3+\tfrac12T^{(4)}-A^{(6)}_4+A^{(7)}_{11},\nn
T^{(6)}_{5ac}=&-3+3\zeta_3+\tfrac12T^{(4)}+2A^{(7)}_{11},\nn
T^{(6)}_{5bc}=&\tfrac{39}{5}-12\zeta_3+\tfrac32T^{(4)}-14A^{(6)}_4+12A^{(7)}_{10}-4A^{(7)}_{11},\nn
T^{(6)}_{5be}=&\tfrac{21}{5}-9\zeta_3+\tfrac52T^{(4)}-16A^{(6)}_4+12A^{(7)}_{10}-6A^{(7)}_{11},\nn
T^{(6)}_{5cd}=&-\tfrac{39}{5}+12\zeta_3-T^{(4)}+14A^{(6)}_4-12A^{(7)}_{10}+8A^{(7)}_{11}.
\end{align}
We see that the effect of imposing symmetry has been to reduce the freedom in the $a$-function coefficients from the original six parameters to two.

\section{Conclusions}
We have shown how scheme changes in $\phi^4$ theory may be analysed within a compact and efficient framework. In particular we have derived the full set of scheme invariants up to four loop order and shown that their number is consistent with general expectations, though considerably higher than might be expected from a naive counting. In particular we have identified the existence of quadratic invariants which would be missed in a naive counting. Furthermore, we have shown that in the context of the Hopf algebra approach to renormalisation, each invariant is associated with a cocommutative combination of graphs. We have also considered the construction of the $a$-function generating the $\beta$-functions up to four-loop order via a gradient flow equation. In particular we have analysed the consistency conditions which guarantee this construction, again showing that their number is as expected and furthermore that, as expected, they may be expressed in terms of linear combinations of the scheme invariants. Finally we have considered one-vertex reducible diagrams and shown that there is a natural family of schemes in which these do not contribute to the $\beta$-function.

Future work might explore the Hopf algebra connection further. Furthermore, at higher orders than we have yet considered there might be the possibility of cubic and higher order invariants. The extension of the analysis presented here to gauge theories might present additional challenges. 
\vskip 20pt
\noindent{\Large{\bf Acknowledgments}}

\noindent We are very grateful to Hugh Osborn for his early involvement in the project, for many discussions and helpful comments, and for a careful reading of the manuscript.

\begin{appendix}	

\section{General results}
For a theory with couplings $g^I$, the corresponding $\beta$-functions are defined by
\be
\beta^I(g)=\mu\frac {d}{d\mu}g^I
\ee
and the $\beta$-functions in a new renormalisation group scheme defined by $g^{\prime I}(g)$ are given by
\be
\beta^{\prime I}(g')=\beta(g)_gg^{\prime I},
\ee
where for any vector $V$ in coupling space,
\be
V_g\equiv V^J\frac{\pa}{\pa g^J}.
\ee
We choose to parametrise the redefined coupling as 
\be
g'=e^{v_g}g.
\label{gpdef}
\ee
We then find using the easily proved result 
\be
f(e^{v_g}h)=e^{v_g}f(h)
\ee
that
\be
\beta'(g)=e^{-v_g}\beta_g(g)e^{v_g}g.
\ee
Then using 
\be
[v_g,V_g]=({\cal L}_vV)_g, \quad {\cal L}_vV=v_gV-V_gv,
\ee
together with
\be
e^ABe^{-A}=B+[A,B]+\tfrac12[A,[A,B]]+\ldots
\ee
we find 
\be
\beta'(g)=\sum_{n=0}^{\infty}\frac{(-1)^n}{n!}{{\cal L}_v}^n\beta(g).
\ee
For our purposes it is useful to use this result in the form
\be
\delta\beta(g)=\beta'(g)-\beta(g)=-{\cal L}_v\hhbet,
\label{delba}
\ee
where 
\be
\hhbet=\beta-\tfrac{1}{2!}{\cal L}_v\beta+\tfrac{1}{3!}{{\cal L}_v}^2\beta+\ldots
=\beta-\tfrac{1}{2}{\cal L}_v(\beta-\tfrac{1}{3}{\cal L}_v(\beta-\ldots))
\label{delbb}
\ee

\section{Symmetric Hopf co-product}
In this Appendix we give the full results for the co-commutative expressions on the right-hand sides of Eqs.~\eqref{hopffour}, \eqref{hopffourq}, \eqref{hopffive} and \eqref{hopffiveq}. For the combinations correponding to four-loop linear invariants in Eq.~\eqref{hopffour}, we have
\begin{align}
C_1^{(4)L}=&0,\nn
C_2^{(4)L}=&2g_{\lambda}\!{}^1\otimes_s g_{\lambda}\!{}^{3a}+2g_{\lambda}\!{}^1\otimes_s g_{\lambda}\!{}^{3c}+4g_{\lambda}\!{}^2\otimes g_{\lambda}\!{}^2+g_{\lambda}\!{}^{2R}\otimes g_{\lambda}\!{}^{2R},\nn
C_3^{(4)L}=&g_{\lambda}\!{}^1\otimes_s g_{\lambda}\!{}^{3a}+g_{\lambda}\!{}^1\otimes_s g_{\lambda}\!{}^{3c}+2g_{\lambda}\!{}^1\otimes_s g_{\lambda}\!{}^{3e}+g_{\lambda}\!{}^2\otimes_s g_{\lambda}\!{}^{2R}+2g_{\lambda}\!{}^2\otimes g_{\lambda}\!{}^2,\nn
C_4^{(4)L}=&2g_{\lambda}\!{}^1\otimes_s g_{\lambda}\!{}^{3e}+2g_{\lambda}\!{}^2\otimes_s g_{\lambda}\!{}^{2R}
-4g_{\lambda}\!{}^2\otimes g_{\lambda}\!{}^2
-2g_{\lambda}\!{}^1\otimes_s g_{\lambda}\!{}^1g_{\lambda}\!{}^2-(g_{\lambda}\!{}^1)^2\otimes_s g_{\lambda}\!{}^{2R},\nn
C_5^{(4)L}=&2g_{\lambda}\!{}^1\otimes_s g_{\lambda}\!{}^{3b}+g_{\lambda}\!{}^1\otimes_s g_{\gamma}\!{}^{3}+2g_{\lambda}\!{}^2\otimes_s g_{\gamma}\!{}^{2},\nn
C_6^{(4)L}=&g_{\lambda}\!{}^1\otimes_s g_{\lambda}\!{}^{3a}+g_{\lambda}\!{}^1\otimes_s g_{\lambda}\!{}^{3d}-g_{\lambda}\!{}^1\otimes_s g_{\lambda}\!{}^{3bR}+g_{\lambda}\!{}^2\otimes g_{\lambda}\!{}^2,\nn
C_7^{(4)L}=&2g_{\lambda}\!{}^1\otimes_s g_{\lambda}\!{}^{3c}-2g_{\lambda}\!{}^1\otimes_s g_{\lambda}\!{}^{3d}+g_{\lambda}\!{}^1\otimes_s g_{\lambda}\!{}^{3e}
+g_{\lambda}\!{}^1\otimes_s g_{\lambda}\!{}^{3aR}
+g_{\lambda}\!{}^1\otimes_s g_{\lambda}\!{}^{3bR}\nn&+2g_{\lambda}\!{}^2\otimes_s g_{\lambda}\!{}^{2R}
-2g_{\lambda}\!{}^2\otimes g_{\lambda}\!{}^2+g_{\lambda}\!{}^{2R}\otimes g_{\lambda}\!{}^{2R},\nn
C_8^{(4)L}=&g_{\lambda}\!{}^2\otimes_s g_{\lambda}\!{}^{2R}-4 g_{\lambda}\!{}^2\otimes g_{\lambda}\!{}^2
-2g_{\lambda}\!{}^1\otimes_s g_{\lambda}\!{}^1g_{\lambda}\!{}^2
+g_{\lambda}\!{}^1\otimes_s g_{\lambda}\!{}^1g_{\lambda}\!{}^{2R},\nn
C_9^{(4)L}=&2g_{\lambda}\!{}^2\otimes_sg_{\gamma}\!{}^{2}+g_{\lambda}\!{}^1\otimes_s g_{\lambda}\!{}^{3b}+\tfrac12g_{\lambda}\!{}^1\otimes_s g_{\gamma}\!{}^{3}+g_{\lambda}\!{}^1\otimes_s g_{\lambda}\!{}^1g_{\gamma}\!{}^{2},\nn
C_{10}^{(4)L}=&g_{\lambda}\!{}^1\otimes_s g_{\lambda}\!{}^{3f},\nn
C_{11}^{(4)L}=&2g_{\lambda}\!{}^{2R}\otimes_s g_{\gamma}\!{}^{2}+2g_{\lambda}\!{}^1\otimes_s g_{\lambda}\!{}^{3b}+g_{\lambda}\!{}^1\otimes_s g_{\gamma}\!{}^{3}+2g_{\lambda}\!{}^1\otimes_s g_{\lambda}\!{}^1g_{\gamma}\!{}^{2},\nn
C_{12}^{(4)L}=&2g_{\lambda}\!{}^1\otimes_s g_{\lambda}\!{}^{3aR}-2g_{\lambda}\!{}^1\otimes_s g_{\lambda}\!{}^{3bR}+2g_{\lambda}\!{}^2\otimes g_{\lambda}\!{}^2
+3g_{\lambda}\!{}^{2R}\otimes g_{\lambda}\!{}^{2R}-2g_{\lambda}\!{}^2\otimes_s g_{\lambda}\!{}^{2R},\nn
C_{13}^{(4)L}=&2g_{\lambda}\!{}^2\otimes g_{\lambda}\!{}^2
- g_{\lambda}\!{}^{2R}\otimes g_{\lambda}\!{}^{2R}+2g_{\lambda}\!{}^1\otimes_s g_{\lambda}\!{}^1g_{\lambda}\!{}^2+(g_{\lambda}\!{}^1)^2\otimes_s g_{\lambda}\!{}^{2R},\nn
C_{14}^{(4)L}=&4g_{\lambda}\!{}^2\otimes g_{\lambda}\!{}^2-g_{\lambda}\!{}^2\otimes_s g_{\lambda}\!{}^{2R}-2g_{\lambda}\!{}^1\otimes_s g_{\lambda}\!{}^{3e}
-g_{\lambda}\!{}^1\otimes_s g_{\lambda}\!{}^1g_{\lambda}\!{}^{2R}
-2(g_{\lambda}\!{}^1)^2\otimes_s g_{\lambda}\!{}^2.
\end{align}
For the combinations correponding to four-loop quadratic invariants in Eq.~\eqref{hopffourq}, we have
\begin{align}
C_1^{(4)Q}=&2g_{\lambda}\!{}^1\otimes_s g_{\gamma}\!{}^{4c}-g_{\lambda}\!{}^1\otimes_s g_{\gamma}\!{}^{4d}
+g_{\gamma}\!{}^{2}\otimes_s g_{\lambda}\!{}^{3d}
+2g_{\lambda}\!{}^2\otimes_s g_{\lambda}\!{}^{3b}+2g_{\lambda}\!{}^1\otimes_s g_{\lambda}\!{}^2g_{\gamma}\!{}^{2}\nn&+2g_{\lambda}\!{}^2\otimes_s g_{\lambda}\!{}^1g_{\gamma}\!{}^{2}
+2g_{\lambda}\!{}^1g_{\lambda}\!{}^2\otimes_s g_{\gamma}\!{}^{2}-(g_{\lambda}\!{}^1)^2\otimes_s g_{\lambda}\!{}^{3b}-(g_{\lambda}\!{}^1)^3\otimes_sg_{\gamma}\!{}^{2},\nn
C_2^{(4)Q}=&g_{\lambda}\!{}^1\otimes_s g_{\lambda}\!{}^{4eR}-g_{\lambda}\!{}^{2R}\otimes_s g_{\lambda}\!{}^{3b}
-g_{\gamma}\!{}^{2}\otimes_s g_{\lambda}\!{}^{3bR}
+(g_{\lambda}\!{}^1)^2\otimes_s g_{\lambda}\!{}^{3b}-g_{\lambda}\!{}^1\otimes_s g_{\lambda}\!{}^1g_{\lambda}\!{}^{3b}\nn
&-g_{\lambda}\!{}^1g_{\gamma}\!{}^{2}\otimes_s g_{\lambda}\!{}^2-g_{\lambda}\!{}^2g_{\gamma}\!{}^{2}\otimes_s g_{\lambda}\!{}^1-g_{\lambda}\!{}^{2R}g_{\gamma}\!{}^{2}\otimes_s g_{\lambda}\!{}^1-g_{\lambda}\!{}^1g_{\gamma}\!{}^{2}\otimes_s (g_{\lambda}\!{}^1)^2,\nn
C_3^{(4)Q}=&g_{\lambda}\!{}^1\os(g_{\lambda}\!{}^{4dR}-2g_{\lambda}\!{}^{4gR})+2g_{\lambda}\!{}^2
\os g_{\lambda}\!{}^{3aR}-g_{\lambda}\!{}^{2R}\os g_{\lambda}\!{}^{3d}\nn
&+(g_{\lambda}\!{}^1)^2\os g_{\lambda}\!{}^{3c}+g_{\lambda}\!{}^1\os g_{\lambda}\!{}^1g_{\lambda}\!{}^{3c}-2(g_{\lambda}\!{}^1)^2\os g_{\lambda}\!{}^{3e}
-2g_{\lambda}\!{}^1\os g_{\lambda}\!{}^1g_{\lambda}\!{}^{3e}\nn
&+g_{\lambda}\!{}^{2R}\os g_{\lambda}\!{}^1g_{\lambda}\!{}^{2R}+4g_{\lambda}\!{}^{2R}\os g_{\lambda}\!{}^1g_{\lambda}\!{}^{2}-2g_{\lambda}\!{}^{2}\os g_{\lambda}\!{}^1g_{\lambda}\!{}^{2R}+4g_{\lambda}\!{}^{1}\os g_{\lambda}\!{}^2g_{\lambda}\!{}^{2R}\nn
&+2g_{\lambda}\!{}^1\os g_{\lambda}\!{}^1g_{\lambda}\!{}^{3d}-2g_{\lambda}\!{}^1\os g_{\lambda}\!{}^1g_{\lambda}\!{}^{3aR}-2(g_{\lambda}\!{}^1)^2\os g_{\lambda}\!{}^1g_{\lambda}\!{}^{2}+5(g_{\lambda}\!{}^1)^2\os g_{\lambda}\!{}^1g_{\lambda}\!{}^{2R}.
\end{align}
For the combination corresponding to the five-loop linear invariant in Eq.~\eqref{hopffive}, we have
\begin{align}
C_1^{(5)L}=&-2g_{\gamma}\!{}^{2}\otimes_s g_{\lambda}\!{}^{3aR}+4g_{\gamma}\!{}^{2}\otimes_s g_{\lambda}\!{}^{3bR}
-2g_{\lambda}\!{}^1\otimes_s g_{\lambda}\!{}^{4eR}-g_{\lambda}\!{}^{2R}\otimes_s g_{\gamma}\!{}^{3}\nn
&+2g_{\gamma}\!{}^{2}\otimes_s g_{\lambda}\!{}^1g_{\lambda}\!{}^{2R}-4g_{\lambda}\!{}^1g_{\lambda}\!{}^2\otimes_s g_{\gamma}\!{}^{2}+(g_{\lambda}\!{}^1)^2\otimes_s g_{\gamma}\!{}^{3}+2g_{\lambda}\!{}^1\otimes_s g_{\lambda}\!{}^1g_{\lambda}\!{}^{3b}
\end{align}
Finally, for the combinations corresponding to five-loop quadratic invariants in Eq.~\eqref{hopffiveq}, we have
\begin{align}
C_1^{(5)Q}=&g_{\lambda}\!{}^1\os g_{\lambda}\!{}^{5a}+g_{\gamma}\!{}^2\os g_{\lambda}\!{}^{4p}+g_{\lambda}\!{}^{3b}\os g_{\lambda}\!{}^{3f}\nn
&+g_{\gamma}\!{}^2\os g_{\lambda}\!{}^1g_{\lambda}\!{}^{3f}+g_{\lambda}\!{}^{3f}\os
g_{\lambda}\!{}^1g_{\gamma}\!{}^2+g_{\lambda}\!{}^1\os g_{\gamma}\!{}^2g_{\lambda}\!{}^{3f},\nn
C_2^{(5)Q}=&g_{\lambda}\!{}^1\otimes_s(g_{\gamma}\!{}^{5g}-2g_{\gamma}\!{}^{5k})-g_{\gamma}\!{}^2\otimes_sg_{\lambda}\!{}^{4c}+g_{\gamma}\!{}^3\otimes_sg_{\lambda}\!{}^{3b}\nn&
-(g_{\gamma}\!{}^3+2g_{\lambda}\!{}^{3b})\otimes_s g_{\lambda}\!{}^1g_{\gamma}\!{}^2
+ 2g_{\lambda}\!{}^1g_{\gamma}\!{}^2\otimes  g_{\lambda}\!{}^1g_{\gamma}\!{}^2,\nn
C_3^{(5)Q}=&g_{\lambda}\!{}^1\otimes_s(g_{\lambda}\!{}^{5aR}-g_{\lambda}\!{}^{5bR})+g_{\gamma}\!{}^2\otimes_s(-g_{\lambda}\!{}^{4aR}+g_{\lambda}\!{}^{4cR})\nn
&+g_{\lambda}\!{}^{3b}\otimes_s(-g_{\lambda}\!{}^{3aR}+g_{\lambda}\!{}^{3bR})
+(g_{\lambda}\!{}^1)^2\otimes_s g_{\lambda}\!{}^{4eR}+g_{\lambda}\!{}^1\otimes_sg_{\lambda}\!{}^1g_{\lambda}\!{}^{4eR}\nn
&-2g_{\lambda}\!{}^1\otimes_sg_{\lambda}\!{}^{3b}g_{\lambda}\!{}^{2R}-g_{\lambda}\!{}^1g_{\lambda}\!{}^{3b}\otimes_sg_{\lambda}\!{}^{2R}+g_{\lambda}\!{}^{3b}\otimes_sg_{\lambda}\!{}^1(g_{\lambda}\!{}^{2R}-g_{\lambda}\!{}^2)\nn
&+g_{\lambda}\!{}^1\otimes_sg_{\lambda}\!{}^2g_{\lambda}\!{}^{3b}-2g_{\lambda}\!{}^1g_{\gamma}\!{}^2\otimes_sg_{\lambda}\!{}^{3aR}-3g_{\lambda}\!{}^1\otimes_sg_{\lambda}\!{}^{3aR}g_{\gamma}\!{}^2+2g_{\lambda}\!{}^1\otimes_sg_{\lambda}\!{}^{3bR}g_{\gamma}\!{}^2\nn
&+g_{\lambda}\!{}^1g_{\gamma}\!{}^2\otimes_sg_{\lambda}\!{}^{3bR}-3g_{\lambda}\!{}^{2R}\otimes_sg_{\lambda}\!{}^{2R}g_{\gamma}\!{}^2+g_{\lambda}\!{}^2\otimes_sg_{\lambda}\!{}^{2R}g_{\gamma}\!{}^2+g_{\lambda}\!{}^{2R}\otimes_sg_{\lambda}\!{}^2g_{\gamma}\!{}^2\nn
&-g_{\lambda}\!{}^1g_{\lambda}\!{}^{2R}\otimes_sg_{\lambda}\!{}^1g_{\gamma}\!{}^2-(g_{\lambda}\!{}^1)^2\otimes_sg_{\lambda}\!{}^{2R}g_{\gamma}\!{}^2
+g_{\lambda}\!{}^1g_{\lambda}\!{}^2\otimes_sg_{\lambda}\!{}^1g_{\gamma}\!{}^2,\nn
C_{4}^{(5)Q}=&g_{\lambda}\!{}^1\os(g_{\gamma}\!{}^{5d}-g_{\gamma}\!{}^{5g})-2g_{\gamma}\!{}^2\os G_J
-g_{\gamma}\!{}^{3}\otimes g_{\gamma}\!{}^{3}\nn
&-g_{\lambda}\!{}^1\otimes_sg_{\gamma}\!{}^2g_{\gamma}\!{}^3
-2(g_{\lambda}\!{}^1\otimes_sg_{\lambda}\!{}^{3b}g_{\gamma}\!{}^2+g_{\lambda}\!{}^{3b}\otimes_sg_{\lambda}\!{}^1g_{\gamma}\!{}^2)+2g_{\lambda}\!{}^1\otimes_sg_{\lambda}\!{}^1(g_{\gamma}\!{}^2)^2\nn
&+2g_{\lambda}\!{}^1g_{\gamma}\!{}^2\otimes g_{\lambda}\!{}^1g_{\gamma}\!{}^2,\nn
C_5^{(5)Q}=&g_{\lambda}\!{}^1\otimes_s(2g_{\lambda}\!{}^{5bR}-g_{\lambda}\!{}^{5cR})
+g_{\lambda}\!{}^{2R}\otimes_s
G_J\nn
&+g_{\gamma}\!{}^2\otimes_s(g_{\lambda}\!{}^{4aR}-2g_{\lambda}\!{}^{4cR})+g_{\gamma}\!{}^3\otimes_sg_{\lambda}\!{}^{3bR}+g_{\lambda}\!{}^1\os g_{\lambda}\!{}^1g_{\lambda}\!{}^{4c}-(g_{\lambda}\!{}^1)^2\os g_{\lambda}\!{}^{4c}\nn
&-(g_{\lambda}\!{}^1)^2\os g_{\lambda}\!{}^{4e}
+(g_{\lambda}\!{}^1)^2\os g_{\lambda}\!{}^{4j}-(g_{\lambda}\!{}^1)^2\os g_{\lambda}\!{}^{4eR}
-2g_{\lambda}\!{}^1\os g_{\lambda}\!{}^1g_{\lambda}\!{}^{4eR}\nn
&+2g_{\lambda}\!{}^2\os g_{\lambda}\!{}^1g_{\lambda}\!{}^{3b}
+2g_{\lambda}\!{}^1g_{\lambda}\!{}^2\os g_{\lambda}\!{}^{3b}+2g_{\lambda}\!{}^1\otimes_sg_{\lambda}\!{}^{2R}g_{\lambda}\!{}^{3b}+g_{\lambda}\!{}^{2R}\otimes_sg_{\lambda}\!{}^1
g_{\lambda}\!{}^{3b}\nn&+2g_{\lambda}\!{}^{3b}\otimes_sg_{\lambda}\!{}^1g_{\lambda}\!{}^{2R}
+2g_{\lambda}\!{}^{3aR}\os g_{\lambda}\!{}^1g_{\gamma}\!{}^2+2g_{\gamma}\!{}^2g_{\lambda}\!{}^{3aR}\os g_{\lambda}\!{}^1
-2g_{\gamma}\!{}^2g_{\lambda}\!{}^{3bR}\os g_{\lambda}\!{}^1\nn&-2g_{\lambda}\!{}^{3bR}\os g_{\lambda}\!{}^1g_{\gamma}\!{}^2+2g_{\lambda}\!{}^1g_{\lambda}\!{}^{3bR}\os g_{\gamma}\!{}^2
+g_{\lambda}\!{}^1\otimes_sg_{\lambda}\!{}^{2R}g_{\gamma}\!{}^3+g_{\lambda}\!{}^2\os g_{\lambda}\!{}^1g_{\gamma}\!{}^3\nn&+g_{\lambda}\!{}^1\os g_{\lambda}\!{}^2g_{\gamma}\!{}^3-2g_{\lambda}\!{}^{2R}\otimes_sg_{\lambda}\!{}^{2}g_{\gamma}\!{}^2
-2g_{\lambda}\!{}^2\otimes_sg_{\lambda}\!{}^{2R}g_{\gamma}\!{}^2\nn&+3g_{\gamma}\!{}^2g_{\lambda}\!{}^{2R}
\os g_{\lambda}\!{}^{2R}
-2(g_{\lambda}\!{}^1)^3\os g_{\lambda}\!{}^{3b}+2(g_{\lambda}\!{}^1)^2\os g_{\lambda}\!{}^{3b}g_{\lambda}\!{}^1+2g_{\lambda}\!{}^1\os g_{\lambda}\!{}^{3b}(g_{\lambda}\!{}^1)^2\nn
&+(g_{\lambda}\!{}^1)^2\os g_{\lambda}\!{}^1g_{\gamma}\!{}^3
+2(g_{\lambda}\!{}^1)^2\os g_{\lambda}\!{}^2g_{\gamma}\!{}^2+(g_{\lambda}\!{}^1)^2\os g_{\lambda}\!{}^{2R}g_{\gamma}\!{}^2+2(g_{\lambda}\!{}^1)^3\os g_{\lambda}\!{}^1g_{\gamma}\!{}^2,\nn
C_{6}^{(5)Q}=&g_{\lambda}\!{}^1\os(g_{\gamma}\!{}^{5c}+g_{\gamma}\!{}^{5e}
-2g_{\gamma}\!{}^{5h})
-g_{\gamma}\!{}^2\os(g_{\lambda}\!{}^{4b}-2g_{\lambda}\!{}^{4d}-g_{\lambda}\!{}^{4aR}+2g_{\lambda}\!{}^{4cR})\nn
&+g_{\lambda}\!{}^{2R}\os G_J
-g_{\gamma}\!{}^3\os (g_{\lambda}\!{}^{3c}-g_{\lambda}\!{}^{3d})
\nn
&-(g_{\lambda}\!{}^1)^2\os g_{\lambda}\!{}^{4c}-(g_{\lambda}\!{}^1)^2\os g_{\lambda}\!{}^{4e}+(g_{\lambda}\!{}^1)^2\os g_{\lambda}\!{}^{4j}
-2g_{\lambda}\!{}^1\os g_{\lambda}\!{}^1g_{\lambda}\!{}^{4eR}\nn
&-(g_{\lambda}\!{}^1)^2\os g_{\lambda}\!{}^{4eR}+2g_{\lambda}\!{}^{1}\os g_{\lambda}\!{}^{2R}g_{\lambda}\!{}^{3b}+g_{\lambda}\!{}^{2R}\os g_{\lambda}\!{}^1g_{\lambda}\!{}^{3b}
+2g_{\lambda}\!{}^{3b}\os g_{\lambda}\!{}^1g_{\lambda}\!{}^{2R}\nn
&-2g_{\lambda}\!{}^1\os (g_{\lambda}\!{}^{3c}-g_{\lambda}\!{}^{3d}+g_{\lambda}\!{}^{3bR})g_{\gamma}\!{}^2+2g_{\lambda}\!{}^{3aR}\os g_{\lambda}\!{}^1g_{\gamma}\!{}^2+2g_{\lambda}\!{}^1\os g_{\gamma}\!{}^2g_{\lambda}\!{}^{3aR}\nn
&-2g_{\lambda}\!{}^{3bR}\os g_{\lambda}\!{}^1g_{\gamma}\!{}^2+g_{\gamma}\!{}^3\os g_{\lambda}\!{}^1g_{\lambda}\!{}^{2R}
+2g_{\lambda}\!{}^{2}\os g_{\lambda}\!{}^{2}g_{\gamma}\!{}^{2}
-2g_{\lambda}\!{}^{2R}\os g_{\lambda}\!{}^{2}g_{\gamma}\!{}^{2}\nn
&-2g_{\lambda}\!{}^{2}\os g_{\lambda}\!{}^{2R}g_{\gamma}\!{}^{2}+2g_{\lambda}\!{}^{2R}\os g_{\lambda}\!{}^{2R}g_{\gamma}\!{}^{2}
-2g_{\lambda}\!{}^1g_{\lambda}\!{}^{2R}\os g_{\lambda}\!{}^1g_{\gamma}\!{}^{2}
+(g_{\lambda}\!{}^1)^2\os g_{\lambda}\!{}^{2R}g_{\gamma}\!{}^2\nn
&-(g_{\lambda}\!{}^1)^3\os (g_{\gamma}\!{}^3+2g_{\lambda}\!{}^{3b})
+2(g_{\lambda}\!{}^1)^3\os g_{\lambda}\!{}^1g_{\gamma}\!{}^2,\nn
C_{7}^{(5)Q}=&g_{\lambda}\!{}^1\os(2g_{\gamma}\!{}^{5b}-g_{\gamma}\!{}^{5e})+g_{\gamma}\!{}^2\os(
g_{\lambda}\!{}^{4l}-g_{\lambda}\!{}^{4o}+2g_{\lambda}\!{}^{4r})\nn&
+(2g_{\lambda}\!{}^2-g_{\lambda}\!{}^{2R})\os G_J
+4g_{\lambda}\!{}^{3b}\os g_{\lambda}\!{}^{3e}
+g_{\lambda}\!{}^{3c}\os g_{\gamma}\!{}^3\nn
&+2g_{\lambda}\!{}^{3b}\os g_{\lambda}\!{}^1
g_{\lambda}\!{}^2+4g_{\lambda}\!{}^2\os g_{\lambda}\!{}^1
g_{\lambda}\!{}^{3b}+6g_{\lambda}\!{}^1\os g_{\lambda}\!{}^2g_{\lambda}\!{}^{3b}-g_{\lambda}\!{}^{3b}
\os g_{\lambda}\!{}^1g_{\lambda}\!{}^{2R}\nn
&-g_{\lambda}\!{}^1\os g_{\lambda}\!{}^{2R}g_{\lambda}\!{}^{3b}
-g_{\lambda}\!{}^{3c}\os g_{\lambda}\!{}^1g_{\gamma}\!{}^2
+g_{\lambda}\!{}^1\os g_{\gamma}\!{}^2g_{\lambda}\!{}^{3c}+4g_{\gamma}\!{}^2\os g_{\lambda}\!{}^1g_{\lambda}\!{}^{3e}\nn
&+4g_{\lambda}\!{}^{3e}\os g_{\lambda}\!{}^1g_{\gamma}\!{}^{2}
+4g_{\lambda}\!{}^1\os g_{\gamma}\!{}^2g_{\lambda}\!{}^{3e}
+2g_{\gamma}\!{}^3\os g_{\lambda}\!{}^1g_{\lambda}\!{}^2+2g_{\lambda}\!{}^1\os g_{\lambda}\!{}^2g_{\gamma}\!{}^3\nn
&+2g_{\lambda}\!{}^2\os g_{\lambda}\!{}^1g_{\gamma}\!{}^3
-g_{\gamma}\!{}^3\os g_{\lambda}\!{}^1g_{\lambda}\!{}^{2R}
+2g_{\lambda}\!{}^2\os g_{\lambda}\!{}^{2}g_{\gamma}\!{}^2
+g_{\lambda}\!{}^1g_{\lambda}\!{}^{2R}\os g_{\lambda}\!{}^1g_{\gamma}\!{}^2\nn
&+2g_{\lambda}\!{}^1g_{\lambda}\!{}^2\os g_{\lambda}\!{}^1g_{\gamma}\!{}^2
+4(g_{\lambda}\!{}^1)^2\os g_{\lambda}\!{}^2g_{\gamma}\!{}^2,
\nn
C_{8}^{(5)Q}=&g_{\lambda}\!{}^1\os(2g_{\gamma}\!{}^{5e}+g_{\gamma}\!{}^{5f}
-4g_{\gamma}\!{}^{5i})
-g_{\gamma}\!{}^2\os(2g_{\lambda}\!{}^{4b}-3g_{\lambda}\!{}^{4aR}+4g_{\lambda}\!{}^{4cR})\nn&
+(5g_{\lambda}\!{}^{2R}-8g_{\lambda}\!{}^2)\os G_J\nn
&+g_{\gamma}\!{}^{3}\os(2g_{\lambda}\!{}^{3a}-2g_{\lambda}\!{}^{3c}
-g_{\lambda}\!{}^{3aR}+2g_{\lambda}\!{}^{3bR})+2g_{\lambda}\!{}^1\os g_{\lambda}\!{}^1g_{\lambda}\!{}^{4c}\nn
&+2g_{\lambda}\!{}^1\os g_{\lambda}\!{}^1g_{\lambda}\!{}^{4e}
-2g_{\lambda}\!{}^1\os g_{\lambda}\!{}^1g_{\lambda}\!{}^{4j}+(g_{\lambda}\!{}^1)^2\os g_{\gamma}\!{}^{4c}+2g_{\lambda}\!{}^1\os g_{\lambda}\!{}^1g_{\gamma}\!{}^{4c}\nn
&+8g_{\lambda}\!{}^{3aR}\os g_{\lambda}\!{}^1g_{\gamma}\!{}^2+6g_{\lambda}\!{}^1\os g_{\gamma}\!{}^2g_{\lambda}\!{}^{3aR}-8g_{\lambda}\!{}^{3bR}\os g_{\lambda}\!{}^1g_{\gamma}\!{}^2-4g_{\lambda}\!{}^1\os g_{\gamma}\!{}^2g_{\lambda}\!{}^{3bR}\nn
&-5g_{\lambda}\!{}^{3a}\os g_{\lambda}\!{}^1g_{\gamma}\!{}^2-g_{\lambda}\!{}^1\os g_{\gamma}\!{}^2g_{\lambda}\!{}^{3a}-g_{\gamma}\!{}^2\os g_{\lambda}\!{}^1g_{\lambda}\!{}^{3a}-8g_{\lambda}\!{}^1\os g_{\lambda}\!{}^2g_{\lambda}\!{}^{3b}\nn
&-8g_{\lambda}\!{}^{3b}\os g_{\lambda}\!{}^1g_{\lambda}\!{}^2+5g_{\lambda}\!{}^1\os g_{\lambda}\!{}^{2R}g_{\lambda}\!{}^{3b}+5g_{\lambda}\!{}^{3b}\os g_{\lambda}\!{}^1g_{\lambda}\!{}^{2R}
-4g_{\lambda}\!{}^1\os g_{\gamma}\!{}^2g_{\lambda}\!{}^{3c}\nn 
&-g_{\lambda}\!{}^1\os g_{\lambda}\!{}^{2}g_{\gamma}\!{}^3-7g_{\gamma}\!{}^3\os g_{\lambda}\!{}^1g_{\lambda}\!{}^{2}-g_{\lambda}\!{}^2\os g_{\lambda}\!{}^1g_{\gamma}\!{}^3
+4g_{\gamma}\!{}^3\os g_{\lambda}\!{}^1g_{\lambda}\!{}^{2R}\nn
&-4g_{\lambda}\!{}^{2R}\os g_{\lambda}\!{}^{2} g_{\gamma}\!{}^2-4g_{\lambda}\!{}^{2}\os g_{\lambda}\!{}^{2R} g_{\gamma}\!{}^2+7g_{\lambda}\!{}^{2R}\os g_{\lambda}\!{}^{2R} g_{\gamma}\!{}^2
+(g_{\lambda}\!{}^1)^2\os g_{\lambda}\!{}^1g_{\gamma}\!{}^3\nn
&+2(g_{\lambda}\!{}^1)^2\os g_{\lambda}\!{}^1g_{\lambda}\!{}^{3b}
-2(g_{\lambda}\!{}^1)^2\os g_{\lambda}\!{}^2g_{\gamma}\!{}^2
+8g_{\lambda}\!{}^1g_{\lambda}\!{}^2\os g_{\lambda}\!{}^1g_{\gamma}\!{}^2\nn
&-6g_{\lambda}\!{}^1g_{\lambda}\!{}^{2R}\os g_{\lambda}\!{}^1g_{\gamma}\!{}^2
+2g_{\lambda}\!{}^1\os g_{\lambda}\!{}^1g_{\lambda}\!{}^2g_{\gamma}\!{}^2
-3g_{\lambda}\!{}^1\os g_{\lambda}\!{}^1g_{\lambda}\!{}^{2R}g_{\gamma}\!{}^2\nn
&-2(g_{\lambda}\!{}^1)^2\os (g_{\lambda}\!{}^1)^2g_{\gamma}\!{}^2.
\end{align}

\section{Differential operators for scheme changes}
Following the general considerations of Ref.~\cite{jo} we may define differential operators 
\be
Y=\sum_{l',s}(\delta_{l,s}Y^{\lambda l's}+\epsilon_{l's}Y^{\gamma l's}),
\ee
where
\be
Y^{\lambda l's}=\sum_{l,r}(c_{lr}\D^{\lambda lr,\lambda l's}+d_{lr}\D^{\gamma lr,\lambda l's}),
\quad Y^{\gamma l's}=\sum_{l,r}(c_{lr}\D^{\lambda lr,\gamma l's}+d_{lr}\D^{\gamma lr,\gamma l's}),
\label{Ydef}
\ee
which generate scheme changes according to
\be
\{c_{lr},d_{lr}\}\go \hbox{exp}(Y)\{c_{lr},d_{lr}\}.
\ee
Here $\{r,s\}$ label the $\beta$ or $\gamma$ function coefficients at each loop order $\{l,l'\}$. The operators 
$\D^{\lambda lr,\lambda l's}$, etc satisfy 
\be
\D^{\lambda lr,\lambda l's}=-\D^{\lambda l's,\lambda lr},\quad
\D^{\lambda lr,\gamma l's}=-\D^{\gamma l's,\lambda lr},\quad
\D^{\gamma lr,\gamma l's}=-\D^{\gamma l's,\gamma lr},
\ee
Scheme invariants are then determined as polynomial functions $F(\{c_{lr},d_{lr}\}$ such that
\be
Y^{\lambda lr}F=Y^{\gamma lr}F=0
\label{schinv}
\ee
for all $\lambda$, $r$. 

In the case of $\phi^4$ theory we find at lowest order
\begin{align}
\D^{\lambda1,\lambda2}=&-2\frac{\pa}{\pa c_{ 3a}}+2\frac{\pa}{\pa c_{ 3c}}+2\frac{\pa}{\pa c_{ 3d}},\nn
 \D^{\lambda{ 1},\lambda 2R}=&2\frac{\pa}{\pa c_{ 3a}}-2\frac{\pa}{\pa c_{ 3c}}+\frac{\pa}{\pa c_{ 3aR}}+2\frac{\pa}{\pa c_{3bR}},\nn
 \D^{\lambda{ 1},\gamma  2}=&-2\frac{\pa}{\pa c_{ 3b}}+6\frac{\pa}{\pa  d_{ 3}},
\label{Done}
\end{align}
and at next-to-leading order
\begin{align}
\D^{\lambda1,\lambda3a}=&4\frac{\pa}{\pa c_{4a}}+2\frac{\pa}{\pa c_{4b}}+2\frac{\pa}{\pa c_{4d}}-2\frac{\pa}{\pa c_{4f}},\nn
\D^{\lambda{ 1},\lambda 3b}=&6\frac{\pa}{\pa c_{4c}}-2\frac{\pa}{\pa c_{4e}}+\frac{\pa}{\pa c_{4j}},\nn
\D^{\lambda{ 1},\lambda 3c}=&-2\frac{\pa}{\pa c_{4b}}+2\frac{\pa}{\pa c_{4f}}+3\frac{\pa}{\pa c_{4g}}+2\frac{\pa}{\pa c_{4k}}+\frac{\pa}{\pa c_{4o}},\nn
\D^{\lambda{ 1},\lambda 3d}=&-2\frac{\pa}{\pa c_{4d}}-2\frac{\pa}{\pa c_{4h}}-2\frac{\pa}{\pa c_{4i}}+2\frac{\pa}{\pa c_{4o}}
+2\frac{\pa}{\pa c_{4r}},\nn
\D^{\lambda{ 1},\lambda 3e}=&-4\frac{\pa}{\pa c_{4a}}+2\frac{\pa}{\pa c_{4h}}+2\frac{\pa}{\pa c_{4i}}+\frac{\pa}{\pa c_{4k}}
+2\frac{\pa}{\pa c_{4l}}+\frac{\pa}{\pa c_{4r}},\nn
\D^{\lambda{ 1},\lambda 3f}=&\frac{\pa}{\pa c_{4p}}-\frac{\pa}{\pa c_{4q}},\nn
\D^{\lambda{ 1},\lambda 3aR}=&-2\frac{\pa}{\pa c_{4g}}+2\frac{\pa}{\pa c_{4aR}}+2\frac{\pa}{\pa c_{4cR}}+2\frac{\pa}{\pa c_{4fR}},\nn
\D^{\lambda{ 1},\lambda 3bR}=&2\frac{\pa}{\pa c_{4d}}-2\frac{\pa}{\pa c_{4k}}+4\frac{\pa}{\pa c_{4bR}}+\frac{\pa}{\pa c_{4cR}}+2\frac{\pa}{\pa c_{4dR}}+2\frac{\pa}{\pa c_{4gR}}-\frac{\pa}{\pa c_{4fR}},\nn
\D^{\lambda{ 1},\gamma  3}=&-2\frac{\pa}{\pa c_{4c}}+3\frac{\pa}{\pa  d_{4b}}+2\frac{\pa}{\pa  d_{4c}}+4\frac{\pa}{\pa  d_{4d}},\nn
\D^{\lambda{ 2},\lambda 2R}=&4\frac{\pa}{\pa c_{4a}}-2\frac{\pa}{\pa c_{4g}}-2\frac{\pa}{\pa c_{4l}}-\frac{\pa}{\pa c_{4o}}+2\frac{\pa}{\pa c_{4gR}}+2\frac{\pa}{\pa c_{4fR}},\nn
\D^{\lambda{ 2},\gamma  2}=&-2\frac{\pa}{\pa c_{4e}}-\frac{\pa}{\pa c_{4j}}+6\frac{\pa}{\pa  d_{4c}}+6\frac{\pa}{\pa  d_{4d}},\nn
\D^{\lambda{ 2R},\gamma  2}=&-2\frac{\pa}{\pa c_{4eR}}+6\frac{\pa}{\pa  d_{4b}},\nn
\label{Dtwo}
\end{align}
Note that here we suppress the label $r$ in the case of the one-loop $\beta$-function and the two-loop $\gamma$-function where there is only one coefficient.

The $Y^{\lambda l r}$ and $Y^{\gamma l r}$ defined according to Eq.~\eqref{Ydef} satisfy the commutation relations 
\begin{align}
[Y^{\lambda 1},Y^{\lambda 2}]=&-2Y^{\lambda 3a}+2Y^{\lambda 3c}+2Y^{\lambda 3e},\nn
 {[} Y^{\lambda 1}, Y^{\lambda 2R} ]=&2Y^{\lambda 3a}-2Y^{\lambda 3c}+Y^{\lambda 3aR}+2Y^{\lambda3bR},\nn
 {[} Y^{\lambda 1}, Y^{\gamma 2} ]=&-2Y^{\lambda 3b}+6Y^{\gamma 3},
\label{Yone}
\end{align}
and
\begin{align}
 [ Y^{\lambda 1}, Y^{\lambda 3a} ]=&4Y^{\lambda4a}+2Y^{\lambda4b}+2Y^{\lambda4d}-2Y^{\lambda4f},\nn
{[} Y^{\lambda 1}, Y^{\lambda 3b} ]=&6Y^{\lambda4c}-2Y^{\lambda4e}+Y^{\lambda4j},\nn
{[} Y^{\lambda 1}, Y^{\lambda 3c} ]=&-2Y^{\lambda4b}+2Y^{\lambda4f}+3Y^{\lambda4g}+2Y^{\lambda4k}+Y^{\lambda4o},\nn
{[} Y^{\lambda 1}, Y^{\lambda 3d} ]=&-2Y^{\lambda4d}-2Y^{\lambda4h}-2Y^{\lambda4i}+2Y^{\lambda4o}
+2Y^{\lambda4r},\nn
{[} Y^{\lambda 1}, Y^{\lambda 3e} ]=&-4Y^{\lambda4a}+2Y^{\lambda4h}+2Y^{\lambda4i}+Y^{\lambda4k}
+2Y^{\lambda4l}+Y^{\lambda4r},\nn
{[} Y^{\lambda 1}, Y^{\lambda 3f} ]=&Y^{\lambda4p}-Y^{\lambda4q},\nn
{[} Y^{\lambda 1}, Y^{\lambda 3aR} ]=&-2Y^{\lambda4g}+2Y^{\lambda4aR}+2Y^{\lambda4cR}+2Y^{\lambda4fR},\nn
{[} Y^{\lambda 1}, Y^{\lambda 3bR} ]=&2Y^{\lambda4d}-2Y^{\lambda4k}+4Y^{\lambda4bR}+Y^{\lambda4cR}+2Y^{\lambda4dR}+2Y^{\lambda4gR}-Y^{\lambda4fR},\nn
{[} Y^{\lambda 1}, Y^{\gamma 3} ]=&-4Y^{\lambda4c}+3Y^{\gamma4b}+2Y^{\gamma4c}+4Y^{\gamma4d},\nn
{[} Y^{\lambda 2}, Y^{\lambda 2R} ]=&4Y^{\lambda4a}-2Y^{\lambda4g}-2Y^{\lambda4l}-Y^{\lambda4o}+2Y^{\lambda4gR}+2Y^{\lambda4fR},\nn
{[} Y^{\lambda 2}, Y^{\gamma 2} ]=&-2Y^{\lambda4e}-Y^{\lambda4j}+6Y^{\gamma4c}+6Y^{\gamma4d},\nn
{[} Y^{\lambda 2R}, Y^{\gamma 2} ]=&-2Y^{\lambda4eR}+6Y^{\gamma4b},\nn
\label{Ytwo}
\end{align}
Note that the structure constants appearing in Eqs.~\eqref{Yone}, \eqref{Ytwo} are the same as those in Eqs.~\eqref{Done}, \eqref{Dtwo}, which is a consequence of the Jacobi identities following from the associativity of the graph insertion process as described in Ref.~\cite{jo}.
At the following order we have
\begin{align}
\D^{\lambda3a,\gamma2}=&3\frac{\pa}{\pa d_{5i}},\nn
\D^{\lambda3b,\gamma2}=&3\frac{\pa}{\pa d_{5k}},\nn
\D^{\lambda3c,\gamma2}=&6\frac{\pa}{\pa d_{5e}}+3\frac{\pa}{\pa d_{5j}},\nn
\D^{\lambda3d,\gamma2}=&6\frac{\pa}{\pa d_{5h}},\nn
\D^{\lambda3e,\gamma2}=&12\frac{\pa}{\pa d_{5b}}+6\frac{\pa}{\pa d_{5c}}+3\frac{\pa}{\pa d_{5h}},\nn
\D^{\lambda3f,\gamma2}=&2\frac{\pa}{\pa d_{5a}},\nn
\D^{\lambda3aR,\gamma2}=&6\frac{\pa}{\pa d_{5f}}-2\frac{\pa}{\pa c_{5aR}},\nn
\D^{\lambda3bR,\gamma2}=&3\frac{\pa}{\pa d_{5i}}+3\frac{\pa}{\pa d_{5j}}-2\frac{\pa}{\pa c_{5bR}},\nn
\D^{\lambda1,\gamma4a}=&6\frac{\pa}{\pa d_{5d}}+2\frac{\pa}{\pa d_{5g}}+2\frac{\pa}{\pa d_{5k}}\nn
\D^{\lambda1,\gamma4b}=&4\frac{\pa}{\pa d_{5f}}+2\frac{\pa}{\pa d_{5i}}+2\frac{\pa}{\pa d_{5j}},\nn
\D^{\lambda1,\gamma4c}=&4\frac{\pa}{\pa d_{5b}}+2\frac{\pa}{\pa d_{5e}}+2\frac{\pa}{\pa d_{5h}}
+\frac{\pa}{\pa d_{5i}},\nn
\D^{\lambda1,\gamma4d}=&4\frac{\pa}{\pa d_{5c}}+2\frac{\pa}{\pa d_{5h}}+2\frac{\pa}{\pa d_{5j}},\nn
\D^{\lambda1,\lambda4eR}=&\frac{\pa}{\pa c_{5aR}}+2\frac{\pa}{\pa c_{5bR}}+6\frac{\pa}{\pa c_{5cR}},\nn
\D^{\lambda2,\gamma3}=&4\frac{\pa}{\pa d_{5b}}+4\frac{\pa}{\pa d_{5c}}+2\frac{\pa}{\pa d_{5h}}+2\frac{\pa}{\pa d_{5i}}+\frac{\pa}{\pa d_{5j}},\nn
\D^{\lambda2R,\gamma3}=&2\frac{\pa}{\pa d_{5e}}+3\frac{\pa}{\pa d_{5f}}+2\frac{\pa}{\pa d_{5j}}-2\frac{\pa}{\pa c_{5cR}},\nn
\D^{\gamma2,\gamma3}=&\frac{\pa}{\pa d_{5g}}+2\frac{\pa}{\pa d_{5k}}-3\frac{\pa}{\pa d_{5d}},
\label{Dthree}
\end{align}
with, correspondingly, the commutation relations
\begin{align}
 {[}Y^{\lambda3a},Y^{\gamma2} ]=&3 Y^{\gamma 5i},\nn
 {[}Y^{\lambda3b},Y^{\gamma2} ]=&3 Y^{\gamma 5k},\nn
 {[}Y^{\lambda3d},Y^{\gamma2} ]=&6 Y^{\gamma 5e}+3 Y^{\gamma 5j},\nn
 {[}Y^{\lambda3d},Y^{\gamma2} ]=&6 Y^{\gamma 5h},\nn
 {[}Y^{\lambda3e},Y^{\gamma2} ]=&12 Y^{\gamma 5b}+6 Y^{\gamma 5d}+3 Y^{\gamma 5h},\nn
 {[}Y^{\lambda3f},Y^{\gamma2} ]=&2 Y^{\gamma 5a},\nn
 {[}Y^{\lambda3aR},Y^{\gamma2} ]=&6 Y^{\gamma 5f}-2Y^{\gamma 5aR},\nn
 {[}Y^{\lambda3bR},Y^{\gamma2} ]=&3 Y^{\gamma 5i}+3 Y^{\gamma 5j}-2Y^{\lambda 5bR},\nn
 {[}Y^{\lambda1},Y^{\gamma4a} ]=&6 Y^{\gamma 5d}+2 Y^{\gamma 5g}+2 Y^{\gamma 5k}\nn
 {[}Y^{\lambda1},Y^{\gamma4b} ]=&4 Y^{\gamma 5f}+2 Y^{\gamma 5i}+2 Y^{\gamma 5j},\nn
 {[}Y^{\lambda1},Y^{\gamma4c} ]=&4 Y^{\gamma 5b}+2 Y^{\gamma 5e}+2 Y^{\gamma 5h}
+2 Y^{\gamma 5i},\nn
 {[}Y^{\lambda1},Y^{\gamma4d} ]=& Y^{\gamma 5c}+2 Y^{\gamma 5h}+2 Y^{\gamma 5j},\nn
 {[}Y^{\lambda1},Y^{\lambda4eR} ]=&Y^{\lambda 5aR}+2Y^{\lambda 5bR}+6Y^{\lambda 5cR},\nn
 {[}Y^{\lambda2},Y^{\gamma3} ]=&4 Y^{\gamma 5b}+4 Y^{\gamma 5c}+2 Y^{\gamma 5h}+2 Y^{\gamma 5i}+ Y^{\gamma 5j},\nn
 {[}Y^{\lambda2R},Y^{\gamma3} ]=&2 Y^{\gamma 5e}+3 Y^{\gamma 5f}+2 Y^{\gamma 5j}-2Y^{\lambda 5cR},\nn
 {[}Y^{\gamma2},Y^{\gamma3} ]=& Y^{\gamma 5g}+2 Y^{\gamma 5k}-3 Y^{\gamma 5d},\nn
\end{align}
It is readily verified using Eqs.~\eqref{Ydef}, \eqref{Done}, \eqref{Dtwo}, \eqref{Dthree} that the linear and quadratic invariants constructed in previous sections satisfy Eq.~\eqref{schinv}.
\end{appendix}

\end{document}